\documentclass[11pt]{article}

\usepackage{amsthm}
\usepackage{amsmath}
\usepackage{amssymb}
\usepackage{graphicx}
\usepackage[round]{natbib}
\usepackage{bbm}
\usepackage{textcomp}
\usepackage{url}
\usepackage[margin=2cm]{geometry}
\usepackage[usenames,dvipsnames]{xcolor}
\usepackage{subcaption}
\usepackage{mathtools}
\usepackage{mathrsfs}
\usepackage{extarrows}
\usepackage{comment}
\usepackage{enumitem} 
\usepackage{stmaryrd, scalerel} 

\usepackage{xr-hyper}
\usepackage{hyperref}
\hypersetup{colorlinks=true,linkcolor=blue,filecolor=blue, citecolor = blue}

\newcommand{\A}{\mathcal{A}} 
\newcommand{\F}{\mathcal{F}} 
\newcommand{\C}{\mathcal{C}} 
\newcommand{\G}{\mathcal{G}} 
\renewcommand{\L}{\mathcal{L}} 

\newcommand{\X}{\mathcal{X}} 

\newcommand{\prob}{\mathbb{P}} 
\newcommand{\E}{\mathbb{E}} 
\newcommand{\R}{\mathbb{R}} 
\newcommand{\N}{\mathbb{N}} 

\newcommand{\one}{\mathbbm{1}} 

\newcommand{\diff}{\,\mathrm{d}} 

\newcommand*{\Swarrow}{\rotatebox[origin=c]{-45}{\(\Downarrow\)}} 
\newcommand*{\Searrow}{\rotatebox[origin=c]{45}{\(\Downarrow\)}} 
\newcommand*{\Sehookarrow}{\rotatebox[origin=c]{135}{\(\hookleftarrow\)}} 

\DeclareMathOperator*{\argmin}{arg\,min} %
\DeclarePairedDelimiter{\pare}{(}{)} 

\newcommand{\supp}{\operatorname{supp}}

\renewcommand{\subset}{\subseteq}

\theoremstyle{plain}  
\newtheorem{thm}{Theorem} 
\newtheorem{lemma}[thm]{Lemma} 
\newtheorem{prop}[thm]{Proposition} 
\newtheorem{cor}[thm]{Corollary} 

\theoremstyle{definition} 
\newtheorem{defn}{Definition}

\newtheorem{exmp}{Example}

\theoremstyle{remark} 
\newtheorem{rem}{Remark}[section]

\newcommand{\johanna}[1]{{\color{Emerald} Johanna:``#1''\\}}

\makeatletter
\let\@fnsymbol\@arabic
\makeatother

\begin{document}

\title{In-sample calibration yields conformal calibration guarantees}
\date{}
\author{Sam Allen\thanks{Corresponding author. Address: Institute of Statistics, Karlsruhe Institute of Technology, Bl{\"u}cherstrasse 17, 76185 Karlsruhe, Germany. Email: \url{sam.allen@kit.edu}} \and Georgios Gavrilopoulos\thanks{Seminar for Statistics, ETH Zurich, Zurich, Switzerland.} \and Alexander Henzi\thanks{Department of Statistics and Data Science, Tsinghua University, China} \and Gian-Reto Kleger\thanks{HOCH Health Ostschweiz, Kantonsspital St. Gallen, Klinik f{\"u}r Intensivmedizin, St. Gallen, Switzerland.} \and Johanna Ziegel\footnotemark[2]}

\maketitle

\vspace{-0.5cm}

\begin{abstract}
    Conformal prediction produces a set of predictions that has out-of-sample calibration guarantees by construction, under the assumption of exchangeability. In this work, we study the calibration properties of conformal predictive systems, which issue sets of predictive distributions for real-valued outcomes. We demonstrate that conformal predictive systems implicitly exploit prediction methods that are in-sample calibrated to construct sets that are guaranteed to contain a calibrated predictive distribution out-of-sample. This allows us to take any prediction method that is in-sample calibrated, and \emph{conformalize} it to obtain a predictive system with out-of-sample calibration guarantees. While the satisfied notion of calibration is typically that prediction intervals derived from the predictive distribution have the correct marginal coverage, we show that this line of reasoning can be extended to stronger conditional notions of calibration that are common in statistical forecasting theory. Using this, we introduce two predictive systems that satisfy stronger out-of-sample calibration guarantees than existing conformal predictive systems. The first method corresponds to a binning of the data, while the second leverages isotonic distributional regression (IDR), a non-parametric distributional regression method under order constraints. We study the theoretical properties of these new predictive systems, and compare their performance in a simulation experiment. They are then applied to two case studies on European temperature forecasts and on predictions for the length of patient stay in Swiss intensive care units. Both approaches are found to outperform existing conformal predictive systems, while conformal IDR additionally provides a natural method for quantifying epistemic uncertainty of the predictions. 
\end{abstract}


\section{Introduction}

When forecasting an unknown outcome $Y$, it is becoming increasingly common to issue forecasts that are probabilistic. Probabilistic forecasts characterize the uncertainty in the outcome, making them more useful for decision making than single-valued point forecasts. When $Y \in \R$, a full probabilistic forecast is a probability distribution over $\R$, thus quantifying the probability of any future event associated with $Y$. A review of probabilistic forecasting can be found in \cite{GneitingKatzfuss2014}. 

A pertinent question to ask is how we can generate \emph{calibrated} forecasts. A probabilistic forecast is calibrated if its predicted event probabilities conform with observed event frequencies, which is necessary for the forecast to be useful for decision making. It is common to assess calibration by checking for uniformity of forecast probability integral transform (PIT) values \citep{DawidEtAl1984,DieboldEtAl1998,GneitingEtAl2007}, which corresponds to the prediction intervals derived from the forecast distribution having the correct marginal coverage. However, this is a fairly weak requirement, and several stronger notions of calibration have therefore also been proposed \citep{Tsyplakov2011,HenziZiegelETAL2021,GneitingResin2023,ArnoldZiegel2023}.

Suppose our goal is to predict an unknown observation $Y_{n+1}$ using previous covariate-observation pairs $(X_1,Y_1),\dots,(X_{n},Y_{n})$ and a new covariate $X_{n+1}$. Conformal prediction provides a framework to generate sets of forecasts for $Y_{n+1}$ that satisfy theoretical calibration guarantees by construction, under the assumption of exchangeability. This has received much attention recently, finding use in a variety of application domains. Textbook introductions can be found in \citet{VovkGammermanETAL2022,AngelopoulosBarberETAL2024}. 

Here, we focus on the case where $(X_1,Y_1),\dots,(X_{n+1},Y_{n+1})$ is an exchangeable sequence with $Y_i \in \mathbb{R}, X_i \in \mathcal{X}$ (for some arbitrary covariate space $\mathcal{X}$), and we are interested in obtaining a full probabilistic forecast for $Y_{n+1}$. In this case, \cite{VovkEtAl2017} introduced \emph{conformal predictive systems} as a means to construct sets of probabilistic forecasts for $Y_{n+1}$ with theoretical calibration guarantees. Conformal predictive systems are \emph{bands} of distribution functions that are guaranteed to contain a calibrated forecast distribution. While conformal predictive systems have received less attention in the literature than conformal prediction methods for regression and binary classification, one could argue that conformal predictive systems are more useful. Just as probabilistic forecasts provide more information than single-value point forecasts since they can be used to obtain prediction intervals at any level, conformal predictive systems analogously allow us to obtain prediction intervals at any level with marginal coverage guarantees. 


However, the general framework proposed by \cite{VovkEtAl2017} yields conformal predictive systems that satisfy a weak unconditional notion of calibration. Unconditional calibration is often insufficient for the forecasts to be useful for decision making. \cite{AllenEtAl2024}, for example, demonstrate that a probabilistic forecast can be unconditionally calibrated without issuing reliable forecasts for extreme events. Instead, we generally want our forecasts to be calibrated conditional on some auxiliary information, with a larger information set corresponding to a stronger notion of calibration. \cite{HolzmannEulert2014} demonstrate that calibration with respect to a larger information set results in a more accurate forecast when assessed using any strictly proper scoring rule, and we therefore desire probabilistic forecasts that are calibrated with respect to as strong a notion of calibration as possible.

 
While the conditional calibration of conformal prediction algorithms is well studied for regression and binary classification tasks, relatively little work has addressed the conditional calibration of conformal predictive systems. \cite{BostromEtAl2021} also mentioned the need for conformal predictive systems to satisfy stronger notions of calibration, and introduced predictive systems that satisfy a form of conditional calibration; the approach works by binning the available data, and generating conformal prediction systems separately for each bin, thereby extending Mondrian conformal prediction for binary outcomes \citep[Section 4]{VovkGammermanETAL2022}. \citet{ChernozhukovWuthrichETAL2021} alternatively proposed conformal predictive systems that satisfy conditional calibration guarantees asymptotically under correct model specification. In this paper, we show that it is possible to obtain predictive systems with stronger calibration guarantees than existing approaches.

We demonstrate that any probabilistic forecasting system (or distributional regression procedure) that satisfies in-sample calibration guarantees can be used to derive predictive systems with (non-asymptotic) conformal calibration guarantees; that is, the resulting predictive systems are guaranteed to contain a calibrated probabilistic forecast out-of-sample. Using this, we introduce predictive systems based on two existing forecasting methods. The first approach, \emph{conformal binning}, corresponds to a binning of the data, modifying the Mondrian predictive systems of \cite{BostromEtAl2021} so that they satisfy a stronger notion of calibration. This is closely related to analog forecasting systems \citep[e.g.][]{HamillWhitaker2006}. The second approach, \emph{conformal isotonic distributional regression} (IDR), derives predictive systems based on IDR \citep{HenziZiegelETAL2021}. IDR is a non-parametric distributional regression technique that finds the optimal forecast distribution under the assumption that there is an isotonic relationship between the covariate $X_i$ and the outcome $Y_i$. The fitted IDR model can be interpreted as a sample estimate of the isotonic conditional law of $Y_i$ given $X_i$ \citep{ArnoldZiegel2023}, and IDR has found particular use as a method to convert point forecasts to probabilistic forecasts, without the need for additional assumptions \citep{WalzEtAl2024}.

Probabilistic forecasts obtained via binning and IDR satisfy in-sample \emph{auto-calibration} and \emph{isotonic calibration}, respectively, which are relatively strong notions of forecast calibration that condition on facets of the predictive distribution. The calibration of IDR holds independently of any isotonicity assumption on the data generating process. The two approaches that we introduce therefore generate predictive systems with stronger calibration guarantees than existing conformal predictive systems, such as the Least Squares Prediction Machine (LSPM) of \cite{VovkEtAl2017}. Since stronger calibration guarantees correspond to more informative predictions, these new predictive systems can be used to obtain more accurate probabilistic forecasts. Similarly, these predictive systems can be used to construct prediction intervals and prediction sets with strong out-of-sample calibration guarantees. This suggests that by conditioning on the predictive distribution itself, rather than the covariates used to construct it, we can circumvent well-known impossibility results related to the conditional calibration of conformal prediction sets \citep{LeiWasserman2014,BarberCandesETAL2021}.


Conformal IDR has the additional appealing property that it provides a natural method for quantifying \emph{epistemic uncertainty} of the predictions. While probabilistic predictions quantify the (aleatoric) uncertainty of the future outcome $Y_{n+1}$, there is also uncertainty in the probabilistic prediction itself, since we never know the true conditional distribution of $Y_{n+1}$ given $X_{n+1}$. This is the epistemic uncertainty of the forecast. Several studies have argued the need to quantify epistemic forecast uncertainty (also called second-order probabilities or forecast ambiguity), see for example \citet{Seo2009,EckelEtAl2012,KendallGal2017,LofstromLofstromETAL2024}. 
Predictive systems generate bands that are guaranteed to contain a calibrated predictive distribution, and we refer to the distance between the lower and upper bands as the \emph{thickness} of the predictive system; the calibration guarantee is only useful if the thickness is not too large. Standard conformal predictive system exhibit a thickness that depends only on $n$, the number of previous covariate-observation pairs from which the predictive system is constructed, and binning-based conformal predictive systems similarly generate predictions with a fixed thickness in each bin. The thickness of conformal IDR, on the other hand, depends on the covariate value $X_{n+1}$, and how often similar covariate values have been observed in the past. We argue that the thickness of the conformal IDR band can be interpreted in terms of epistemic uncertainty, thereby allowing us to quantify the epistemic and aleatoric uncertainty of the predictions simultaneously, under a non-parametric isotonicity constraint. 

In the following section, we introduce notions of probabilistic forecast calibration, and demonstrate that distributional regression procedures with in-sample calibration guarantees can be used to generate predictive systems with conformal calibration guarantees. We discuss three examples of predictive systems in Section \ref{sec:examples}: the standard conformal predictive systems proposed by \cite{VovkEtAl2017} based on conformity measures, a conformal binning approach, and conformal IDR; the latter two approaches are constructed using the framework outlined in Section \ref{sec:main}. Section \ref{sec:properties} discusses how predictive systems can be used to generate calibrated prediction intervals. The interpretation of the conformal IDR bands in terms of epistemic uncertainty is discussed in Section \ref{sec:epistemic}. We present simulation results for the three predictive systems in Section \ref{sec:sims}, and compare the approaches in two applications to weather forecasting and epidemiological forecasting in Section \ref{sec:app}. Section \ref{sec:disc} summarizes the results. Proofs are collected in Appendix \ref{app:proofs}, and code to implement the three conformal prediction systems, and to reproduce the results herein, is available at \url{https://github.com/sallen12/ConformalIDR}.

\section{Conformalizing probabilistic predictions}\label{sec:main}

\subsection{Calibration of probabilistic predictions}\label{sec:calibration}

Suppose that an unknown future outcome $Y \in \R$ and a predictive distribution function (CDF) $F$ for $Y$ are defined on some common probability space $(\Omega,\F,\prob)$. For probabilistic forecasts to be useful for decision making, they must be calibrated. When forecasting a binary event occurrence, a forecast is traditionally called calibrated if, when the event is predicted to occur with a given probability, the event does indeed transpire with this predicted probability. When the outcome $Y$ is real-valued instead of binary, it is more challenging to define forecast calibration, and many different notions of calibration exist in this case. 
In this section, we review notions of calibration that have been proposed in the literature on forecast evaluation.

Inspired by the case of binary outcomes, we might similarly call a forecast $F$ for real-valued outcomes calibrated if $\prob(Y \le y \mid F) = F(y)$ almost surely for all $y \in \R$. 
That is, given that we have issued $F$ as our forecast, the outcome $Y$ should arise according to the distribution $F$. \cite{Tsyplakov2011} refers to this as \emph{auto-calibration} of the forecast $F$. 


Auto-calibration can be relaxed by conditioning on $\sigma$-lattices instead of $\sigma$-algebras. \cite{ArnoldZiegel2023} say that a forecast is \emph{isotonically calibrated} if $\mathbb{P}(Y > y \mid \mathcal{A}(F)) = 1 - F(y)$ almost surely for all $y \in \R$. 
Note that this is not equivalent to requiring that $\mathbb{P}(Y \le y \mid \mathcal{A}(F)) = F(y)$ almost surely for all $y \in \R$, since we generally do not have that $1 - \mathbb{P}(Y > y \mid \C) = \mathbb{P}(Y \le y \mid \C)$ for some $\sigma$-lattice $\C \subset \F$ \citep[Remark 1]{ArnoldZiegel2023}. Nonetheless, isotonic calibration is similar in essence to auto-calibration. The results of this article do not require an understanding of $\sigma$-lattices, though isotonic calibration is discussed further in Appendix \ref{app:isotoniccal}, and additional details can be found in \cite{ArnoldZiegel2023} and references therein. 


Relevant weaker notions of calibration arise by conditioning only on particular functionals of the forecast distribution, such as threshold exceedances or quantiles \citep{GneitingResin2023}. The forecast $F$ is \emph{threshold calibrated} if $\prob(Y \le y \mid F(y)) = F(y)$ almost surely for all $y \in \R$ \citep{HenziZiegelETAL2021}, and \emph{quantile calibrated} if $\prob(Y < F^{-1}(\alpha) \mid F^{-1}(\alpha)) \leq \alpha \leq \prob(Y \leq F^{-1}(\alpha) \mid F^{-1}(\alpha))$ almost surely for all $\alpha \in (0, 1)$ \citep{ArnoldZiegel2023}. Threshold calibration and quantile calibration are equivalent under continuity assumptions on $F$ \citep[Theorem 2.16]{GneitingResin2023}.

These notions can be weakened further by requiring only that these criteria hold unconditionally. The forecast $F$ is called \emph{probabilistically calibrated} for $Y$ if 
\begin{equation}\label{eq:probcal}
\prob(F(Y) \le \alpha) \le \alpha \le \mathbb{P}(F(Y-) < \alpha), \quad \text{for all $\alpha \in (0,1)$,}
\end{equation}
where $F(x-) = \lim_{z \uparrow x} F(z)$. If $F$ is continuous, probabilistic calibration implies that the forecast PIT value $F(Y)$ follows a standard uniform distribution; this notion of calibration is most commonly assessed in practice. If $F$ is deterministic, the above condition implies that $F$ is the CDF of $Y$; see Proposition \ref{prop:PIT1} in Appendix \ref{app:PIT}. 

\begin{rem}\label{rem:probcal}
The definition of probabilistic calibration at \eqref{eq:probcal} slightly differs from the literature. \citet[Definition 2.7]{GneitingRanjan2013} call a forecast probabilistically calibrated if 
\begin{equation}\label{eq:probcal2}
\prob(F(Y-) + V(F(Y) - F(Y-)) \le \alpha) = \alpha \quad \text{for all $\alpha \in (0,1)$},
\end{equation}
where $V$ is a standard uniform random variable that is independent of $(F,Y)$. These, and other definitions of calibration, are compared in \citet[Proposition 2.7]{ResinEtAl2026}. When $F$ is continuous, the two definitions are equivalent. Equation \eqref{eq:probcal2} implies $\eqref{eq:probcal}$ since $\{F(Y) \le \alpha\} \subset \{F(Y-) + V(F(Y) - F(Y-)) \le \alpha\} $ and $\{F(Y-) + V(F(Y) - F(Y-)) < \alpha\} \subset \{F(Y-) < \alpha\}$. Conversely, while \eqref{eq:probcal} also implies that there exists a randomization variable $V$ taking values in $[0,1]$ such that the expression at \eqref{eq:probcal2} is standard uniform, $V$ may now depend on $(F,Y)$; see Proposition \ref{prop:PIT2} in Appendix \ref{app:PIT}. In other words, condition \eqref{eq:probcal} means that one can define a generalized probability integral transform that is standard uniform. We therefore consider \eqref{eq:probcal} a sensible definition of probabilistic calibration that preserves the desirable properties of \eqref{eq:probcal2} but is often easier to handle with discrete predictive CDFs, and is not restricted to the construction of a specific randomization variable $V$. 
\end{rem}

\begin{figure}
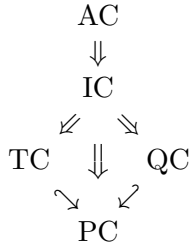

    \centering
    \[
    \begin{array}[c]{c c c}
        & \mathrm{AC} & \\
        & \Downarrow & \\
        & \mathrm{IC} & \\
        \multicolumn{3}{c}{\Swarrow \quad \Searrow} \\
        \mathrm{TC} & \big\Downarrow & \mathrm{QC} \\
        \multicolumn{3}{c}{\Sehookarrow \quad \Swarrow} \\
        & \mathrm{PC} & \\
    \end{array}
    \]
    \caption{Hierarchies of the notions of forecast calibration. Auto-calibration (AC) implies isotonic calibration (IC), which itself implies threshold calibration (TC), quantile calibration (QC), and probabilistic calibration (PC). QC implies PC, and TC also implies PC under smoothness assumptions, denoted by a hooked arrow.}
    \label{fig:implications}
\end{figure}

Auto-calibration implies isotonic calibration, and isotonic calibration implies threshold, quantile, and probabilistic calibration. Isotonic calibration is also equivalent to auto-calibration when there is an isotonic relationship between $F$ and $Y$. Under smoothness assumptions, threshold calibration and quantile calibration both imply probabilistic calibration \citep{GneitingResin2023,ArnoldZiegel2023}. These implications are displayed in Figure \ref{fig:implications}, see also \citet[Figure 1]{ArnoldEtAl2024}. Forecasts that satisfy stronger notions of calibration generally lead to better performance when assessed using proper scoring rules \citep{HolzmannEulert2014}, and we therefore prefer forecast distributions that are auto-calibrated or isotonically calibrated to forecasts that only satisfy weaker notions.

\subsection{Distributional regression procedures}

Probabilistic forecasts are typically generated using procedures that take a training data set and a new covariate as inputs, and output a predictive distribution; these are often referred to as \emph{distributional regression procedures}. Let $z_1=(x_1,y_1),\dots,z_m=(x_m,y_m) \in \mathcal{Z} = \mathcal{X}\times \mathbb{R}$, and define the empirical distribution as $\hat{\mathbb{P}} = \hat{\mathbb{P}}_m = (1/m)\sum_{i=1}^m \delta_{z_i}$. In this section, we will consider properties that hold \emph{in-sample}, so it is helpful to think of $z_1,\dots,z_m$ as $m$ deterministic points in $\mathcal{Z}$, which can be interpreted as a fixed training data set. If $(X,Y)$ is a random variable generated by a random draw from $(x_1,y_1),\dots,(x_m,y_m)$ (with multiplicity), then it has distribution $\hat{\mathbb{P}}$.

Let $G$ be a procedure that assigns a CDF $G^{\hat{\mathbb{P}}}_{x_i}$ to any empirical distribution $\hat{\mathbb{P}}$ and $x_i$, $i = 1,\dots,m$. That is, 
\[
G^{\hat{\mathbb{P}}}_{x_i}: \mathbb{R} \to [0,1], \quad y \mapsto G^{\hat{\mathbb{P}}}_{x_i}(y)
\]
is an increasing, right-continuous function with limits $1$ and $0$ at $\pm\infty$, respectively.
The procedure $G$ can be interpreted as a distributional regression procedure, where $x_i \in \mathcal{X}$ represents the (new) covariates, and the covariate-observation pairs in $\hat{\mathbb{P}}$ represent the data used to train $G$. The procedure may depend on further parameters that we suppress here in the notation. A simple example is a regression model whereby $G_{x_i}^{\hat{\mathbb{P}}}$ is a parametric distribution function with parameters that are learned from the data in $\hat{\mathbb{P}}$ and applied to the covariate $x_i$. Further examples of regression procedures are given in Examples \ref{ex:procedure1} and \ref{ex:procedure2}, Proposition \ref{prop:bin_ac}, and Equations \ref{eq:G0} and \ref{eq:G_IDR}.


\begin{defn}\label{def:Gcal}
    Let $(X,Y)$ have distribution $\hat{\mathbb{P}}$.
    
    \begin{enumerate}
        \item[a)] The procedure $G$ is \emph{probabilistically calibrated} if, for any empirical distribution $\hat{\mathbb{P}}$, $G^{\hat{\mathbb{P}}}_X$ is probabilistically calibrated for $Y$, that is,
        \[
        \hat{\mathbb{P}}\big(G^{\hat{\mathbb{P}}}_X(Y) \le \alpha\big) \le \alpha \le \hat{\mathbb{P}}\big(G^{\hat{\mathbb{P}}}_X(Y-) < \alpha\big) \quad \text{for all $\alpha \in (0,1)$.}
        \]

        \item[b)] The procedure $G$ is \emph{isotonically calibrated} if, for any empirical distribution $\hat{\mathbb{P}}$, $G^{\hat{\mathbb{P}}}_X$ is isotonically calibrated for $Y$, that is,
        \[
        \hat{\mathbb{P}}\big(Y > y \mid \mathcal{A}\big(G^{\hat{\mathbb{P}}}_X\big)\big) = 1- G^{\hat{\mathbb{P}}}_X(y) \quad \text{for all $y \in \mathbb{R}$}.
        \]
        
        \item[c)] The procedure $G$ is \emph{auto-calibrated} if, for any empirical distribution $\hat{\mathbb{P}}$, $G^{\hat{\mathbb{P}}}_X$ is auto-calibrated for $Y$, that is,
        \[
       \hat{\mathbb{P}}\big(Y \le y \mid G^{\hat{\mathbb{P}}}_X\big) = G^{\hat{\mathbb{P}}}_X(y) \quad \text{for all $y \in \mathbb{R}$}.
        \]
    \end{enumerate}
\end{defn}


\begin{rem}
    The notions of calibration in Definition \ref{def:Gcal} correspond to \emph{in-sample} calibration, in the sense that if the pair $(X, Y) \sim \hat{\mathbb{P}}$ is a random draw from the empirical distribution used to train the distributional regression procedure $G$, then plugging the covariate $X$ into $G$ yields a calibrated forecast for $Y$. Consequently, the implications between notions of calibration explained in Section \ref{sec:calibration} and summarized in Figure \ref{fig:implications} carry over to the calibration of procedures $G$.
\end{rem}

\begin{exmp}\label{ex:procedure1}
    Suppose that $\mathcal{X} = \R^p$. Let \(\hat{y}_1,\ldots,\hat{y}_m\) be any regression fit of \((y_1,\ldots,y_m)\) on \((x_1,\dots,x_m)\) (for example, weighted least squares), and let \(\epsilon_i=y_i-\hat{y}_i\) be the residuals. Consider a distributional regression procedure based on the weighted empirical distribution of the residuals centered at the regression fit, that is, for $k=1,\dots,m$,
    \[
    G^{\hat{\mathbb{P}}}_{x_k}(y) = \frac{1}{m}\sum_{j=1}^m\one\{\hat{y}_k + \epsilon_j \le y\}, \quad y \in \R.
    \]
    This procedure is probabilistically calibrated, see Proposition \ref{prop:proof_example_2.1}. This has been studied in more detail in the context of conformal predictive systems by \cite{AllenEtAl2026a}.
\end{exmp}

\begin{exmp}\label{ex:procedure2}
    Suppose that for given values $x_1,\dots,x_m$, we have a clustering algorithm that gives us a partition $B_1,\dots,B_\ell$ of $\{1,\dots,m\}$ grouping similar covariate values together. Then, we can define a binning procedure as
    \[
    G^{\hat{\mathbb{P}}}_{x_k}(y) = \frac{1}{|B_i|}\sum_{j \in B_i} \one\{y_j \le y\}, \quad \text{for $k \in B_i$}.
    \]
    Binning procedures are auto-calibrated, and these are studied in more detail in Section \ref{sec:conf_bin}.
\end{exmp}

\subsection{Predictive systems}

Ideally, we would like our distributional regression procedures to issue out-of-sample forecast distributions that are calibrated by construction. Since this is generally too difficult to achieve, it is convenient to relax the definition of a forecast CDF to allow for a whole predictive system, that is, a ``band'' of predictive distributions \citep{VovkGammermanETAL2022}.

\begin{defn}\label{def:ps}
A \emph{predictive system} is a set $\Pi \subseteq \R \times [0,1]$ of the form
\[
\Pi = \{(y,\tau) \in \R \times [0,1] \mid \Pi_\ell(y) \le \tau \le \Pi_u(y)\},
\]
where the lower and upper bounds $\Pi_\ell:\R \to [0,1]$, $\Pi_u:\R \to [0,1]$ are increasing, satisfy $\Pi_\ell(y) \le \Pi_u(y)$, $y \in \R$, and
\[
    \lim_{y \to -\infty} \Pi_\ell(y) = 0, \quad \lim_{y \to \infty} \Pi_u(y) = 1.
\]
The \emph{thickness} of a predictive system $\Pi$ is defined as 
    \[
    \mathrm{th}(\Pi) = \inf\{\varepsilon > 0 \mid \Pi_u(y) - \Pi_\ell(y) \le \varepsilon \; \text{for all but finitely many $y \in \R$}\}.
    \]
\end{defn}

A predictive system essentially corresponds to a set of predictive distribution functions. Any predictive distribution that lies between the lower and upper bounds $\Pi_\ell$ and $\Pi_u$ is within the predictive system. Figure \ref{fig:2} provides a graphical illustration of a predictive system and a predictive distribution that lies within its bands. We show in the following section how we can use in-sample calibrated distributional regression procedures to obtain predictive systems that are guaranteed to contain an out-of-sample calibrated predictive distribution. Clearly, the band has to be narrow enough to make this calibration guarantee useful; the thickness of a predictive system provides a measure of how narrow the band is. As far as we are aware, this paper is the first attempt to thoroughly connect (conformal) predictive systems with the extant literature on probabilistic forecast calibration. 

\begin{rem}
    Definition \ref{def:ps} is slightly different to the definition of a predictive system given by \citet{VovkGammermanETAL2022}, who parametrize all possible CDFs within the predictive system with one parameter $\tau \in [0,1]$. Instead, we allow for a free choice of any CDF within the bounds $\Pi_\ell$, $\Pi_u$, which corresponds to a more general definition of a prediction set for probabilistic forecasts. This relates to Remark \ref{rem:probcal} regarding different randomizations at the CDF's jump points. 
\end{rem}

\subsection{In-sample calibration guarantees lead to conformal calibration guarantees}\label{sec:insample}

For a given procedure $G$, empirical distribution $\hat{\mathbb{P}}$, and covariate $x \in \mathcal{X}$, we can define a predictive system $\Pi$ by the bounds
\begin{align}
\Pi_\ell[\hat{\mathbb{P}},x](y) &= \inf\{G^{\hat{\mathbb{P}} + (x,y')}_x(y) \mid y' \in \mathbb{R}\},\label{eq:PB1}\\
\Pi_u[\hat{\mathbb{P}},x](y) &= \sup\{G^{\hat{\mathbb{P}} + (x,y')}_x(y) \mid y' \in \mathbb{R}\},\label{eq:PB2}
\end{align}
where $\hat{\mathbb{P}}+(x,y')$ is an abbreviation for the empirical distribution of $(x_1, y_1), \dots, (x_m, y_m), (x, y')$. These predictive systems essentially assume that the new outcome will be equal to $y' \in \R$, augment the training data with the assumed covariate-outcome pair $(x, y')$, fit the distributional regression procedure using this augmented training data set, and then repeat this for all possible outcomes $y'$. The bounds of the predictive system then correspond to the pointwise infimum and supremum of the resulting predictive distributions. For many procedures, it is possible to calculate the lower and upper bounds without retraining for all possible $y'$. This is the crux of conformal prediction.

The following theorem is our main result. It demonstrates that if the regression procedure is in-sample calibrated, then the resulting predictive system generated using \eqref{eq:PB1} and \eqref{eq:PB2} is guaranteed to contain a calibrated predictive distribution. We prove a more general version of the theorem in Appendix \ref{app:proofs}.

\begin{thm}\label{thm:master}
Let $(X_1,Y_1),\dots,(X_n,Y_n),(X_{n+1},Y_{n+1})$ be an exchangeable sequence, and let $\Pi$ be the predictive system defined by the bounds \eqref{eq:PB1} and \eqref{eq:PB2}, for some procedure $G$, $\hat{\mathbb{P}} = \hat{\mathbb{P}}_n =  (1/n)\sum_{j=1}^n \delta_{(X_j,Y_j)}$, and $x = X_{n+1}$. 
\begin{enumerate}
    \item[(i)] If $G$ is probabilistically calibrated, then $\Pi$ contains a probabilistically calibrated predictive CDF for $Y_{n+1}$. 

    \item[(ii)] If $G$ is isotonically calibrated, then $\Pi$ contains an isotonically calibrated predictive CDF for $Y_{n+1}$. 
    
    \item[(iii)] If $G$ is auto-calibrated, then $\Pi$ contains an auto-calibrated predictive CDF for $Y_{n+1}$. 
\end{enumerate}
\end{thm}


\begin{rem}
     We say that a predictive system $\Pi$ \emph{contains} a calibrated predictive CDF for $Y_{n+1}$ if there exists a calibrated forecast $G$ such that $\Pi_\ell(y) \leq G(y) \leq \Pi_{u}(y)$ for all $y \in \R$ almost surely. For probabilistic calibration, this is equivalent to the standard definition of validity of a conformal predictive system, as given in \cite{VovkGammermanETAL2022}, updated to use the new definition of probabilistic calibration at \eqref{eq:probcal}. We generally say that a predictive system has \emph{out-of-sample} or \emph{conformal} calibration guarantees if it is guaranteed to contain a calibrated predictive CDF for $Y_{n+1}$, for any specified notion of calibration. 
\end{rem}

The proof of Theorem \ref{thm:master} uses the result that if the procedure $G$ is in-sample calibrated, then the predictive CDF $G^{\hat{\mathbb{P}}_{n+1}}_{X_{n+1}}$ with $\hat{\mathbb{P}}_{n+1} = \hat{\mathbb{P}}_n +(X_{n+1},Y_{n+1})$ provides a calibrated probabilistic forecast for $Y_{n+1}$, and this falls within the bounds defining $\Pi$. While $G^{\hat{\mathbb{P}}_{n+1}}_{X_{n+1}}$ is not available at the time of forecasting, since it depends on $Y_{n+1}$, the predictive system $\Pi$ is. If the thickness of the predictive system is close to zero, the out-of-sample calibration guarantees allow us to issue a predictive distribution that is `close' to the calibrated predictive CDF $G^{\hat{\mathbb{P}}_{n+1}}_{X_{n+1}}$. In the following sections, we consider concrete procedures that satisfy different in-sample calibration guarantees, leading to predictive systems with different out-of-sample calibration guarantees.

If $G$ is calibrated, then we obtain a predictive system with corresponding calibration guarantees regardless of how the procedure $G$ uses the information contained in the data in $\hat{\mathbb{P}}_n$. If $G$ produces predictive distributions that are highly dispersed and uninformative, then the prediction may not be useful, even if it is calibrated. On the other hand, even if $G$ yields very concentrated distributions, the thickness of the resulting predictive system could still be large. We discuss these effects in more detail in Section \ref{sec:epistemic}, and interpret them with regard to epistemic uncertainty.

\begin{rem}[Splitting approach]
The guarantees of Theorem \ref{thm:master} for the bounds $\Pi_\ell$, $\Pi_u$ defined at \eqref{eq:PB1}, \eqref{eq:PB2} remain valid if the procedure $G$ depends on parameters that are estimated using an independent sample of data. More precisely, suppose that the training data $(X_1,Y_1),\dots,(X_n,Y_n)$ is split into two sets: an \emph{estimation set} $\mathcal{D}_0$ containing the data points with indices $i \le n_0$, and a \emph{calibration set} $\mathcal{D}_1$ containing the data points with indices $n_0 < i \le n$. Let $n_1 = n - n_0$ denote the size of the calibration set. The parameters of the procedure $G$ can be estimated using the data in $\mathcal{D}_0$. Assuming that the data $(X_{n_0+1},Y_{n_0+1}),\dots$, $(X_n,Y_n)$, $(X_{n+1},Y_{n+1})$ is exchangeable given $\mathcal{D}_0$, the bounds in \eqref{eq:PB1}, \eqref{eq:PB2} can then be computed using the data in the calibration set $\mathcal{D}_1$. We call this a splitting approach. The guarantees of Theorem \ref{thm:master} continue to hold: \citet[Section 3.4.1]{AngelopoulosBarberETAL2024} provide the same arguments for classical conformal prediction. A splitting approach is useful to reduce computational costs, and is typically also applied in existing conformal predictive systems \citep{VovkEtAl2018}, see Remark \ref{rem:splitCM}. In particular, it is beneficial to estimate a procedure $G$ on $\mathcal{D}_0$ such that $G^{\hat{\mathbb{P}}_n + (x,y')}_{x}(y)$ is monotone in $y'$ for each $y$, because then the bounds in \eqref{eq:PB1} and \eqref{eq:PB2} can typically be computed using only two evaluations of the procedure, see Remark \ref{rem:splitIDR}.
\end{rem}

\section{Examples of predictive systems}\label{sec:examples}

In this section, we describe the standard approach to generate conformal predictive systems proposed by \cite{VovkEtAl2017}, and we then use Theorem \ref{thm:master} to introduce two new predictive systems with stronger out-of-sample calibration guarantees. 

\subsection{Probabilistic calibration: CM predictive systems}\label{sec:CPS}

The results in this section are not new and can be found with somewhat different notation and arguments in the textbook \citet{VovkGammermanETAL2022}. We include them here for completeness and in order to facilitate understanding of our new results in the subsequent sections. 


\cite{VovkEtAl2017} introduce predictive systems using \emph{conformity measures}. A conformity measure $A$ is a function that quantifies how similar a covariate-outcome pair $z \in \mathcal{Z}$ is to a set of such pairs (denoted in the following by the corresponding empirical distribution $\hat{\mathbb{P}}$). That is, $A(\hat{\mathbb{P}},z)$ quantifies how much the point $z$ conforms with the data in $\hat{\mathbb{P}}$. Conformal predictive systems are then defined via the bounds
\begin{equation}\label{eq:CM_low}
    \Pi_\ell[\hat{\mathbb{P}}, x](y) = \frac{1}{m + 1} \sum_{i = 1}^m \one\{A(\hat{\mathbb{P}} + (x, y), (x_i, y_i)) < A(\hat{\mathbb{P}} + (x, y),(x, y))\},
\end{equation}
\begin{equation}\label{eq:CM_upp}
    \Pi_u[\hat{\mathbb{P}}, x](y) = \frac{1}{m + 1} \left( 1 + \sum_{i = 1}^m \one\{A(\hat{\mathbb{P}} + (x, y),(x_i, y_i)) \leq A(\hat{\mathbb{P}} + (x, y),(x, y)) \} \right).
\end{equation}
It is sufficient to specify $A(\hat{\mathbb{P}},z)$ for $z$ in the support of $\hat{\mathbb{P}}$ in order to define conformity measures. The bounds defined at \eqref{eq:CM_low} and \eqref{eq:CM_upp} are not necessarily increasing in $y$ for any $\hat{\mathbb{P}}$, $x \in \mathcal{X}$, meaning these bounds do not necessarily define a valid predictive system; Lemma \ref{lem:monoPS} in Appendix \ref{app:proofs} gives a sufficient condition on the conformity measure $A$ such that the bounds are a predictive system according to Definition \ref{def:ps} \citep[see also][Proposition 7.2]{VovkGammermanETAL2022}. In this case, the thickness of the predictive system is at most $1/(m+1)$. Since we present several possibilities to construct predictive systems, we call the predictive systems that are constructed using conformity measures, as in \eqref{eq:CM_low} and \eqref{eq:CM_upp}, \emph{conformity measure based (CM) predictive systems}.

\begin{exmp}[Least Squares Prediction Machine]\label{ex:lspm}
    Consider the conformity measure
    \[
     A(\hat{\mathbb{P}}, (x_k, y_k)) = \frac{y_{k} - \hat{y}_{k}}{\sqrt{1 - \bar{h}_{k}}}, \quad k = 1, \dots, m,
    \]
    where $\hat{y}_{k}$ is the weighted least squares prediction for $y_{k}$ constructed using $\hat{\mathbb{P}}$ as training data, and $\bar{h}_{k}$ is the leverage of $(x_{k}, y_{k})$ in the regression fit. Using this conformity measure within \eqref{eq:CM_low} and \eqref{eq:CM_upp} recovers the (studentized) Least Squares Prediction Machine (LSPM), see \citet{VovkEtAl2017}, \citet[Section 7.3]{VovkGammermanETAL2022}.
\end{exmp}

With $\Pi_\ell$ and $\Pi_u$ defined as at \eqref{eq:CM_low} and \eqref{eq:CM_upp}, given an exchangeable sequence $(X_1,Y_1),\dots,(X_n,Y_n)$,
$(X_{n+1},Y_{n+1})$, the bounds $\Pi_\ell[\hat{\mathbb{P}}_n, X_{n+1}]$ and $\Pi_u[\hat{\mathbb{P}}_n, X_{n+1}]$ are guaranteed to contain a probabilistically calibrated forecast distribution for $Y_{n+1}$ \citep{VovkEtAl2017}; this property is often referred to as the \emph{validity} of conformal predictive systems.

These CM predictive systems generally do not directly correspond to the bounds at \eqref{eq:PB1} and \eqref{eq:PB2} defined using a certain forecasting procedure. However, they are closely related to a simple calibrated forecasting procedure. Firstly, consider the following forecasting procedure that does not use covariates:
\begin{equation}\label{eq:G0}
G_0^{\hat{\mathbb{P}}}(y) = \frac{1}{m}\sum_{i=1}^m \one\{y_i \le y\}.
\end{equation}
It is a standard exercise in probability to show that $G_0$ is probabilistically calibrated. In fact, Proposition \ref{prop:PIT2} in Appendix \ref{app:PIT} shows that this procedure is the only possible probabilistically calibrated procedure when there is no covariate information. For $y' \in \R$ (and arbitrary $x \in \mathcal{X}$), we have
\[
G_0^{\hat{\mathbb{P}} + (x, y')}(y) = \frac{1}{m + 1}\left(\sum_{i = 1}^m \one\{y_i \le y\} + \one\{y' \le y\}\right),
\]
and hence the bounds at \eqref{eq:PB1} and \eqref{eq:PB2} become
\begin{equation*}
    \Pi_\ell[\hat{\mathbb{P}}, x](y) = \frac{1}{m + 1}\sum_{i=1}^m \one\{y_i \le y\}, \quad
    \Pi_u[\hat{\mathbb{P}}, x](y) = \frac{1}{m + 1}\left(\sum_{i=1}^m \one\{y_i \le y\} + 1\right) = \Pi_\ell[\hat{\mathbb{P}},x](y) + \frac{1}{m+1}.
\end{equation*}

The classical conformal predictive systems introduced by \cite{VovkEtAl2017} use conformity measures to elegantly exploit this one-dimensional probabilistically calibrated procedure $G_0$ and obtain probabilistically calibrated predictive distributions using covariate information. 
For $(x,y)\in \{z_1, \dots, z_m\}$, we define
\begin{equation}\label{eq:classical_cps}
    G_x^{\hat{\mathbb{P}}}(y) = G_0^{\hat{\mathbb{P}}^A} (A(\hat{\mathbb{P}}, (x, y))) = \frac{1}{m} \sum_{i=1}^{m} \one\{A(\hat{\mathbb{P}}, (x_i, y_i)) \le A(\hat{\mathbb{P}}, (x, y))\},
\end{equation}
where $\hat{\mathbb{P}}^A$ denotes the empirical distribution of the conformity scores $A(\hat{\mathbb{P}}, (x_1, y_1)), \dots, A(\hat{\mathbb{P}}, (x_m, y_m))$. 
This resembles a distributional regression procedure that quantifies how well each point in the training data conforms with $(x, y)$, as quantified using $A$, and then assigns higher probability to training points with higher conformity scores. However, the function $G$ defined at \eqref{eq:classical_cps} is not necessarily increasing in $y$ for any $\hat{\mathbb{P}}$, $x \in \mathcal{X}$, meaning $G$ may not output valid CDFs and is thus not a distributional regression procedure. Lemma \ref{lem:monoPS} in Appendix \ref{app:proofs} again gives sufficient conditions on the conformity measure $A$ such that $G$ outputs valid CDFs. 
Nonetheless, it is always possible to use $G$ to obtain prediction sets with out-of-sample coverage guarantees due to the following corollary to Theorem \ref{thm:master}.

\begin{cor}\label{cor:CM}
Let $(X_1,Y_1),\dots,(X_n,Y_n),(X_{n+1},Y_{n+1})$ be an exchangeable sequence with empirical distribution $\hat{\mathbb{P}}_{n+1}$. 
Then, $G$ defined at \eqref{eq:classical_cps} satisfies
\begin{equation*}
    \prob\Big(G_{X_{n+1}}^{\hat{\mathbb{P}}_{n+1}}(Y_{n+1}) < u\Big) \le u \le \prob\Big(G_{X_{n+1}}^{\hat{\mathbb{P}}_{n+1}}(Y_{n+1}-) \le u\Big) \quad \text{for all $u \in (0,1)$}.
\end{equation*}   
\end{cor}
\begin{proof}
    We apply part 1 of Theorem \ref{thm:master} with $G_0$ in place of $G$, a constant instead of the $X_i$, and $A(\hat{\mathbb{P}}_{n+1},(X_i,Y_i))$ instead of the $Y_i$. 
\end{proof}




\begin{rem}[Split conformal prediction]\label{rem:splitCM}
    In the split conformal setting, the conformal predictive systems of \cite{VovkEtAl2017} correspond to the bounds at \eqref{eq:PB1} and \eqref{eq:PB2} defined using the regression procedure defined at \eqref{eq:classical_cps}. 
    Conformal predictive systems are typically calculated using a splitting approach \citep{VovkEtAl2018}. \emph{Split} CM predictive systems estimate a conformity measure $A$ using the data in $\mathcal{D}_0$ that does not depend on $\hat{\mathbb{P}}$, that is, $A$ is constant in the first argument;
    in the splitting approach, $\hat{\mathbb{P}}$ is the empirical distribution of the calibration data $\mathcal{D}_1$. The data in $\mathcal{D}_1$ is then used to produce a CM predictive system using $A$. \emph{Full} CM predictive systems, on the other hand, use the full dataset to construct the CM predictive system with a conformity measure $A$ that non-trivially depends on $\hat{\mathbb{P}}$, which means that many potentially costly evaluations of $A$ are necessary. 
\end{rem}

\begin{rem}\label{rem:conf_order}
    Instead of a conformity measure $A$, one can obtain CM predictive systems by specifying a \emph{conformity order}, which is a total order $\le_{\hat{\mathbb{P}}}$ on $\mathcal{Z}$ that depends on $\hat{\mathbb{P}}$. Since only the values comprising $\hat{\mathbb{P}}$ are compared with respect to $\le_{\hat{\mathbb{P}}}$, it suffices to specify $\le_{\hat{\mathbb{P}}}$ for these elements. A conformity measure $A$ induces a conformity order by setting $z \le_{\hat{\mathbb{P}}} z'$ if $A(\hat{\mathbb{P}},z) \le A(\hat{\mathbb{P}},z')$. Conversely, for each conformity order, there are often many conformity measures that lead to the same predictive system.
\end{rem}


\subsection{Auto-calibration: Conformal binning}\label{sec:conf_bin}

While CM conformal predictive systems satisfy probabilistic calibration guarantees, Theorem \ref{thm:master} demonstrates that predictive systems with stronger calibration guarantees can be obtained from more strongly calibrated distributional regression procedures. The following proposition characterizes auto-calibrated distributional regression procedures; they all rely on some form of binning of the data.

\begin{prop}\label{prop:bin_ac}
    Suppose that $G$ is an auto-calibrated procedure. Then there exists a partition $B_1,\dots,B_{m'}$ of $\{1,\dots,m\}$ (that may depend on $\hat{\mathbb{P}}$) such that, for all $y \in \R$,
    \begin{equation}\label{eq:bin_ac}
    G_{x_k}^{\hat{\mathbb{P}}}(y) = \frac{1}{| B_j |} \sum_{i \in B_j} \one\{y_i \le y\}, \quad k \in B_j.
    \end{equation}
\end{prop}

\begin{rem}
    Proposition 2.4 shows that every auto-calibrated distributional regression procedure is a binning procedure. Using the same arguments, it is straightforward to prove the converse, that every binning procedure is auto-calibrated (in-sample). Hence, from Theorem \ref{thm:master}, by choosing any binning of the covariate-outcome pairs, we can construct a predictive system that is guaranteed to contain an auto-calibrated predictive distribution.
\end{rem}

While auto-calibration is a strong notion of calibration, a distributional regression procedure can be auto-calibrated without being informative. For example, the procedure $G_{0}$ at \eqref{eq:G0} is auto-calibrated, despite not containing any covariate information; the resulting distributional regression prediction is simply the empirical distribution of $y_1,\dots,y_m$. To exploit the information contained in the covariates, the binning should be such that pairs $(x_{i}, y_{i})$ in the same bin correspond to pairs with similar relationships between the covariates and outcomes. In practice, this is often simplified to identifying bins of similar covariate values.

\begin{exmp}\label{ex:kmeans}
    If it is possible to calculate the distance between elements in $\X$, then the covariates can be binned using clustering methods. In Sections \ref{sec:sims} and \ref{sec:app}, $\X \subseteq \R^d$, and bins are constructed using $k$-means clustering applied to the covariates, using the squared Euclidean distance as metric. A split conformal framework is adopted in which the clusters are determined from the estimation set $\mathcal{D}_{0}$. That is, let $k \in \N$ such that $k \leq n_0$. Then, given the covariate values $X_{1}, \dots, X_{n_0}$ from the covariate-observation pairs in $\mathcal{D}_{0}$, the partition $B_{1}, \dots, B_{k}$ of $\{1, \dots, n_{0}\}$ is found that satisfies
    \[
    \argmin_{\{B_{1}, \dots, B_{k}\}} \sum_{i=1}^{k} \sum_{j \in B_{i}} | X_{j} - \mu_{i} |^{2},
    \]
    where $\mu_{i}$ is the mean of the values $X_{j}$ for $j \in B_{i}$, $i = 1, \dots, k$. This results in mutually exclusive bins for which the squared distance between covariate values in different bins is maximised. New covariates, including those in $\mathcal{D}_{1}$, can be assigned to the bin whose mean $\mu_{i}$ is closest, and a prediction can then be made using the procedure described in \eqref{eq:bin_ac} (Proposition \ref{prop:bin_ac}).
\end{exmp}

\begin{exmp}\label{ex:IVAPD}
    \citet{NouretdinovVolkhonskiyETAL2018} propose an extension of Venn-Abers prediction to regression problems. They convert the covariates to some univariate \emph{score} or \emph{index}, and then perform isotonic mean regression for $Y_{1}, \dots, Y_{n_{0}}$ using the scores as covariates (in the estimation set $\mathcal{D}_{0}$). The isotonic regression fit yields bins of the covariate space corresponding to where the regression line is flat. As above, these bins can be used within Proposition \ref{prop:bin_ac} to generate an auto-calibrated predictive system.
\end{exmp}

\begin{rem}
    \cite{BostromEtAl2021} introduce Mondrian conformal predictive systems, which are similarly based on a form of binning of the data. However, Mondrian conformal predictive systems are not auto-calibrated. Instead, they are probabilistically calibrated conditional on the bins, which is weaker than auto-calibration. We therefore omit these predictive systems from the rest of our analysis.
\end{rem}

\subsection{Isotonic calibration: Conformal isotonic distributional regression}

Isotonic distributional regression (IDR) is a non-parametric distributional regression approach that is isotonically calibrated in-sample. Hence, using results from Section \ref{sec:insample}, we can obtain a ``conformalized'' version of IDR that yields a predictive system with isotonic calibration guarantees. While the construction of IDR is motivated by an assumption of isotonicity between the covariates and the outcome, its in-sample calibration guarantees do not depend on any assumptions on the data generating process, since they hold conditionally on the training sample. Furthermore, the calibration properties also remain valid if the order on the covariate space $\mathcal{X}$ is estimated either on the training sample or from different data, making it applicable within a splitting approach \citep[Section 4]{HenziKlegerETAL2023}.

For any $(x_1,y_1),\dots,(x_m,y_m) \in \mathcal{Z}$, with empirical distribution $\hat{\mathbb{P}}$, assume that $\le$ is a partial order on $\mathcal{X}$. Let $G_{x}^{\hat{\mathbb{P}}}$ be the IDR CDF at $x \in \mathcal{X}$, that is,
\begin{equation}\label{eq:G_IDR}
(G_{x_i}^{\hat{\mathbb{P}}}(y))_{i=1}^m = \argmin_{(\theta_1,\dots,\theta_m)} \sum_{i=1}^m (\theta_i - \one\{y_i \le y\})^2, \quad y \in \mathbb{R},
\end{equation}
where the argmin is taken over all $m$-tuples $(\theta_1,\dots,\theta_m) \in [0,1]^m$ such that $x_i \le x_j$ implies $\theta_j \le \theta_i$, and $x_i = x_j$ implies $\theta_j = \theta_i$. Then, for any choice of the partial order, the procedure $G$ is isotonically calibrated \citep[see][Section 4.2]{ArnoldZiegel2023}, and therefore also threshold calibrated, quantile calibrated, and probabilistically calibrated; see \citet[Theorem 2]{HenziZiegelETAL2021}, \citet[Proposition 5.2]{ArnoldZiegel2023}.

We define \emph{conformal IDR} as the predictive band $\Pi$ with bounds as at \eqref{eq:PB1}, \eqref{eq:PB2}, where $G$ is the IDR procedure described above. The following corollary is immediate from Theorem \ref{thm:master}.

\begin{cor}
Let $(X_1,Y_1),\dots,(X_n,Y_n),(X_{n+1},Y_{n+1})$ be an exchangeable sequence. Then, the conformal IDR predictive system $\Pi$ contains an isotonically calibrated predictive CDF for $Y_{n+1}$. More precisely,
\[
\Pi_{\ell}[\hat{\mathbb{P}}_n,X_{n+1}](y) \le 1-\mathbb{P}\left(Y_{n+1} > y \Big| \mathcal{A}\left(X_{n+1}\right)\right) \le \Pi_{u}[\hat{\mathbb{P}}_n,X_{n+1}](y), \quad y \in \R.
\]    
\end{cor}

Recall that $\mathcal{A}(X_{n+1})$ denotes the $\sigma$-lattice generated by $X_{n+1}$.

\citet{HenziKlegerETAL2023} introduce \emph{distributional single index models}, which fit IDR to suitable real-valued summary functions with the usual order on $\R$. In other words, and related to Remark \ref{rem:conf_order}, they suggest to estimate a data driven partial or total order on the covariate space. This is achieved by estimating a regression model $\hat{r}(\cdot)$ for the conditional mean (for example) of $Y$ given $X$, and using the order on $\mathcal{X}$ such that $x \le x'$ if and only if $\hat{r}(x) \le \hat{r}(x')$.

\begin{rem}[Splitting approach for IDR]\label{rem:splitIDR}
    In general, the computation of conformal IDR can be costly, and it is therefore recommended to implement it within the splitting approach. For example, if a data-driven order is used to fit IDR, then the real-valued summary function is learned from the estimation set $\mathcal{D}_0$, and does not require re-estimating for each possible outcome. 
    In this case, IDR is monotone in the sample values $y_i$. Let $y_i \le y_i'$ for all $i=1,\dots,m$, and let $\hat{\mathbb{P}}_m$ and $\hat{\mathbb{P}}_m'$ denote the empirical distributions of $(x_1, y_1), \dots, (x_m, y_m)$  and $(x_1, y_1'), \dots, (x_m, y_m')$, respectively. Then, with $G$ defined at \eqref{eq:G_IDR}, $G_{x_i}^{\hat{\mathbb{P}}_m}(y) \le G_{x_i}^{\hat{\mathbb{P}}_m'}(y)$ for all $y \in \R$ and $i=1,\dots,n$ (assuming the same ordering is used for the two IDR fits). Hence, for computing conformal IDR with a splitting approach, it is sufficient to compute two runs of IDR based on 
    \[
    (x_1,y_1),\dots,(x_n,y_n),(x_{n+1},\min_{i=1,\dots,n} y_i - C_1)\; \text{and}\; (x_1,y_1),\dots,(x_n,y_n),(x_{n+1},\max_{i=1,\dots,n} y_i + C_2),
    \]
    for some $C_1, C_2 > 0$, to obtain $\Pi_{\ell}[\hat{\mathbb{P}}_n,X_{n+1}](y)$ and $\Pi_{u}[\hat{\mathbb{P}}_n,X_{n+1}](y)$, respectively, since these will result in the lower and upper bounds of the forecast distributions for $Y_{n+1}$ obtained from $G$. The constants $C_1, C_2$ are somewhat arbitrary, but they allow for the possibility that we do not observe the full support of $Y_{n+1}$ in the training data: the values $\min_{i=1,\dots,n} y_i - C_1$ and $\max_{i=1,\dots,n} y_i + C_2$ are the smallest and largest point of the support of the IDR CDFs, respectively. The IDR solution is unchanged if one first transforms the $y$-values with a strictly increasing function, performs IDR, and then transforms back. The IDR fit therefore does not depend on $C_1$ and $C_2$, so it is not necessary to choose these before computation. 
\end{rem}

\begin{rem}
    Conformal IDR is a generalization of the Venn-Abers predictors of \citet{VovkPetejETAL2015} in the binary and categorical case, and is related to the isotonic regression-based conformal prediction systems proposed by \cite{NouretdinovVolkhonskiyETAL2018}. For binary outcomes, Venn-Abers prediction uses isotonic regression to construct conformal predictions with conditional calibration guarantees, extending the seminal work of \cite{ZadroznyElkan2002}. For real-valued outcomes, \cite{NouretdinovVolkhonskiyETAL2018} propose converting an outcome to a sequence of binary exceedances, depending on whether the outcome exceeds a sequence of thresholds. Venn-Abers prediction can then be performed at each threshold to obtain a sequence of probabilities containing the probability that the outcome will exceed each threshold, which constitutes a predictive distribution function for the real-valued outcome. However, there is no guarantee that the resulting predictive distributions will be valid; that is, the predictive distributions may not be increasing functions of the threshold. \cite{NouretdinovVolkhonskiyETAL2018} argue that this cannot be avoided, and instead introduce an alternative approach, see Example \ref{ex:IVAPD}. We argue that one can indeed obtain valid predictive distributions by performing multiple individual isotonic regression fits at different threshold levels, as motivated by \cite{NouretdinovVolkhonskiyETAL2018}, when formulating the setup slightly differently. This is exactly the conformal IDR approach.
\end{rem}

\begin{rem}
    Under the assumption that the conditional distributions of $Y$ given $X$ are stochastically ordered, it has been shown that IDR is uniformly consistent for the conditional distribution of $Y$ given $X$ (under some regularity assumptions): for an ordinal covariate $X$, the result can be found in \citet{El-BarmiMukerjee2005}; for a real-valued covariate $X \in \R$, the result is due to \citet{Mösching&Dümbgen}; for a higher-dimensional covariate, see \citet{HenziZiegelETAL2021}; and for an estimated partial order, see \citet{HenziKlegerETAL2023,BalabdaouiHenziETAL2024}. The isotonicity assumption for the conditional distributions is not necessary to obtain conformal calibration guarantees with IDR. Interestingly, IDR also remains consistent without the isotonicity assumption. \citet{GavrilopoulosZiegel2025} show that under some regularity assumptions, IDR is uniformly consistent for the isotonic conditional law $\mathcal{L}(Y \mid \mathcal{A}(X))$ \citep{ArnoldZiegel2023}, which can be understood as an isotonic projection of the usual conditional law. This means that the conformal IDR approximates the isotonic conditional law of $Y$ given the $\sigma$-lattice generated by $X$ for large $n$. Clearly, if the isotonicity assumption holds, then the isotonic conditional law is just the usual conditional law and there is some resemblance to the asymptotic conditional guarantees of \citet{ChernozhukovWuthrichETAL2021}. However, in the case of model misspecification, conformal IDR has stronger finite sample and asymptotic calibration guarantees than the methods in \citet{ChernozhukovWuthrichETAL2021}. Under the isotonicity assumption of stochastically ordered conditional distributions of $Y$ given $X$, the consistency of IDR implies that conformal IDR is a universal predictive system in the sense of \citet{Vovk2022}.
\end{rem}

\section{Prediction intervals from predictive systems}\label{sec:properties}

Predictive systems with out-of-sample calibration guarantees can be used to construct prediction intervals with out-of-sample coverage guarantees. Suppose that we wish to issue a $(1 - \alpha)$-level prediction interval for $Y_{n+1}$, for $\alpha \in (0, 1)$. Given a distributional regression procedure $G$, a central $(1 - \alpha)$-prediction interval can be obtained from the $\alpha/2$ and $1-\alpha/2$ quantiles of the predictive distribution $G_{X_{n+1}}^{\hat{\mathbb{P}}_n}$. However, these prediction intervals generally do not satisfy any out-of-sample coverage guarantees. Conformal prediction methods for regression have been proposed to address this, yielding a prediction interval $C_\alpha(X_{n+1})$ that, under the assumption of exchangeability, satisfies
\[
\mathbb{P}(Y_{n+1} \in C_{\alpha}(X_{n+1})) \ge 1 - \alpha.
\]
Strengthening this out-of-sample coverage guarantee has been the focus of a wealth of research in the conformal prediction literature. 

Given a predictive system $\Pi$ generated using the bounds at \eqref{eq:PB1} and \eqref{eq:PB2}, a central $(1 - \alpha)$-prediction interval can be defined as $C_\alpha^\Pi(X_{n+1}) = [L_\alpha^\Pi (X_{n+1}), U_\alpha^\Pi (X_{n+1})]$, where
\begin{align}
    L_\alpha^\Pi (X_{n+1}) & = \sup\{ y \in \R : \Pi_u[\hat{\mathbb{P}}, X_{n+1}](y) \le \alpha/2 \}, \label{eq:pi_lower}\\
    U_\alpha^\Pi (X_{n+1}) & = \inf\{ y \in \R : \Pi_\ell[\hat{\mathbb{P}}, X_{n+1}](y) \ge 1 - \alpha/2 \}. \label{eq:pi_upper}
\end{align}
That is, the lower bound of the prediction interval is the upper $\alpha/2$-quantile of $\Pi_u$, and the upper bound is the lower $1-\alpha/2$-quantile of $\Pi_\ell$. For any predictive system, $\Pi_\ell(y) \le \Pi_u(y)$ for all $y \in \R$, and hence $\Pi_\ell$ is stochastically larger than $\Pi_u$, meaning the quantiles derived from $\Pi_\ell$ are larger than those derived from $\Pi_u$. Defining the lower bound using $\Pi_u$ and the upper bound using $\Pi_\ell$ therefore yields a more conservative prediction interval than if both bounds were taken from one of $\Pi_\ell$ or $\Pi_u$. Since predictive systems do not necessarily have that $\Pi_\ell(y) \to 1$ as $y \to \infty$ and $\Pi_u(y) \to 0$ as $y \to -\infty$, it can be the case that $L_\alpha^\Pi (X_{n+1}) = -\infty$ or $U_\alpha^\Pi (X_{n+1}) = +\infty$, particularly for $\alpha$ close to 0 or 1.

The following theorem shows that predictive systems generated using the bounds at \eqref{eq:PB1} and \eqref{eq:PB2} produce prediction intervals with strong conditional out-of-sample coverage guarantees.

\begin{thm}\label{thm:pred_int}
    Let $(X_1,Y_1),\dots,(X_n,Y_n),(X_{n+1},Y_{n+1})$ be an exchangeable sequence, and let $\Pi$ be the predictive system defined by the bounds \eqref{eq:PB1} and \eqref{eq:PB2}, for some procedure $G$. Let $C_\alpha^{\Pi}(X_{n+1})= [L_\alpha^\Pi (X_{n+1}), U_\alpha^\Pi (X_{n+1})]$ be the corresponding prediction interval defined by \eqref{eq:pi_lower} and \eqref{eq:pi_upper}. 
    \begin{enumerate}
        \item[(i)] If $G$ is probabilistically calibrated, then 
        \[
        \mathbb{P}(Y_{n+1} \in C_\alpha^\Pi(X_{n+1}) ) \ge 1 - \alpha.
        \]
        \item[(ii)] If $G$ is isotonically calibrated, then 
        \[
        \mathbb{P}(Y_{n+1} < L_\alpha^\Pi (X_{n+1})\mid L_\alpha^G (X_{n+1})) \le \frac{\alpha}{2}, \quad \mathbb{P}(Y_{n+1} > U_\alpha^\Pi (X_{n+1})\mid U_\alpha^G (X_{n+1})) \le \frac{\alpha}{2},
        \]
        where $L_\alpha^G(X_{n+1})$ is the upper $\alpha/2$-quantile of $G_{X_{n+1}}^{\hat{\mathbb{P}}_{n+1}}$ and $U_\alpha^G(X_{n+1})$ is the lower $(1-\alpha/2)$-quantile of $G_{X_{n+1}}^{\hat{\mathbb{P}}_{n+1}}$.
        \item[(iii)] If $G$ is auto-calibrated, then 
        \[
        \mathbb{P}(Y_{n+1} \in C_\alpha^\Pi(X_{n+1}) \mid G_{X_{n+1}}^{\hat{\mathbb{P}}_{n+1}} ) \ge 1 - \alpha.
        \]
    \end{enumerate}
\end{thm}

\begin{rem}
    While probabilistic calibration and auto-calibration of $G$ immediately imply that the prediction interval $C_\alpha^\Pi(X_{n+1})$ has out-of-sample coverage guarantees, if $G$ is isotonically calibrated, then it does not necessarily hold that 
    \[
    \mathbb{P}(Y_{n+1} \in C_\alpha^\Pi(X_{n+1}) \mid \mathcal{A}(G_{X_{n+1}}^{\hat{\mathbb{P}}_{n+1}})) \ge 1 - \alpha.
    \]
    Instead, isotonic calibration implies that the bounds of the prediction interval satisfy a form of out-of-sample quantile calibration (see Section \ref{sec:calibration}), which itself implies that the prediction interval has the correct unconditional coverage out-of-sample, that is, $\mathbb{P}(Y_{n+1} \in C_\alpha^\Pi(X_{n+1}) ) \ge 1 - \alpha$. The calibration properties of prediction intervals generated from IDR are discussed in detail by \cite{AllenEtAl2026b}. 
\end{rem}

Theorem \ref{thm:pred_int} demonstrates that if a predictive system contains a calibrated predictive distribution, then prediction intervals derived from the (lower) $\alpha/2$-quantile of the upper bound and the (upper) $1-\alpha/2$-quantile of the lower bound are guaranteed to contain the out-of-sample outcome with the desired (conservative) coverage level. A corollary to this result is that conformal binning yields auto-calibrated prediction intervals, and conformal IDR yields isotonically calibrated prediction intervals. Assuming $G_{X_{n+1}}^{\hat{\mathbb{P}}_{n+1}}$ is an informative prediction, these correspond to strong notions of conditional coverage.

\section{Quantifying epistemic forecast uncertainty}\label{sec:epistemic}

\subsection{Choosing a crisp predictive CDF}\label{sec:crisp}

While predictive systems facilitate the construction of prediction intervals with out-of-sample calibration guarantees, practitioners are often required to issue a single predictive distribution function as their forecast. In this section, we discuss the choice of such a \emph{crisp} CDF that lies between the upper and lower bounds of a predictive system. For example, a straightforward choice is the midpoint between the upper bound $\Pi_u(y)$ and the lower bound $\Pi_\ell(y)$ within some large interval $y \in [D_1,D_2]$ encompassing the support of the training data $y_1,\dots,y_n$. Beyond $[D_1,D_2]$, the predictive CDF can be set to zero and one, respectively, thus constructing a valid CDF. 

For conformal IDR, an alternative approach can be derived by extending the work of \citet{VovkPetejETAL2015}, who considered conformal IDR for binary outcomes under the name of Venn-Abers predictors. In this case, conformal IDR outputs two predictive probabilities, corresponding to the prediction obtained from IDR when the $n+1$-th observation in the training data is assumed to be zero and one respectively. To combine these two probabilities into a single prediction, \cite{VovkPetejETAL2015} suggest to find the probability that minimizes the maximal possible loss under a proper scoring rule. For choosing a crisp CDF within conformal IDR, we could apply the same strategy pointwise to obtain the optimal threshold exceedances probabilities for all thresholds. This is discussed in detail in Appendix \ref{app:crisp}.

However, for conformal binning and conformal IDR, the predictive CDFs obtained directly from $G_{X_{n+1}}^{\hat{\mathbb{P}}_n}$ are also generally guaranteed to fall within the bounds defined by \eqref{eq:PB1} and \eqref{eq:PB2}. That is, the original predictive CDF generated by the distributional regression procedure falls within the corresponding predictive system; this is shown for conformal IDR in Proposition \ref{prop:cidr_crisp} in Appendix \ref{app:crisp}. The intuition behind this is that the lower and upper bounds in \eqref{eq:PB1} and \eqref{eq:PB2} correspond to the distributional regression procedure trained using an additional extreme training data point. When this point is extremely low, the original predictive CDFs are generally shifted downwards, generating the lower bound of the predictive system in \eqref{eq:PB1}. When the new observation is extremely large, the predictive CDFs are generally shifted upwards, corresponding to the upper bound in \eqref{eq:PB2}.

Hence, if a forecaster is to report their entire predictive distribution, then it suffices to provide the original predictive distribution learned from the training data and applied to the new covariate, without the need to construct a predictive system. However, as discussed in the previous section, if the full predictive distribution is desired because the goal is to issue prediction intervals, then the predictive system allows these to be derived with additional out-of-sample guarantees, which is not the case from the original predictive CDFs alone. Furthermore, in the remainder of this section, we argue that the predictive systems generated by conformal IDR provide useful information regarding the epistemic uncertainty in the forecast distribution, and it may often be useful to report this measure of epistemic uncertainty alongside the issued predictive distribution, regardless of how this crisp CDF is constructed.



\subsection{Thickness}\label{sec:thickness}

The out-of-sample calibration guarantees in Theorem \ref{thm:master} are only useful if the thickness of the predictive system is sufficiently small. For a calibration set $\mathcal{D}_1$ of size $n$, or a full conformal approach (without splitting), the thickness of a CM conformal predictive system is $1/(n+1)$. This can be seen from the bounds at \eqref{eq:CM_low} and \eqref{eq:CM_upp}. Similarly, the thickness of a conformal binning predictive system is usually equal to $1/(|B_{k}|+1)$, where $B_{k}$ is the bin in which the new covariate falls. This is true whenever the bins are determined based on the estimation set $\mathcal{D}_0$, or the bins are determined using only the covariate information $X_1,\dots,X_{n+1}$ (without splitting). If the bins correspond to groupings of similar covariate values, then the thickness will be larger when there are fewer previous covariate values that are similar to the new covariate.

\begin{figure}[t!]
\includegraphics[width=\textwidth]{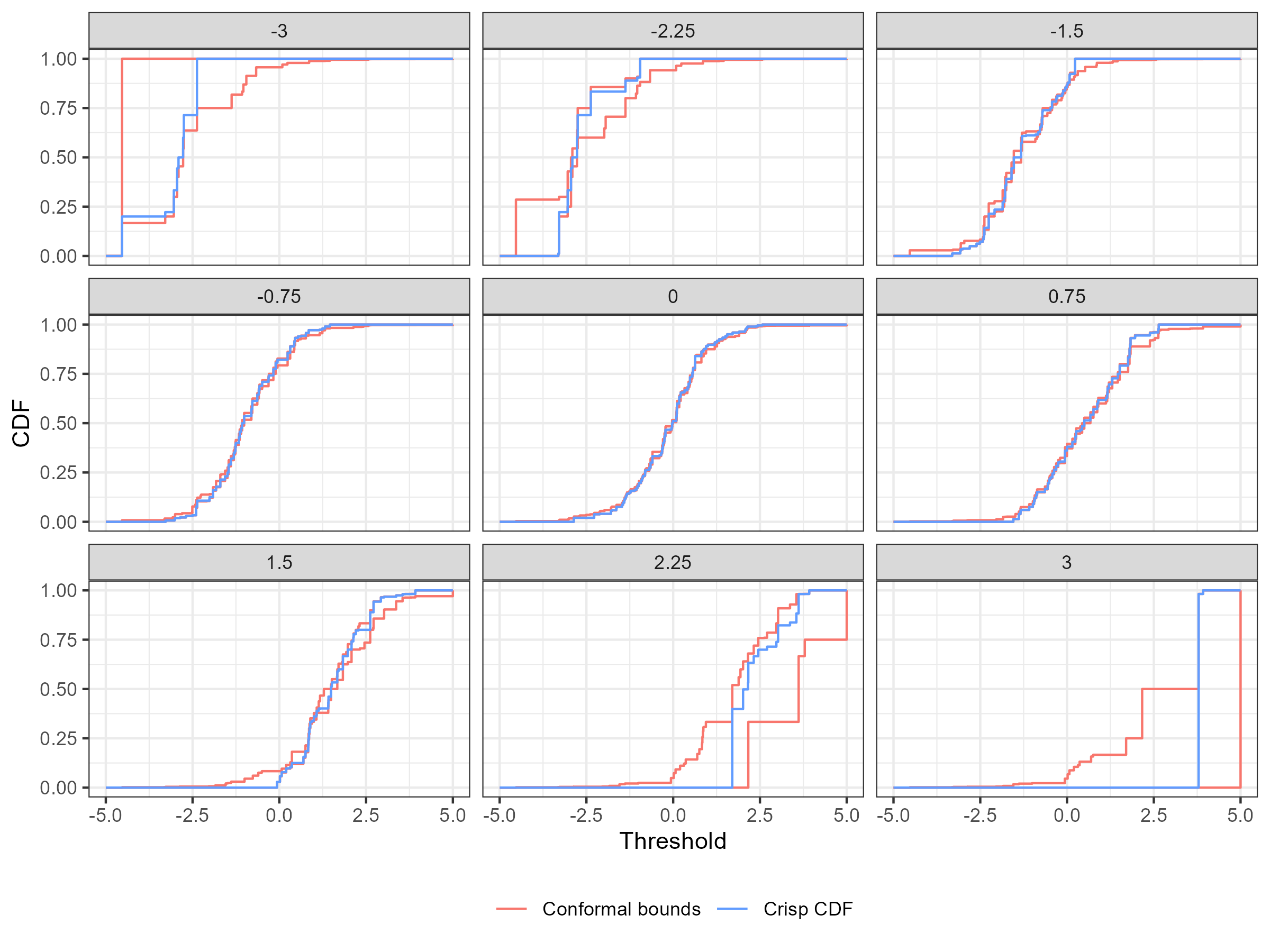}
\caption{Conformal IDR bounds (red) and crisp predictive distribution (blue) for different $X_{n+1}$. \label{fig:2}}
\end{figure}

In contrast, there is no simple bound on the thickness of conformal IDR. A first intuition is given by the following simple simulation example with iid observations $X_i \sim \mathcal{N}(0,1)$, $Y_i \sim \mathcal{N}(X_i,1)$ and sample size $n=512$. In Figure \ref{fig:2}, the conformal IDR bands are plotted in red for different possible values of $X_{n+1}$, while a specific choice for a crisp predictive CDF is given in blue. The thickness is large (even one) when $X_{n+1}$ is at the boundary of the distribution of the covariates, and small when it is in the center of the distribution. So, similarly to the conformal binning approach, and in line with intuition, the thickness depends on the local abundance of training data. We therefore argue that the thickness can be related to the epistemic uncertainty of the prediction, and we discuss this interpretation in Section \ref{sec:epistemic2}.

The following theorem demonstrates that, for a totally ordered covariate space, the expected thickness of the conformal IDR predictive system is bounded above, and that this bound tends to zero as $n$ increases. 

\begin{thm}\label{thm:ConfIDR_thickness}
Let $(X_1,Y_1),\dots,(X_n,Y_n),(X_{n+1},Y_{n+1})$ be an exchangeable sequence with (random) empirical distribution $\hat{\mathbb{P}}_{n+1}$. Conformal IDR has an expected thickness less than or equal to $14n^{-1/6}$, that is,
\[
\mathbb{E}\left(\Pi_u[\hat{\mathbb{P}}_n,X_{n+1}](y) - \Pi_\ell[\hat{\mathbb{P}}_n,X_{n+1}](y)\right) \le 14n^{-1/6}, \quad y \in \R.
\]
\end{thm}

Our proof also yields a lower bound on the expected thickness of conformal IDR of $n^{-1/3}$. The crucial ingredient for the proof of Theorem \ref{thm:ConfIDR_thickness} is a bound on the expected number of jumps of isotonic regression. The current proof uses a bound of $3n^{2/3}$ \citep[Lemma 1]{DimitriadisDumbgenETAL2023} that holds for every realization of isotonic regression when the response is binary, and under no conditions on the data generating process. \citet[Lemma 3.1]{Groeneboom2011} shows that, under some conditions that include isotonicity assumptions, the expected number of jumps of binary isotonic regression is of the order $n^{1/3}$. Under these assumptions, our bound can be improved to the order $n^{-1/3}$.


\subsection{Epistemic uncertainty}\label{sec:epistemic2}

Probabilistic forecasts provide more information than point forecasts since they take the form of entire predictive distributions, thereby quantifying the uncertainty in the future outcome. Intuitively, the spread of the predictions provides information regarding how predictable the future event is. However, these predictive distributions only model the aleatoric uncertainty of the outcome (uncertainty of the outcome, first-order probabilities), and do not account for the additional epistemic uncertainty (prediction uncertainty, second-order probabilities) that arises in the modelling process. While it is widely acknowledged that epistemic uncertainty (or forecast ambiguity) exists, this uncertainty is rarely treated or communicated to forecast users. 

Previous contributions in the econometrics literature on ambiguity and second-order probabilities for binary outcomes include the work of  \citet{Takeoka2007,Seo2009,DillenbergLlerasETAL2014,ChambersLambert2021}. 
\citet{EckelEtAl2012} have suggested ways to estimate and report ambiguity in ensemble weather forecasts, and have considered how valuable such information is to the decision maker. Epistemic uncertainty is also known in mathematical finance, where it is subsumed under the term model risk \citep{Cont2006}. In the machine learning literature, several methods have been proposed to go beyond quantifying only first-order probabilities, and seek to distinguish between epistemic and aleatoric uncertainty \citep{KendallGal2017,AbdarPourpanahETAL2021,HullermeierWaegeman2021}. These are critically discussed by \citet{MeinertGawlikowskiETAL2022,BengsHullermeierETAL2022}.  

We argue that conformal IDR provides a natural way to quantify the epistemic uncertainty of predictive distributions for real-valued outcomes via the thickness of the predictive system. A larger thickness implies that there is larger uncertainty about which distribution function should be issued as the forecast, whereas a small thickness implies small epistemic uncertainty. This interpretation is justified by the consistency of IDR: Using the notation of Theorem \ref{thm:master}, the unachievable target $G_{X_{n+1}}^{\hat{\mathbb{P}}_{n+1}}$ converges to the isotonic conditional law 
$\mathcal{L}\left(Y \mid \mathcal{A} \left(X_{n+1}\right)\right)$, and the thickness of the predictive system is exclusively due to the fact that we do not have the observation $Y_{n+1}$. The thickness (corresponding to epistemic uncertainty) goes to zero with $n \to \infty$ (Theorem \ref{thm:ConfIDR_thickness}), whereas even in the limiting case, given $X_{n+1}$, there is still (aleatoric) uncertainty about which value the realization of $Y_{n+1}$ will take.

While only one predictive distribution will generally be provided to forecast users, this forecast could be communicated to users alongside the thickness of the predictive bands. To ease the interpretation of the epistemic forecast uncertainty, we suggest to partition the thickness into a small number of bins, each corresponding to a different degree of epistemic uncertainty. We use the simple approach to define forecasts as \emph{low uncertainty}, \emph{medium uncertainty}, or \emph{high uncertainty} depending on whether the thickness is smaller than 0.25, between 0.25 and 0.5, or higher than 0.5, respectively. Other thresholds could obviously also be employed. While simple, this provides a traffic-light system that immediately and intuitively informs users how much epistemic uncertainty is present in the forecast, or, in other words, how informative the training data is for the next prediction instance. An example of how such a system can be employed in practice is provided in Section \ref{subsec:cs_weather}.

\section{Simulation results}\label{sec:sims}

We conduct a simulation study to compare the performance of prediction intervals obtained from CM predictive systems, conformal IDR, and conformal binning. The prediction intervals are obtained as described in Section \ref{sec:properties}. The CM predictive system that we study is the LSPM (Example \ref{ex:lspm}). The conformal binning distributions are constructed as described in Example \ref{ex:kmeans}, and are always implemented with $k=10$ bins for simplicity. All methods are implemented in a full conformal setting.

\begin{figure}
    \centering
    \includegraphics[width=0.3\linewidth]{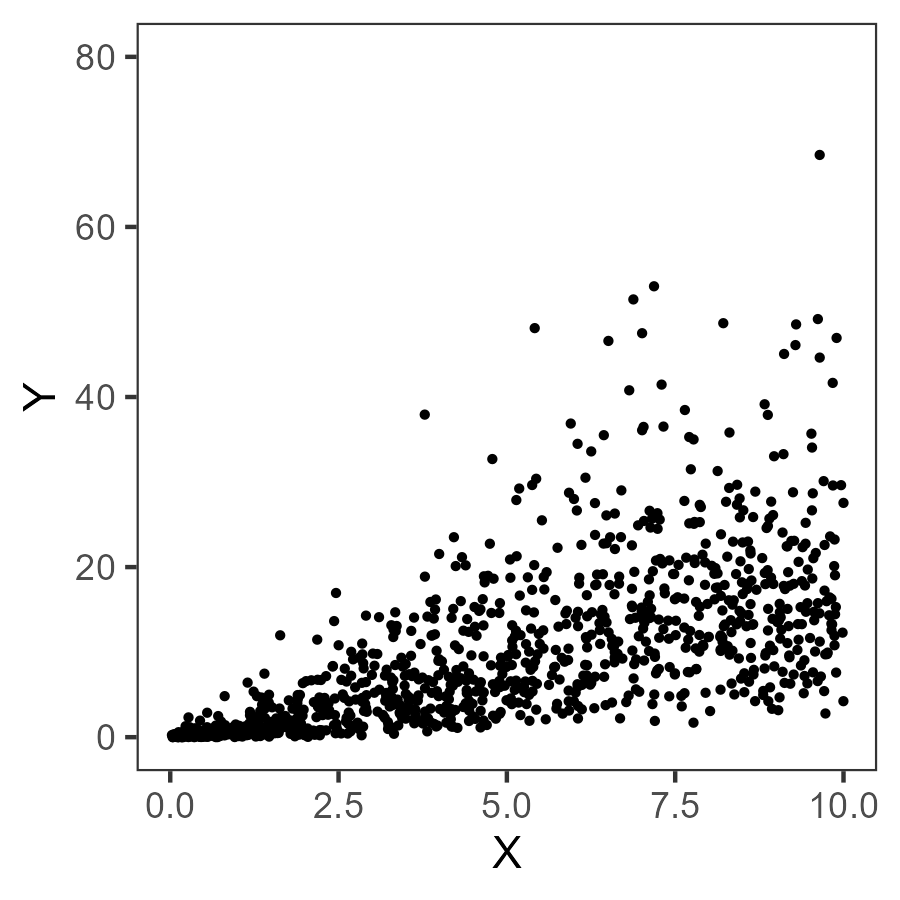}
    \caption{1000 points generated using the model at \eqref{eq:simstudy_data}.}
    \label{fig:simstudy_data}
\end{figure}

Following \cite{HenziZiegelETAL2021}, consider a covariate $X \sim \text{Unif}(0, 10)$, with corresponding outcome
\begin{equation}\label{eq:simstudy_data}
    Y \mid X \sim \text{Gamma}(\text{shape} = \sqrt{X}, \text{scale} = \min \{ \max \{ X, 1\}, 6 \}). 
\end{equation}
This presents a case where the uncertainty in the outcome depends on the covariate; Data drawn from this model are displayed in Figure~\ref{fig:simstudy_data}. We use a training set of $n = 2000$ independent draws from \eqref{eq:simstudy_data}, and the three predictive systems are then evaluated on an independent test set of size $5000$. Results for other choices of $n$ are analogous, as shown in Figure~\ref{fig:simstudy_app} in Appendix \ref{app:sim}.

\begin{figure}[t]
    \centering
    \includegraphics[width=0.49\linewidth]{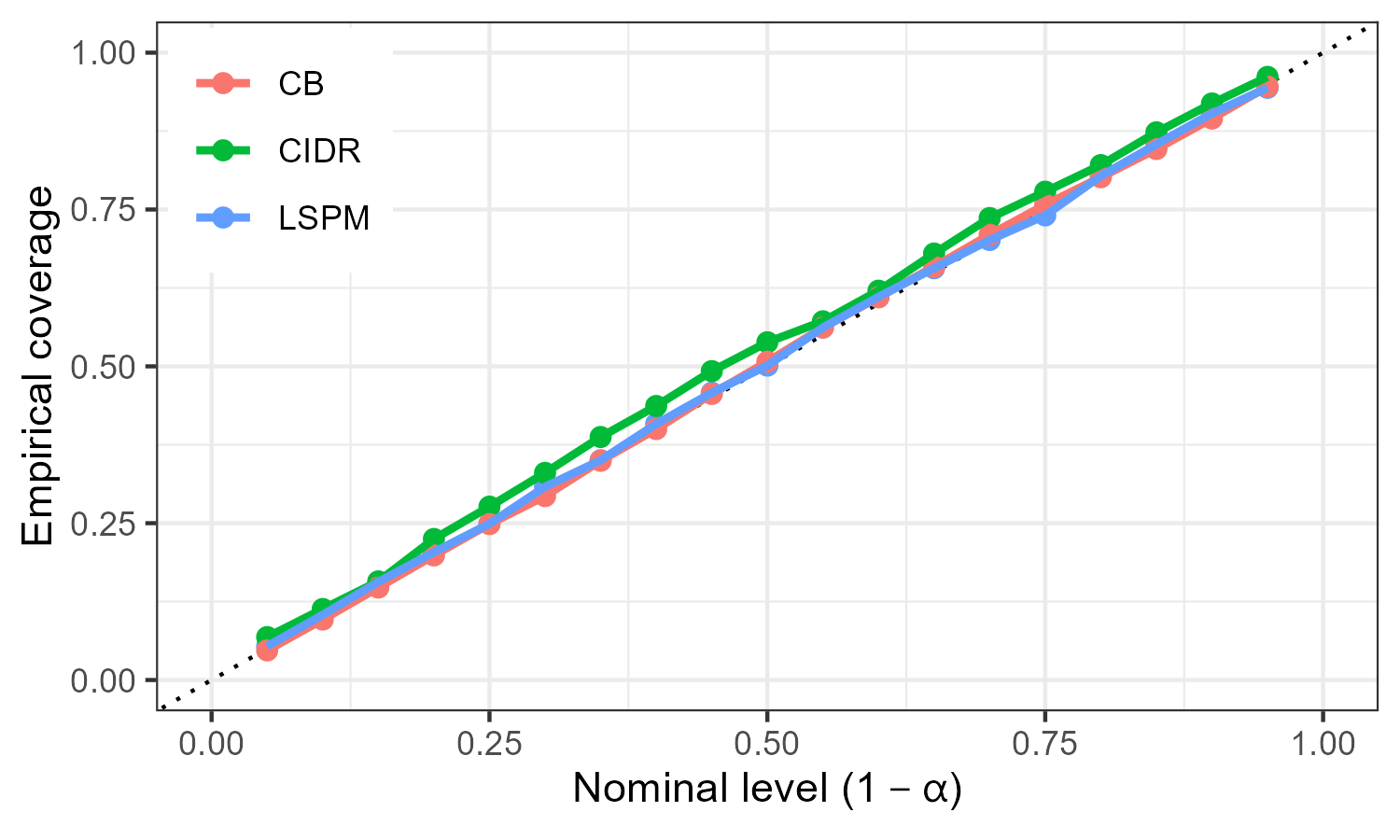}
    \includegraphics[width=0.49\linewidth]{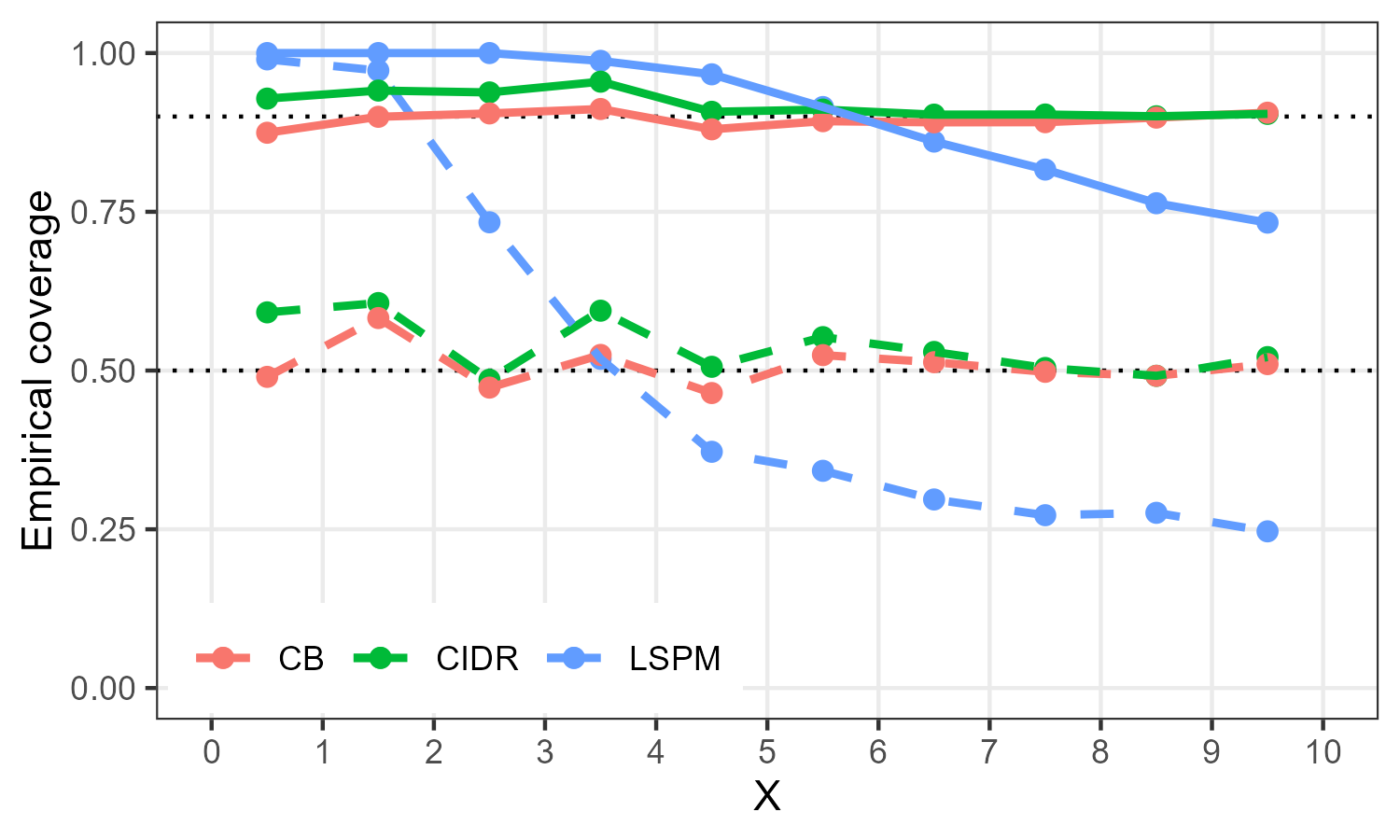}
    \includegraphics[width=0.49\linewidth]{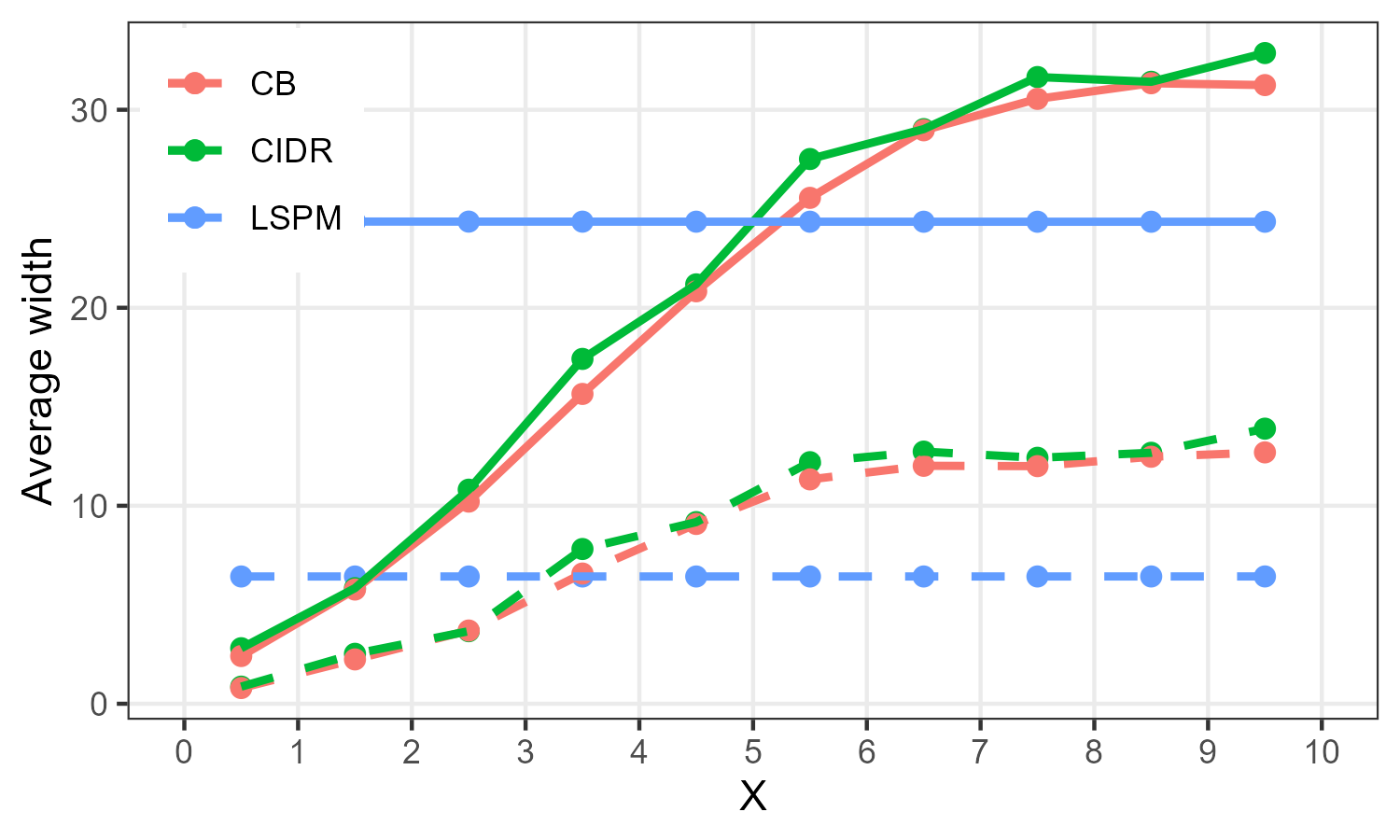}
    \includegraphics[width=0.49\linewidth]{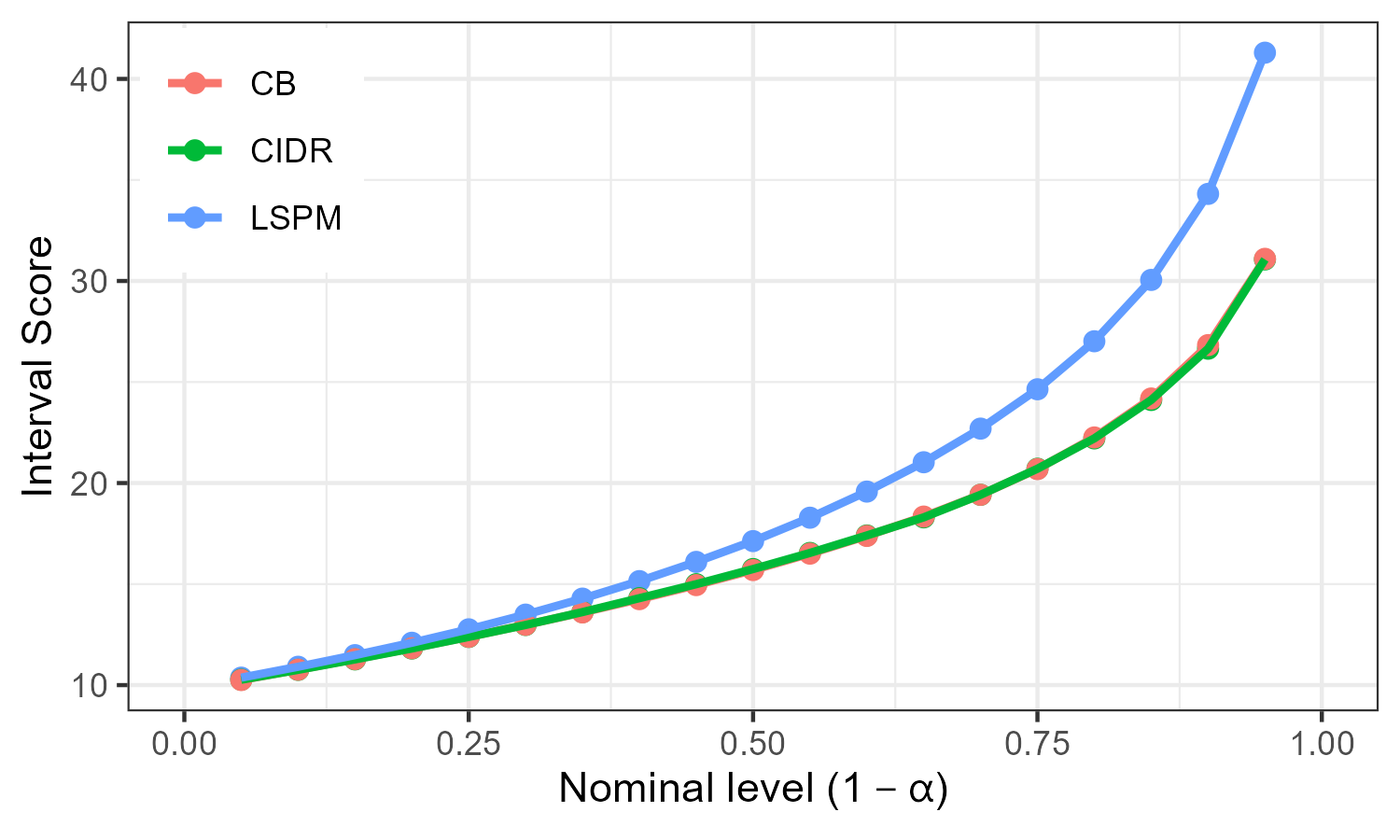}
    \caption{Evaluation of $(1 - \alpha)$-prediction intervals obtained from conformal binning (CB), conformal IDR (CIDR), and the LSPM when predicting the simulated data generated from \eqref{eq:simstudy_data}. Top left: Empirical coverage shown as a function of $\alpha$; the dotted diagonal line is indicative of correct unconditional coverage. Top right: Empirical coverage of the 50\% (dashed) and 90\% (solid) prediction intervals shown as a function of the covariate value $X$; the dotted lines at 0.5 and 0.9 are indicative of correct conditional coverage given $X$. Bottom left: Average width of the 50\% (dashed) and 90\% (solid) prediction intervals shown as a function of the covariate value $X$. Bottom right: Average interval score shown as a function of $\alpha$.}
    \label{fig:simstudy_cov}
\end{figure}

Figure \ref{fig:simstudy_cov} displays diagnostic plots comparing the prediction intervals obtained from the three predictive systems. The top left panel shows the empirical (unconditional) coverage of the $(1 - \alpha)$-prediction intervals obtained from the three predictive systems, for a range of $\alpha$. For all $\alpha$, the empirical coverage of the three predictive systems closely aligns with the nominal coverage level $1 - \alpha$. However, the top right panel of Figure \ref{fig:simstudy_cov} demonstrates that the coverage of the LSPM does not hold conditionally on the covariate information. In particular, this plot shows the empirical coverage of the 50\% and 90\% prediction intervals conditional on each event $X \in [0, 1), X \in [1, 2), \dots, X\in [9, 10]$. While the prediction intervals obtained from conformal binning and conformal IDR exhibit roughly the correct coverage regardless of the covariate value, the LSPM prediction intervals do not adapt depending on $X$, and they therefore over-cover the observation when uncertainty is low, and under-cover when uncertainty is high; these biases cancel out so that the intervals are unconditionally calibrated. This reflects that conformal binning and conformal IDR exhibit stronger calibration guarantees than the LSPM. 

This also manifests in the interval scores assigned to the prediction intervals, displayed in the bottom right panel of Figure \ref{fig:simstudy_cov}. The interval score is a proper scoring rule that quantifies the accuracy of prediction intervals \citep{GneitingRaftery2007}, assessing both the information content of the interval forecasts, as well as their (conditional) calibration. Figure \ref{fig:simstudy_cov} demonstrates that the prediction intervals obtained by conformal binning and conformal IDR receive a lower (better) interval score than those obtained by the LSPM. The reason for this is again that the width of the conformal binning and conformal IDR prediction intervals adapts depending on $X$, whereas it does not for the LSPM. This is illustrated for 90\% prediction intervals in the bottom left Figure \ref{fig:simstudy_cov}.

In \eqref{eq:simstudy_data}, there is a (weak) isotonic relationship between the covariate and the outcome, thereby satisfying the key assumption underlying IDR. However, the calibration guarantees of conformal IDR do not rely on the isotonicity assumption, and we verify this empirically in a second simulation study in Appendix \ref{app:sim}. Moreover, while we evaluate the predictive systems via the prediction intervals obtained from them, since we believe this is where the greatest practical benefit of the calibration guarantees formalized in Section \ref{sec:main} lies, analogous conclusions can also be drawn in other forecasting settings. Figure \ref{fig:simstudy} in Appendix \ref{app:sim} displays probabilistic and threshold calibration diagnostic plots for the crisp CDFs corresponding to the three predictive systems. A similar pattern emerges: all forecasts are unconditionally calibrated, but the LSPM predictive distributions are not conditionally calibrated given the covariate information, and they therefore perform worse when assessed using standard scoring rules for probabilistic forecasts.


\section{Applications}\label{sec:app}


\subsection{ICU patient length of stay forecasts}

Consider an application to forecasts for the patient length of stay (LoS) in Swiss intensive care units (ICUs). As described in \cite{HenziKlegerETAL2023}, the Minimal Dataset of the Swiss Society of Intensive Care Medicine (MDSi) contains patient data regarding admissions to ICUs in Switzerland. We consider a subset of the MDSi dataset that corresponds exactly to the data used in \cite{HenziKlegerETAL2023}. This subset contains data for 18 ICUs in Switzerland, all of which have more than 10,000 admissions. We adopt a split conformal approach, in which the data is partitioned into estimation, calibration, and test sets. The test set contains the most recent 25\% of the data, with the calibration and estimation data comprising a random one and two thirds of of the remaining data respectively. 

As well as the length of stay of all patients, the dataset also contains several variables (including illness severity score, nursing workload score, age and sex of the patient) that may be informative when predicting LoS. These variables are used to obtain a real-valued prediction for the LoS by fitting a linear regression model to the log-transformed LoS in the estimation set. A separate model is fit to each ICU. The data, pre-processing steps, and regression model are discussed in detail in \cite{HenziKlegerETAL2023}. The three conformal prediction methods are applied to the predictions obtained using the regression model, so that $\mathcal{X} = \R_{+}$ and $\mathcal{Y} = \R_{+}$.

\begin{figure}[t]
    \centering
    \includegraphics[width=0.49\linewidth]{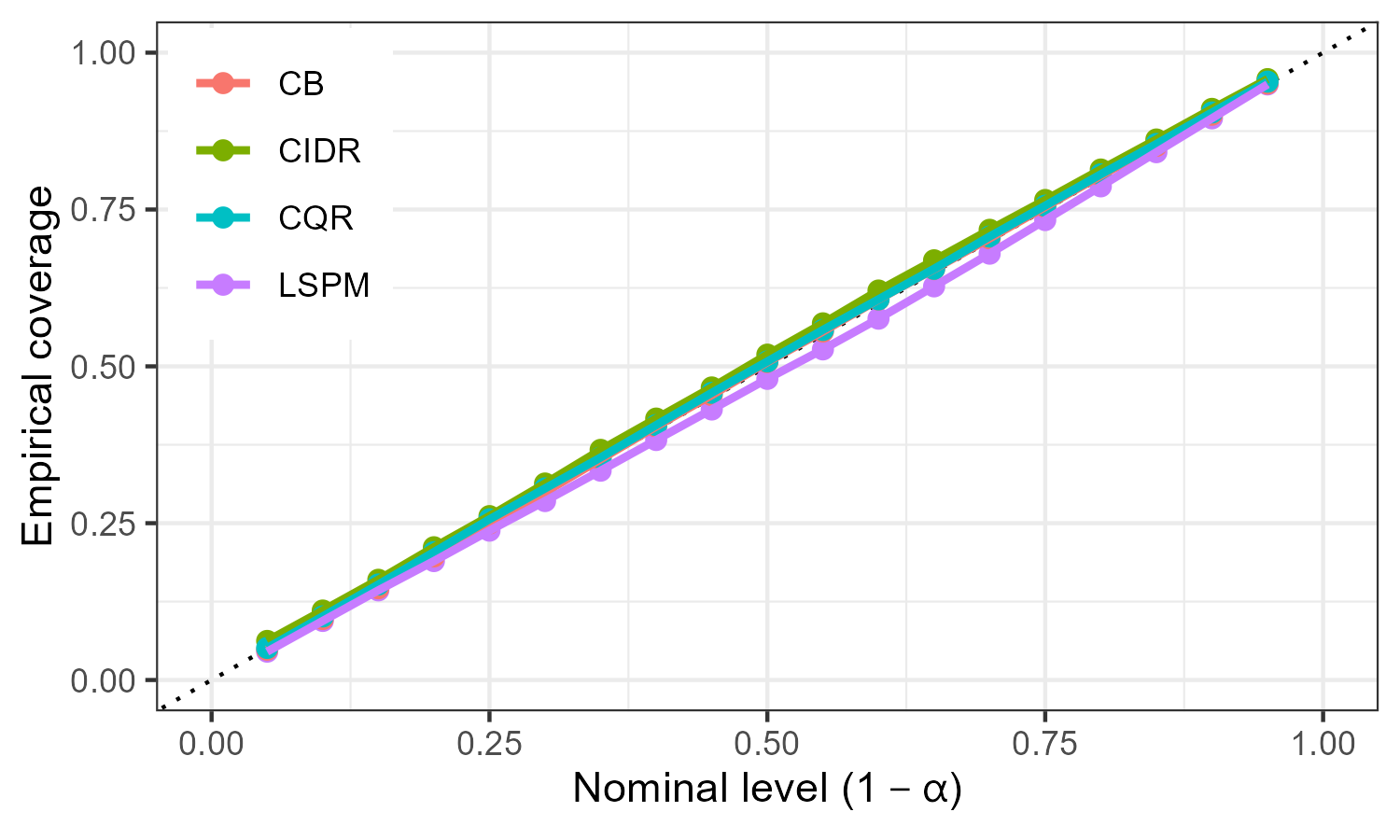}
    \includegraphics[width=0.49\linewidth]{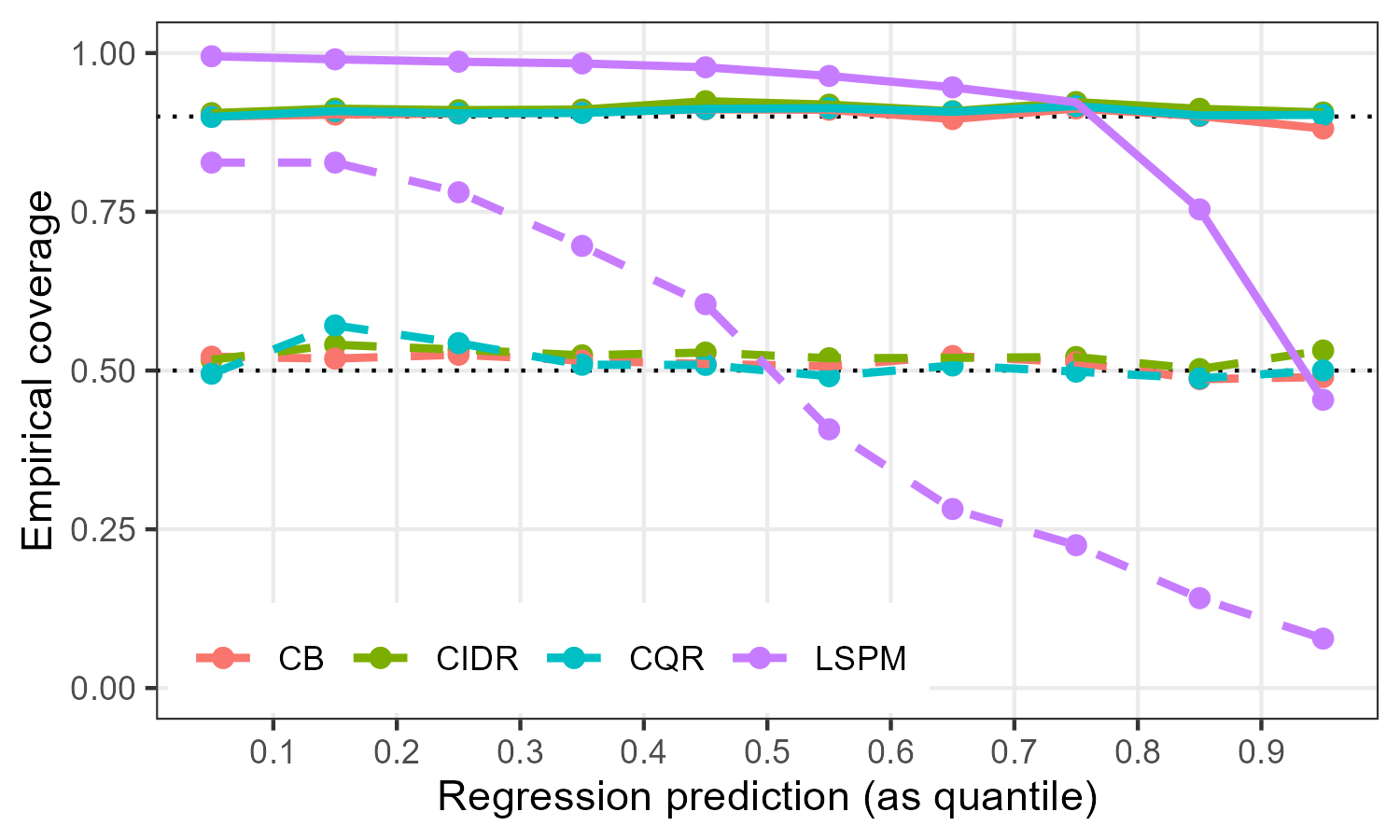}
    \includegraphics[width=0.49\linewidth]{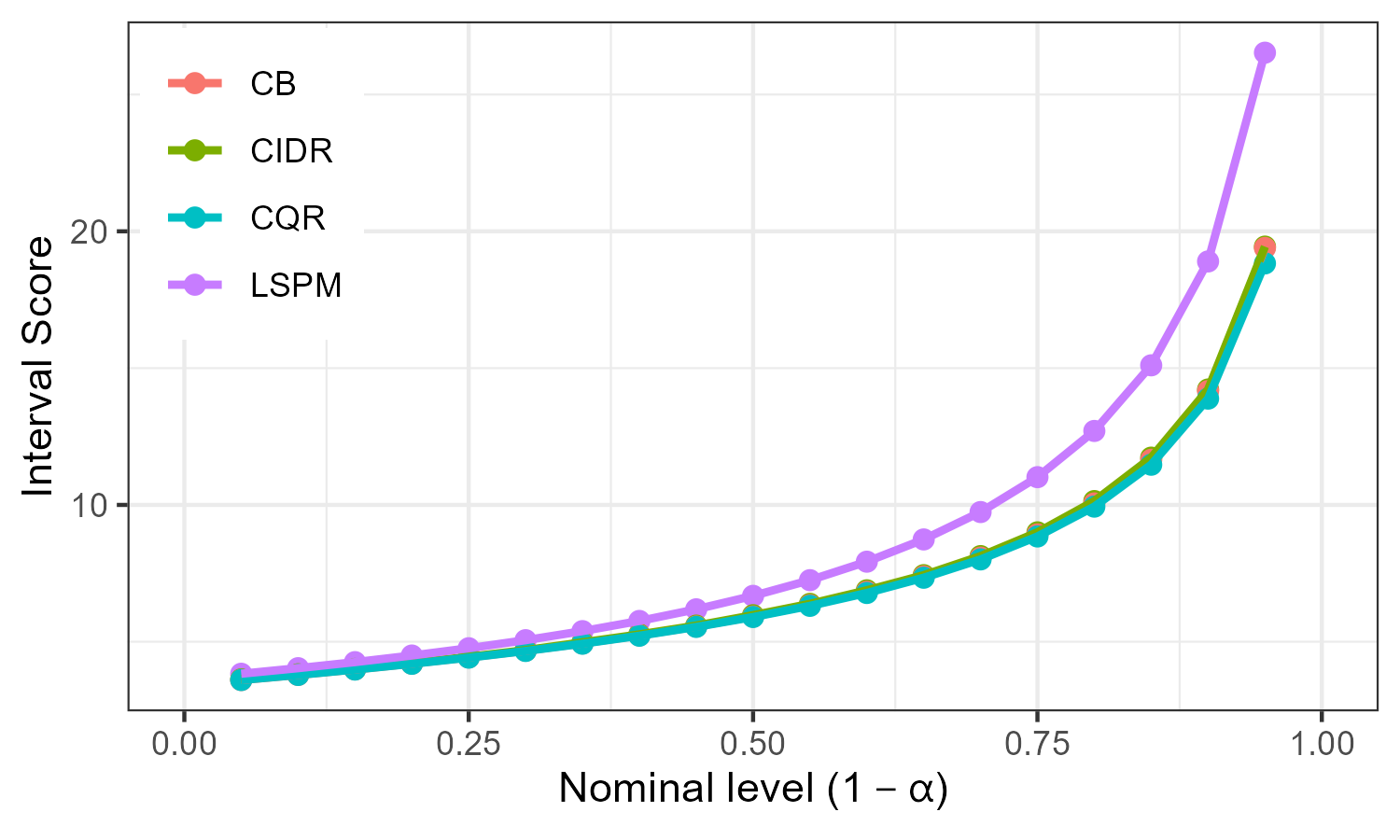}
    \caption{Evaluation of $(1 - \alpha)$-prediction intervals obtained from conformal binning (CB), conformal IDR (CIDR), the LSPM, and conformalized quantile regression (CQR) when predicting the length of stay in ICUs. Results are averaged across all 18 ICUs. Left: Empirical coverage shown as a function of $\alpha$; the dotted diagonal line is indicative of correct unconditional coverage. Right: Empirical coverage of the 50\% (dashed) and 90\% (solid) prediction intervals shown as a function of the covariate value used to generate the predictive systems; the covariate value is converted to a quantile level to account for differences between stations, so that a value between 0 and 0.1 implies the covariate was in the lower decile of covariate values at the relevant ICU; the dotted lines at 0.5 and 0.9 are indicative of correct conditional coverage given the covariate. Bottom: Average interval score shown as a function of $\alpha$.}
    \label{fig:icu_cov}
\end{figure}

CM predictive systems, conformal binning, and conformal IDR are again evaluated via prediction intervals generated from them. Figure \ref{fig:icu_cov} displays the unconditional coverage of the prediction intervals at different levels, as well as conditional coverage of the 50\% and 90\% prediction intervals given the covariate value; in this case, the single covariate is the regression model prediction described above. As for the simulated data in Section \ref{sec:sims}, the three predictive systems all yield prediction intervals that exhibit the correct unconditional coverage at all levels, and the conformal binning and conformal IDR prediction intervals also exhibit the correct coverage conditional on different values of the covariate value. In contrast, the LSPM issues prediction intervals that are not conditionally calibrated, over-covering the length of stay when the covariate value is small, and under-covering when the covariate is large. As a result, when the prediction intervals are evaluated using a proper scoring rule such as the interval score, conformal binning and conformal IDR are clearly preferred (bottom panel of Figure \ref{fig:icu_cov}).

While predictive systems can be used to obtain prediction intervals with out-of-sample calibration guarantees, numerous other conformal prediction methods have also been designed for this purpose. The goal of this work is not to provide a detailed comparison of different approaches. However, for comparison, Figure \ref{fig:icu_cov} additionally presents results for the widely-used conformalized quantile regression (CQR) approach of \cite{RomanoEtAl2019}. CQR is implemented here using cubic regression splines to estimate the interval bounds, estimated on the estimation data; this was found to perform better than using quantile regression forests. CQR does not yield a predictive system, but instead produces a prediction interval at a chosen coverage level $\alpha \in (0, 1)$. It can then be applied at every coverage level of interest. The CQR prediction intervals are guaranteed to exhibit the correct unconditional coverage out-of-sample, under the assumption of exchangeability, and this is verified empirically for the LoS predictions in the left panel of Figure \ref{fig:icu_cov}. While CQR intervals do not necessarily satisfy any conditional calibration properties, the interval bounds are designed to adapt depending on covariate information, and they therefore tend to exhibit reasonable conditional calibration in practice; this also appears to be the case when predicting LoS, as demonstrated in the right panel of Figure \ref{fig:icu_cov}. As a result, the interval score assigned to the CQR intervals is almost indistinguishable from the score assigned to prediction intervals obtained from conformal binning and conformal IDR.

\begin{figure}[t]
    \centering
    \includegraphics[width=0.4\linewidth]{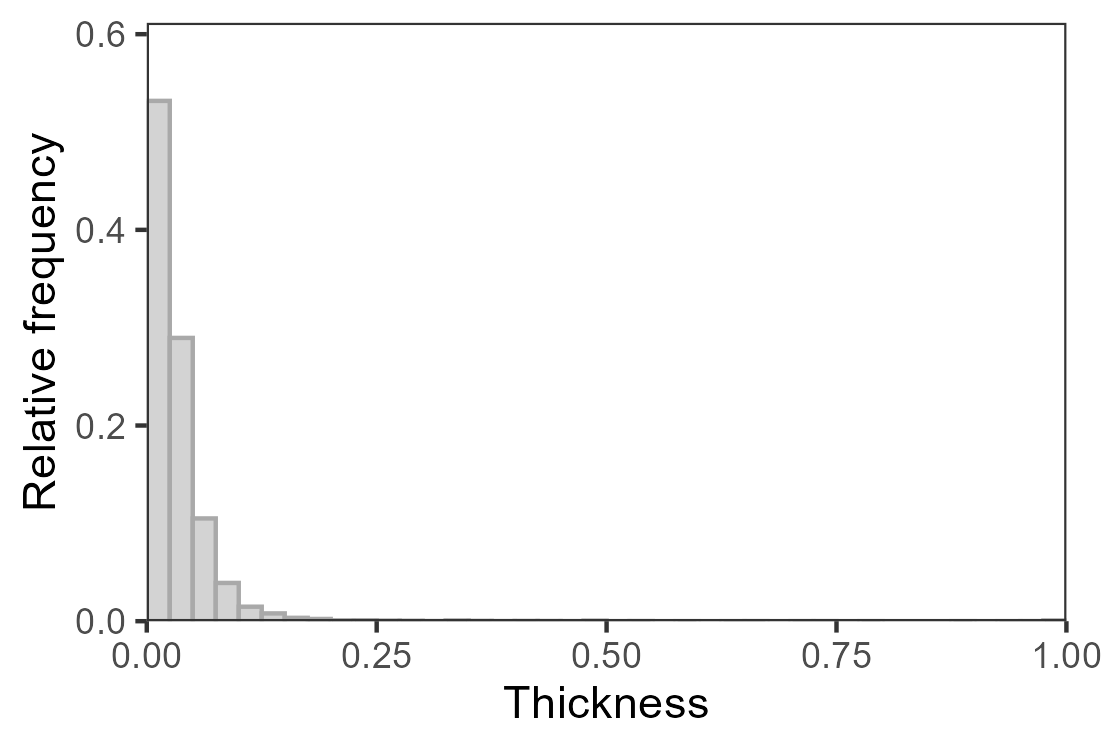}
    \includegraphics[width=0.4\linewidth]{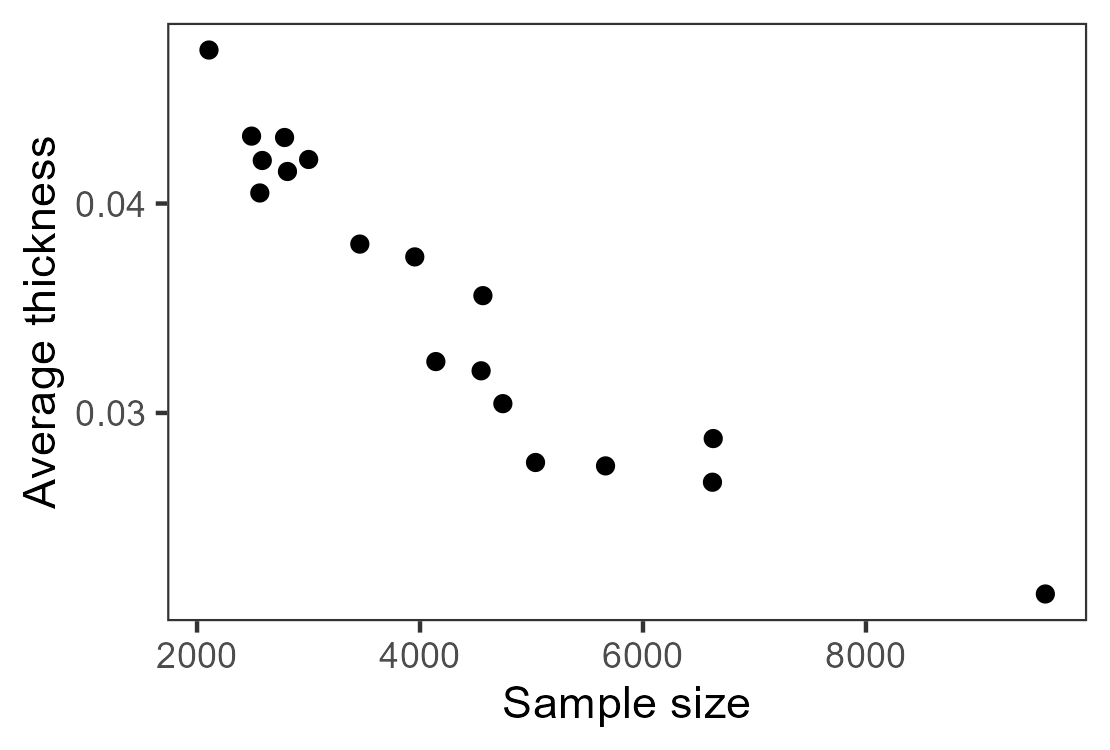}
    \caption{Left: Histogram of the thickness of the conformal IDR bands across all ICUs. Right: Average thickness for each ICU against the corresponding sample size in the training data.}
    \label{fig:ICU_thick}
\end{figure}

Figure \ref{fig:ICU_thick} displays a histogram of the thickness of the conformal IDR predictive bands. The thickness is generally close to zero, suggesting the choice of a crisp CDF would not be too important in this application. There are 54 forecast cases (out of the 77312 in the test data set) where the thickness is equal to one. There seems to be no obvious relationship between the thickness and other covariates, or the observed outcomes. However, there is a strong decreasing relationship between the average thickness of the predictive bands and the amount of data available on which to train the model; Figure \ref{fig:ICU_thick} also shows the average thickness of the forecasts made at each ICU against the corresponding training data size.

\begin{figure}
    \centering
    \includegraphics[width=0.4\linewidth]{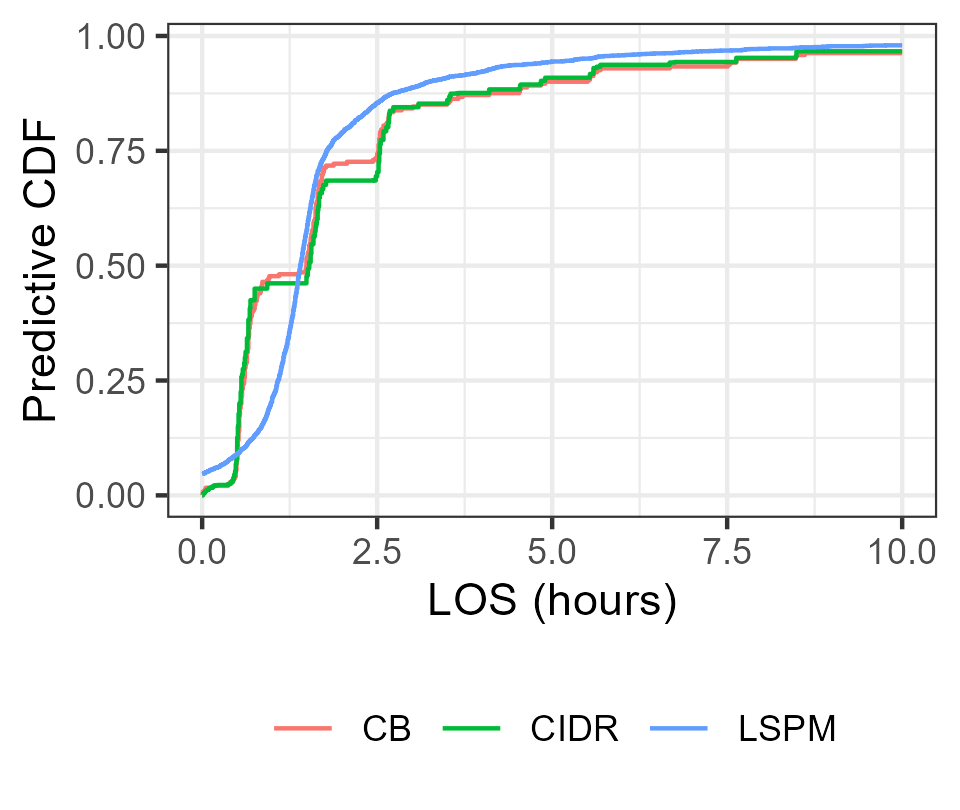}
    \includegraphics[width=0.4\linewidth]{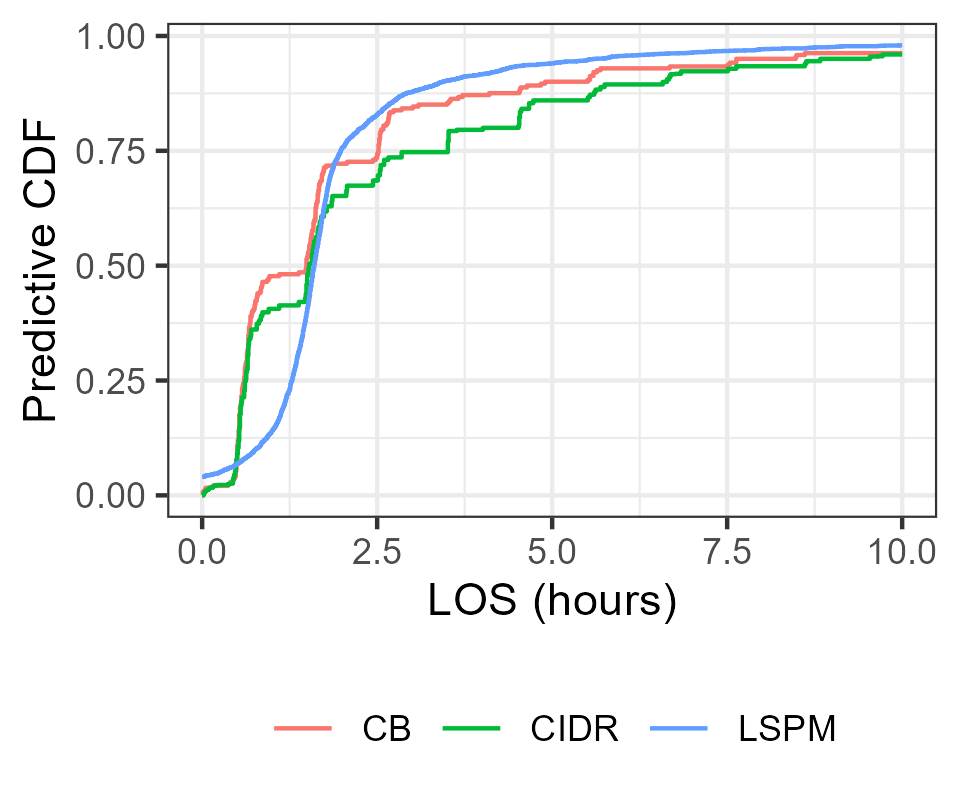}
    \caption{Examples of (crisp) predictive distributions obtained from the LSPM, conformal IDR, and conformal binning predictive systems, for two randomly selected patients.}
    \label{fig:ICU_CDF_ex}
\end{figure}

Finally, Figure \ref{fig:ICU_CDF_ex} displays examples of two crisp CDFs obtained by the LSPM, conformal IDR and conformal binning. A key difference between the CDFs is that conformal IDR and conformal binning more accurately capture the behavior of LoS. Patients are generally discharged from ICU units at similar times of the day, less often at night for example, and the distribution of LoS therefore exhibits several jumps at these times of the day. This behavior is present in the conformal IDR and conformal binning distributions, but not the LSPM CDFs, which are much smoother and hence unrealistic.

\subsection{European temperature forecasts}\label{subsec:cs_weather}

\begin{figure}
    \centering
    \vspace*{-1.5cm}
    \includegraphics[width=0.4\linewidth]{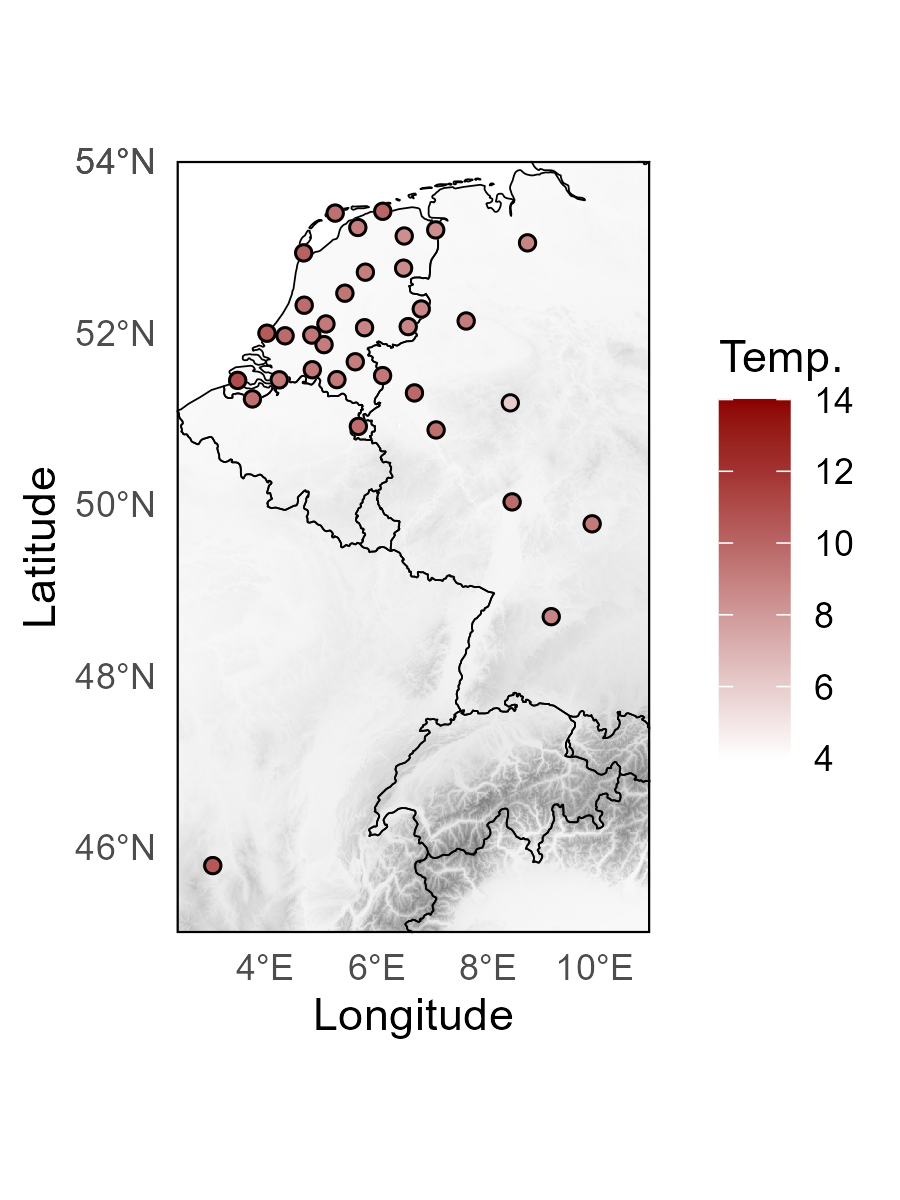}
    \raisebox{1.3cm}{\includegraphics[width=0.4\linewidth]{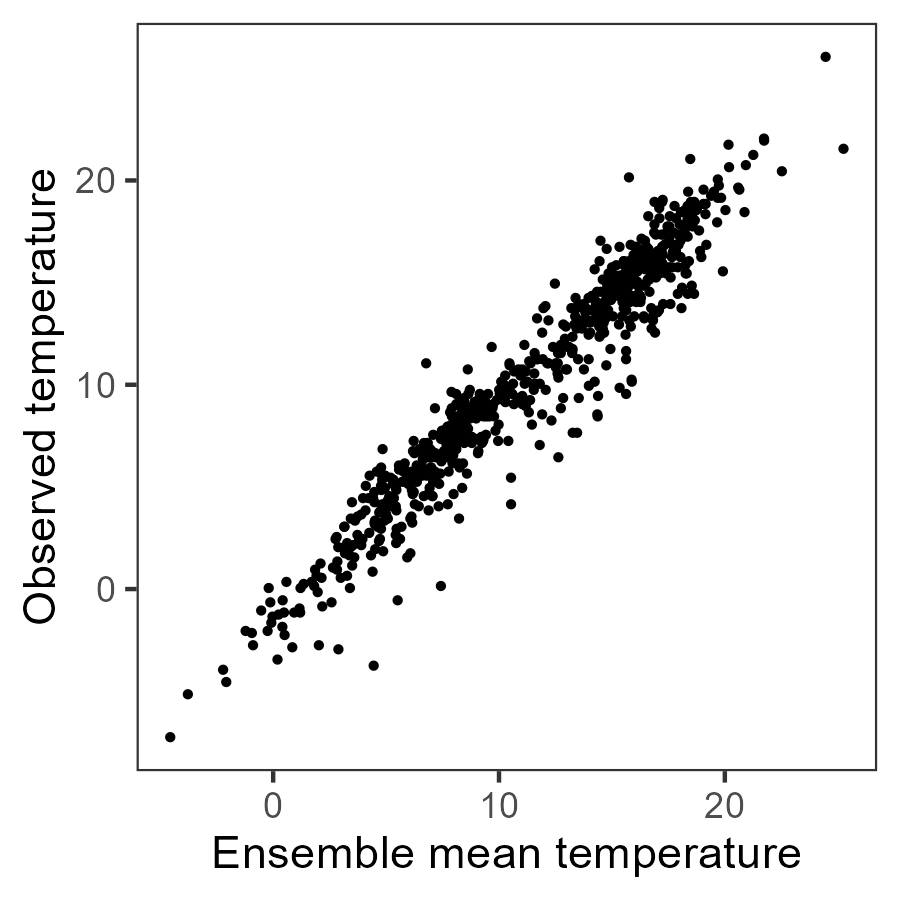}}
    \vspace{-1cm}
    \caption{Left: Stations at which forecasts are considered, with their average temperatures. Right: Ensemble mean temperature forecasts and observations (in degrees celcius) at one station.}
    \label{fig:EUMN_data}
\end{figure}

Consider a second application in the context of weather forecasting. While numerical weather models exist to predict surface weather variables, the output of such models is generally miscalibrated. This output therefore requires statistical post-processing to generate reliable probabilistic forecasts. In this application, we compare the LSPM, conformal IDR, and the conformal binning approach when statistically post-processing daily mean temperature forecasts at 35 weather stations across central Europe, displayed in Figure \ref{fig:EUMN_data}. The forecasts are issued 24 hours in advance.

The data is part of the EUPPBench dataset, which is a benchmark dataset for the comparison of statistical post-processing methods \citep{DemaeyerEtAl2023}. The 35 stations in Figure \ref{fig:EUMN_data} are those used in the EUPPBench dataset that do not have missing observation data. The forecasts take the form of ensembles, i.e. discrete predictive distributions, and we take the mean of these ensemble forecasts as a single univariate real-valued covariate. There is a clear isotonic relationship between the ensemble mean temperature and the corresponding observation (Figure \ref{fig:EUMN_data}). 

The dataset consists of two years of daily forecasts and observations between 2017 and 2018, as well as an archive of forecasts issued in the preceding 20 years; the archive consists of 4180 forecasts and observations at each station \citep[see][for details]{DemaeyerEtAl2023}. The three conformal prediction methods are fit separately at each weather station, and, to account for seasonal trends in temperature, the training data consists of all forecasts in previous years that were issued within a rolling window of 90 days centred around the forecast day. The two years of daily forecasts are then used as test data. A full conformal prediction framework is adopted, in which no splitting of the training data is performed. As described in Example \ref{ex:kmeans}, the bins in the conformal binning approach are obtained using $k$-means clustering. The number of bins is estimated using cross-validation in the training data.

\begin{figure}[t]
    \centering
    \includegraphics[width=0.49\linewidth]{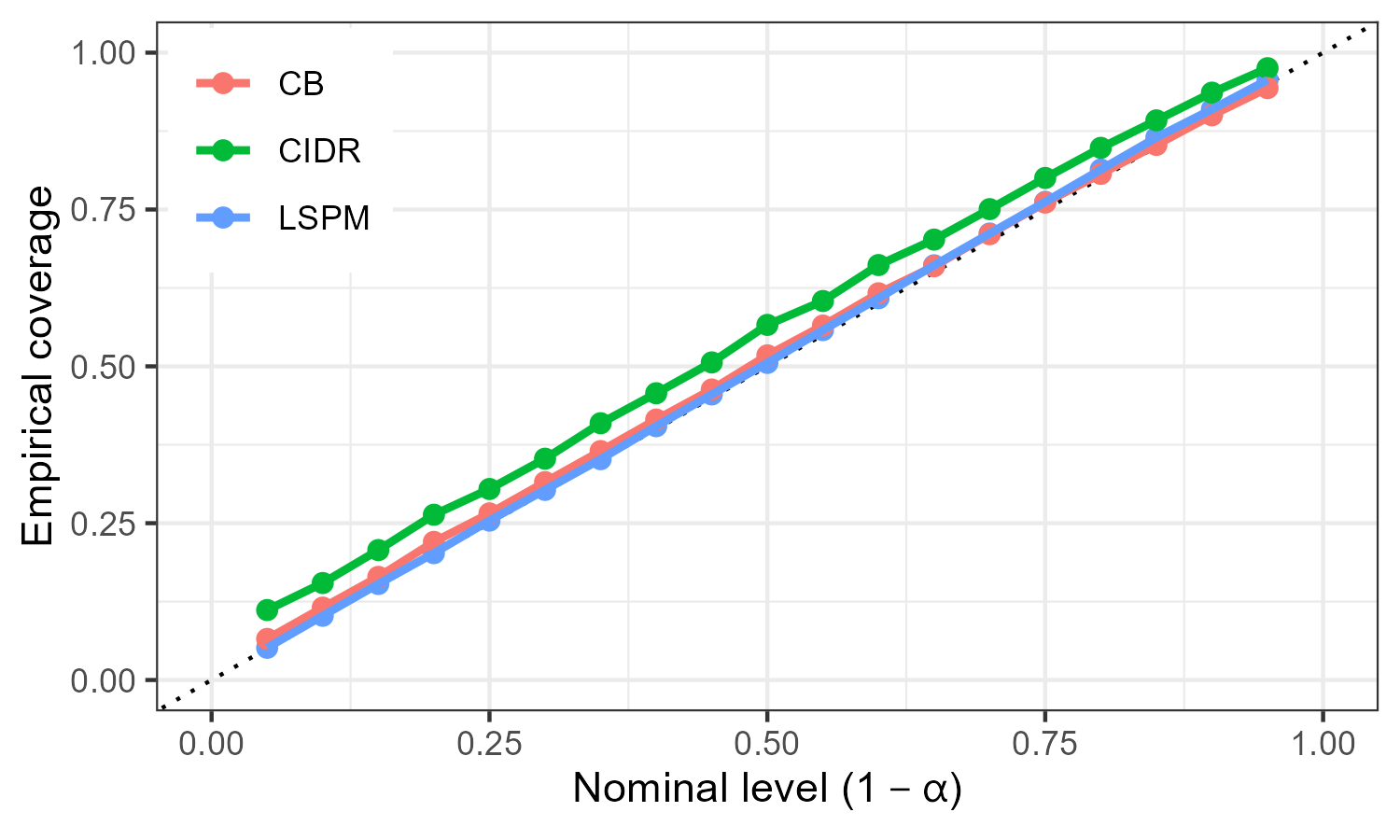}
    \includegraphics[width=0.49\linewidth]{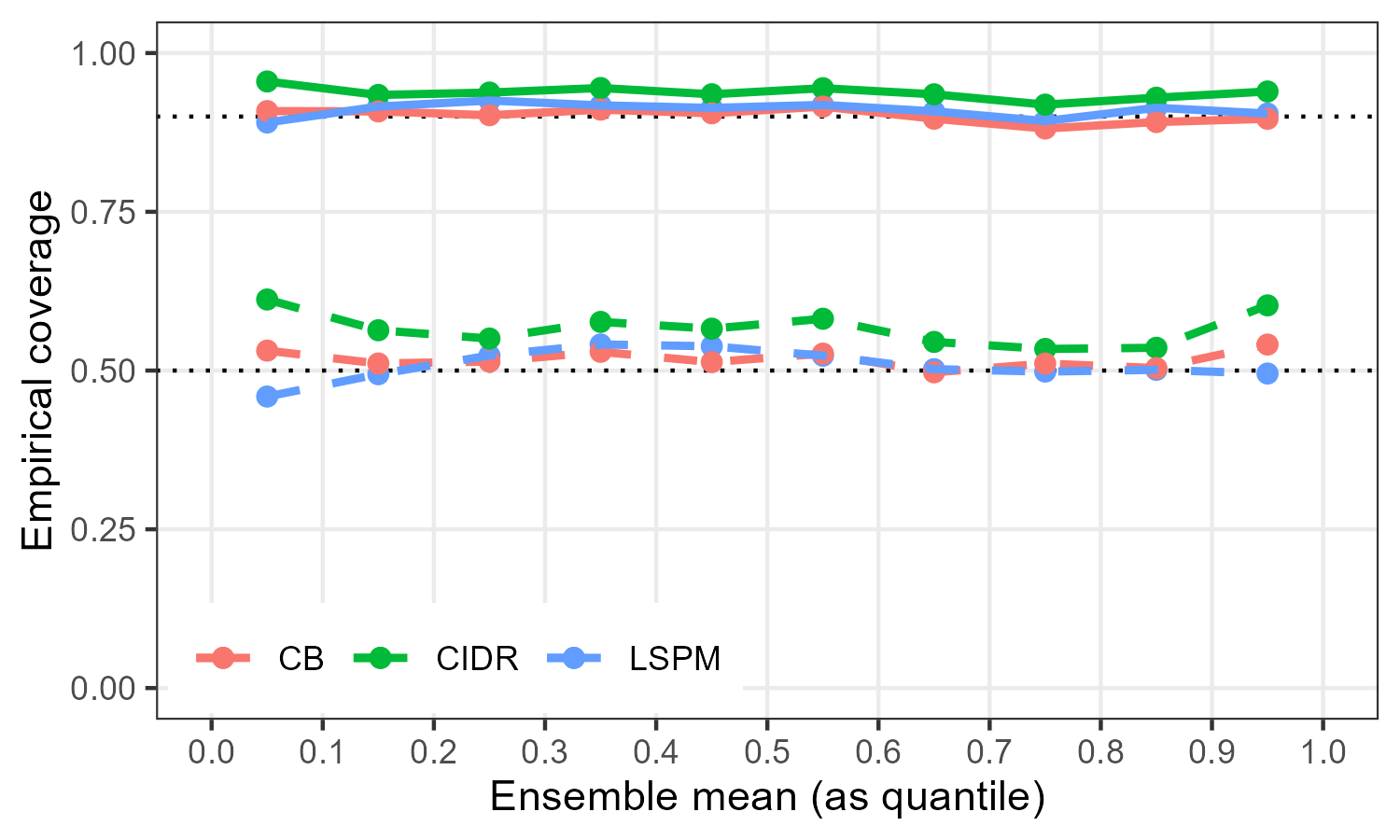}
    \caption{Evaluation of $(1 - \alpha)$-prediction intervals obtained from conformal binning (CB), conformal IDR (CIDR), and the LSPM when predicting European temperatures. Results are averaged across all 35 stations. Left: Empirical coverage shown as a function of $\alpha$. Right: Empirical coverage of the 50\% (dashed) and 90\% (solid) prediction intervals shown as a function of the ensemble mean forecast; the ensemble mean is converted to a quantile level to account for differences between stations, so that a value between 0 and 0.1 implies the ensemble mean forecast was in the lower decile of ensemble mean forecasts at the relevant station.}
    \label{fig:EUMN_comp}
\end{figure}

Unconditional and conditional coverage plots for the prediction intervals generated by the three predictive systems are displayed in Figure \ref{fig:EUMN_comp}, along with the average width and the average interval score over all stations and days. In this application, all three methods perform very similarly. All prediction intervals achieve approximately the correct empirical coverage at all levels, and this does not change when we condition on the ensemble mean forecast value. The average width of the prediction intervals is very similar for all methods, and is slightly larger for conformal IDR; since the conformal IDR predictive systems adapt depending on the covariate value, the resulting thickness is often larger than that of conformal binning and the LSPM, which results in larger prediction intervals. In this case, the distribution of the outcome does not appear to change depending on the ensemble mean forecast (i.e. the covariate) value, and hence there is no benefit of the stronger theoretical guarantees of conformal binning and conformal IDR.

\begin{figure}[t]
    \centering
    \includegraphics[width=0.4\linewidth]{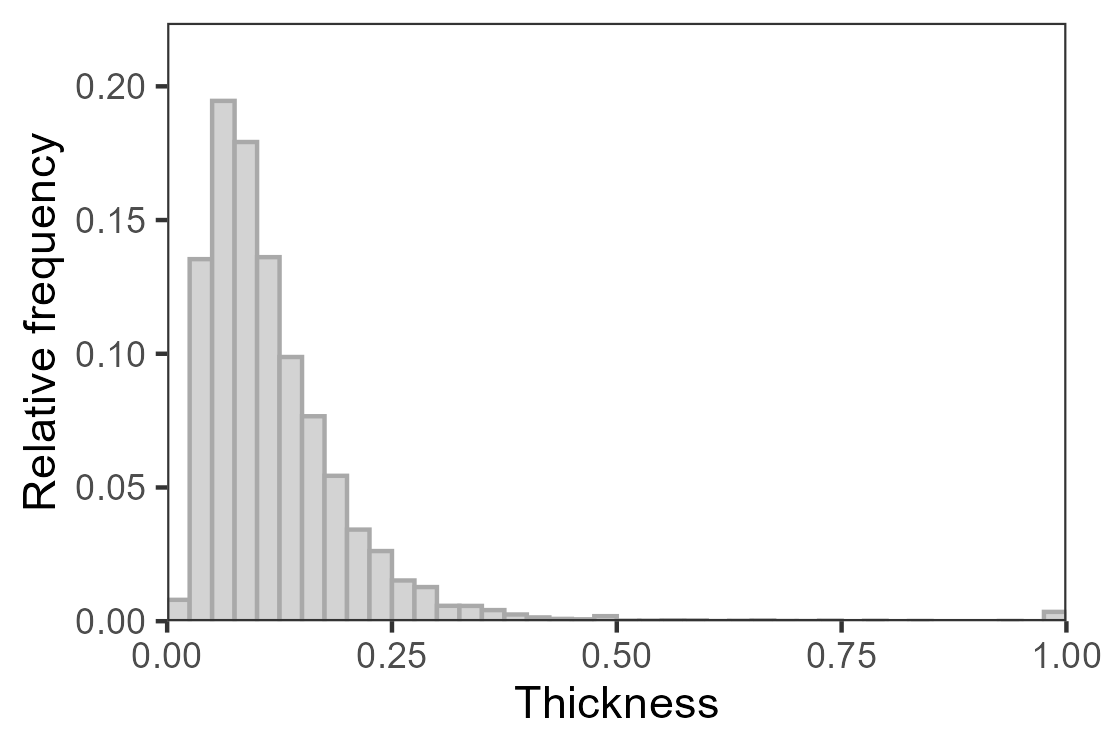}
    \includegraphics[width=0.4\linewidth]{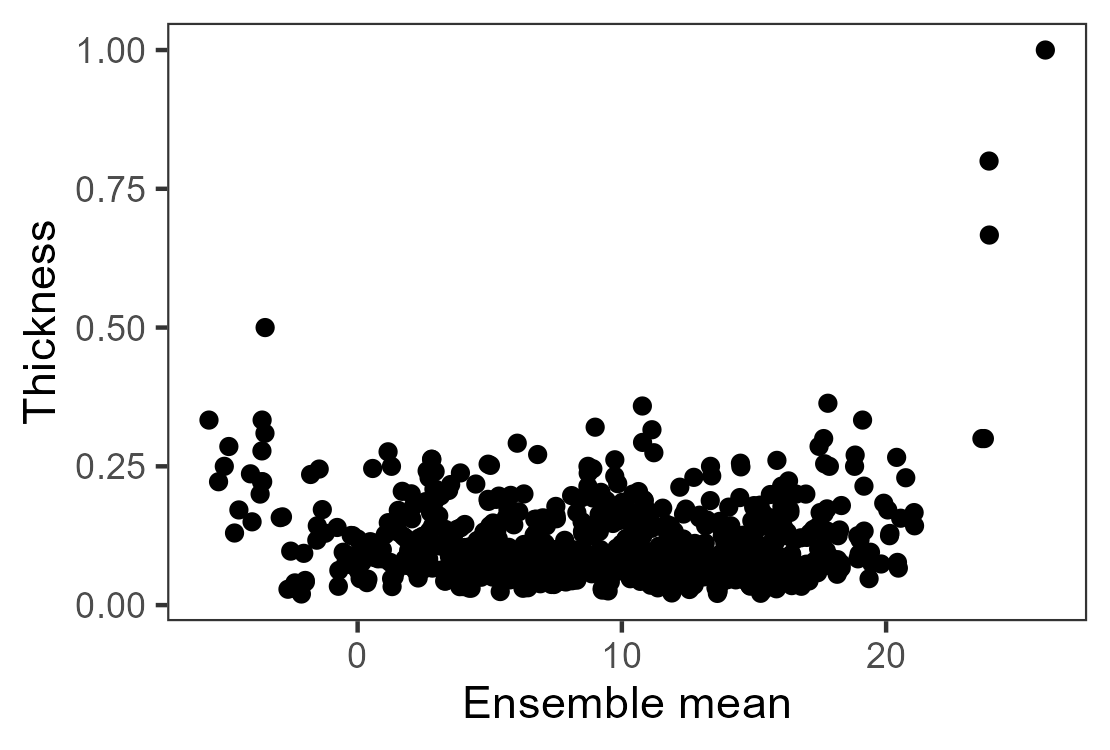}
    \includegraphics[width=0.4\linewidth]{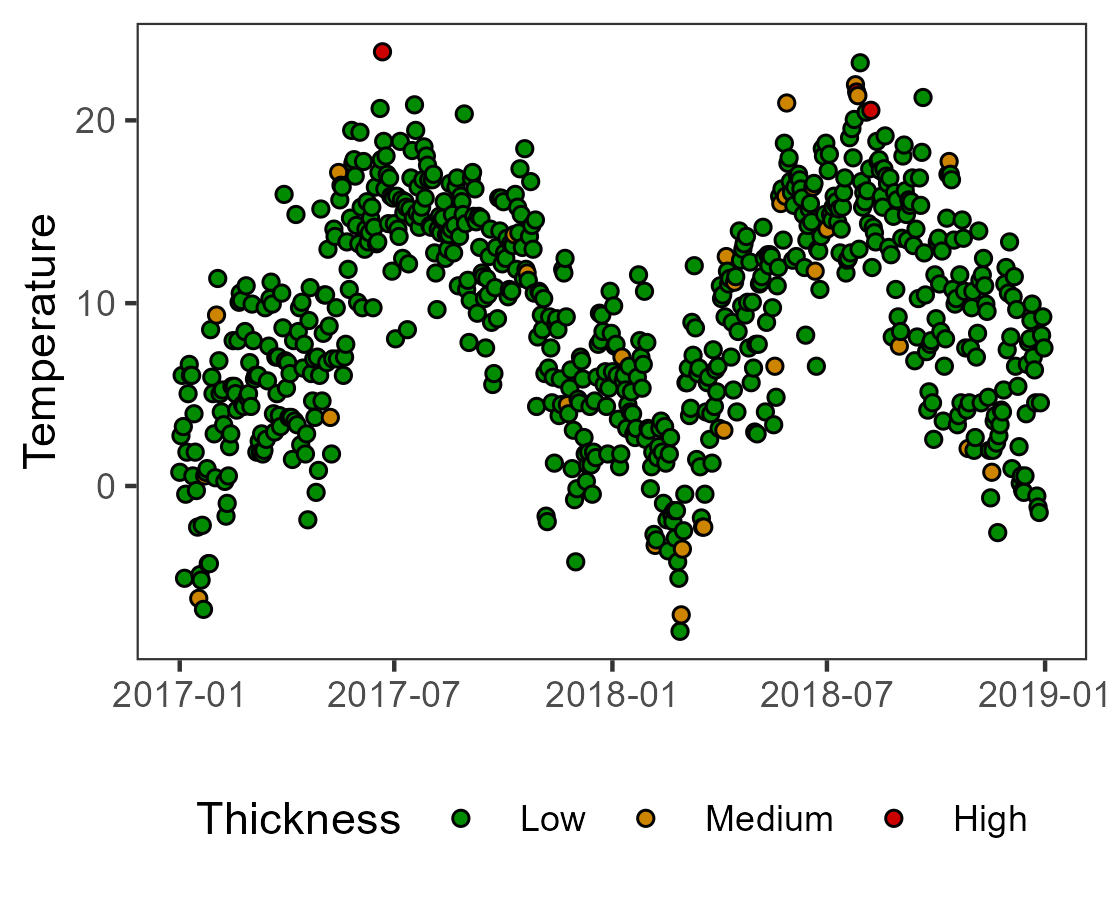}
    \includegraphics[width=0.4\linewidth]{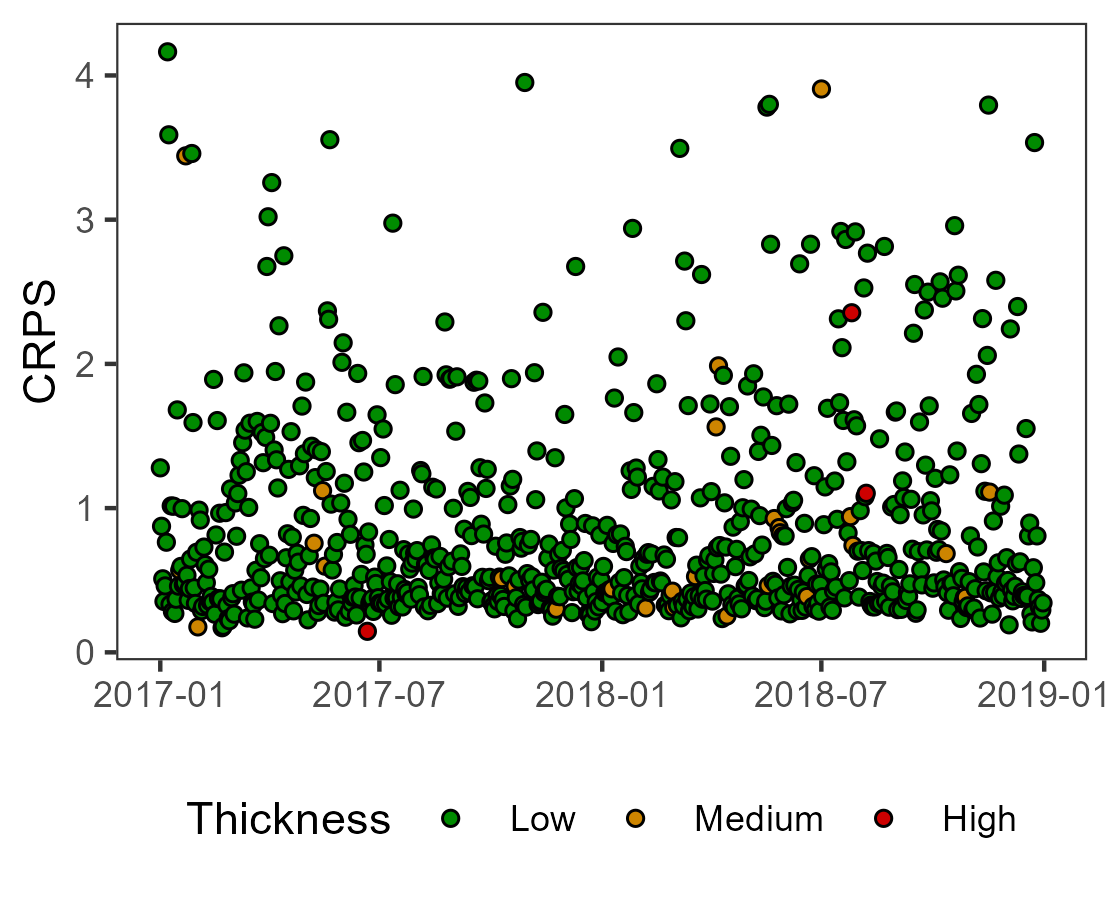}
    \caption{Plots showing the thickness of conformal IDR predictive bands. Results are shown for one randomly selected station, except for the histogram in the top left panel, which is shown for all stations. The bottom row uses the definitions of low, medium and high epistemic uncertainty in Section \ref{sec:epistemic2}.}
    \label{fig:EUMN_thick}
\end{figure}

Nonetheless, the thickness of the conformal IDR predictive systems provides a useful measure of forecast ambiguity. This thickness is displayed in Figure \ref{fig:EUMN_thick}. Results are shown for one station, though similar results are obtained for other stations. The thickness is displayed using the traffic light system described in Section \ref{sec:epistemic}, shown as a function of the observed temperature and the CRPS. The thickness is generally small, with only a handful of forecast cases exhibiting very high epistemic uncertainty. Large thicknesses tend to occur in the winter and summer of 2018, which correspond to when the ensemble mean forecast is relatively extreme (either high or low) compared to its most frequent values. The bottom right panel of Figure \ref{fig:EUMN_thick} demonstrates that the CRPS of the crisp CDF has little association with the thickness of the conformal predictive system. This illustrates the difference between aleatoric and epistemic uncertainty: A predictive system can have a large thickness (i.e. high epistemic uncertainty), even if the crisp CDF is sharp (i.e. low aleatoric uncertainty), and vice versa.

While this is admittedly only a high-level evaluation of the epistemic uncertainty quantification, it should be noted that there are no established methods for the evaluation of whether epistemic forecast uncertainty has been correctly specified by forecasters, see in particular the negative results on second-order proper scoring rules by \citet{BengsHullermeierETAL2023}.

\section{Discussion}\label{sec:disc}


This work demonstrates that probabilistic forecasting methods with in-sample calibration guarantees can be used to construct predictive systems with conformal calibration guarantees under the assumption of exchangeability. That is, they generate bands that contain a calibrated probabilistic forecast for an out-of-sample outcome. Using this, we introduce two novel conformal predictive systems that satisfy stronger notions of calibration than existing approaches. The first approach bins together similar covariates, while the second approach concerns isotonic distributional regression (IDR), a flexible non-parametric distributional regression method that requires almost no implementation choices.

Since IDR and binning methods satisfy relatively strong notions of calibration (in-sample), conformal binning and conformal IDR generates bands that are guaranteed to contain a conditionally calibrated predictive distribution. In contrast, classical CM conformal predictive systems, such as the least squares prediction machine (LSPM), only satisfy probabilistic calibration, which is known to be a weaker, unconditional notion of calibration. We therefore find, in both simulated and real data examples, that when the conditional variance of the target variable depends on the covariate, significant improvements in forecast performance can be obtained using conformal binning or conformal IDR.

While the literature on conformal prediction has focused on classification and regression tasks, we argue that conformal predictive systems are more valuable since they immediately provide prediction intervals with out-of-sample coverage guarantees at all levels. This removes the need to repeatedly run conformal prediction methods for multiple regression tasks. We illustrate in Section \ref{sec:app} how the stronger calibration guarantees of conformal binning and conformal IDR result in more accurate prediction intervals than those obtained from existing predictive systems, such as the LSPM. The performance is also found to be competitive with the widely used conformalized quantile regression approach of \cite{RomanoEtAl2019}, which does not possess the same out-of-sample conditional calibration properties. Several other conformal prediction methods have been proposed, but the aim of this work is not to provide a comprehensive comparison of competing methods; we leave this as a promising avenue for future research.

We additionally study the thickness of the conformal IDR predictive systems. While the thickness of the LSPM and conformal binning procedures are typically determined by the sample size (within bins), the thickness of conformal IDR is found to depend on how familiar the covariate value is; covariate values that have been seen several times before yield low thickness, whereas unusual values result in a large difference between the lower and upper bounds. Therefore, we suggest to interpret the thickness of the conformal IDR as a measure of epistemic uncertainty of the forecast, that is, how uncertain the predictive distribution is. In an application to temperature forecasts, we find that the thickness is generally not associated with the observed outcome nor with the accuracy of the crisp predictive distribution, highlighting the distinction between epistemic and aleatoric uncertainty.

There are several possible extensions to the work considered herein. Conformal IDR can also yield predictive distributions with calibration guarantees under covariate shift using the approach in \cite{TibshiraniBarberETAL2019}. Combining conformal IDR with conformity measures, one can also construct new conformal prediction sets for multivariate or more general types of outcomes. The conformal binning procedure similarly merits further investigation due to its simplicity, good practical performance, and its strong auto-calibration guarantees. For example, while the case studies in Section \ref{sec:app} employ $k$-means clustering to bin the covariates, alternative approaches could also be used for this purpose, such as regression trees. This binning can also be performed easily in higher dimensions, which may provide a means to construct multivariate conformal predictive systems with multivariate calibration guarantees.

\bibliographystyle{apalike}

\bibliography{biblio}


\appendix

\section{Proofs}\label{app:proofs}

\citet{BarberCandesETAL2022} put forward the idea of allowing for deterministic weights in conformal prediction. In situations when the exchangeability assumption is violated, such weights can allow the coverage gap to be bounded. Theorem \ref{thm:master} also holds with weights, and we give this more general result here.

\begin{thm}\label{thm:master_w}
Let $(X_1,Y_1),\dots,(X_n,Y_n),(X_{n+1},Y_{n+1})$ be an exchangeable sequence, let $w_1,\dots,w_n$ be deterministic non-negative weights, and let $\Pi$ be the predictive system defined by the bounds \eqref{eq:PB1} and \eqref{eq:PB2}, for some procedure $G$, $\hat{\mathbb{P}} = \sum_{j=1}^n w_i\delta_{(X_j,Y_j)}/\sum_{j=1}^n w_i$, and $x = X_{n+1}$. Define $\hat{\mathbb{P}}_{n+1} = (\sum_{j=1}^n w_i\delta_{(X_j,Y_j)} + \delta_{(X_{n+1},Y_{n+1})})/(\sum_{j=1}^n w_i+1)$.
\begin{enumerate}
    \item[(i)] If $G$ is probabilistically calibrated, then $\Pi$ contains a probabilistically calibrated predictive CDF for $Y_{n+1}$. 
    More precisely, for all $u \in (0,1)$,
    \begin{equation*}
        \prob\Big(G^{\hat{\mathbb{P}}_{n+1}}_{X_{n+1}}(Y_{n+1}) \le u\Big) \le u \le \prob\Big(G^{\hat{\mathbb{P}}_{n+1}}_{X_{n+1}}(Y_{n+1}-) \le u\Big).
    \end{equation*}

    \item[(ii)] If $G$ is isotonically calibrated, then $\Pi$ contains an isotonically calibrated predictive CDF for $Y_{n+1}$. 
    More precisely, for all $y \in \mathbb{R}$,
    \[
        1 - G^{\hat{\mathbb{P}}_{n+1}}_{X_{n+1}}(y) = \mathbb{P}\Big(Y_{n+1} > y \Big| \mathcal{A}\Big(G^{\hat{\mathbb{P}}_{n+1}}_{X_{n+1}}\Big)\Big).
    \]
    That is, the predictive CDF $G^{\hat{\mathbb{P}}_{n+1}}_{X_{n+1}}$ is isotonically calibrated for $Y_{n+1}$.
    
    \item[(iii)] If $G$ is auto-calibrated, then $\Pi$ contains an auto-calibrated predictive CDF for $Y_{n+1}$. 
    More precisely, for all $y \in \mathbb{R}$,
    \[
        G^{\hat{\mathbb{P}}_{n+1}}_{X_{n+1}}(y) = \mathbb{P}\Big(Y_{n+1} \le y \Big| G^{\hat{\mathbb{P}}_{n+1}}_{X_{n+1}}\Big).
    \]
    That is, the predictive CDF $G^{\hat{\mathbb{P}}_{n+1}}_{X_{n+1}}$ is auto-calibrated for $Y_{n+1}$.
\end{enumerate}
\end{thm}

\begin{proof}
We define \(w_{n+1}\vcentcolon=1\).
    For all three parts of the theorem, we are going to make use of a random variable \(\kappa\), which is independent of \(\left(X_1,Y_1\right),\ldots,\left(X_{n+1},Y_{n+1}\right)\) and takes values in $\{1,\dots,n+1\}$ such that, for all \(i\in \{1,\ldots,n+1\}\),
    \[\mathbb{P}(\kappa=i)=\frac{w_i}{\sum_{j=1}^{n+1}w_j}=\vcentcolon \widetilde{w}_i.\]

For the proof of part (i),
    notice that \(\sum_{i=1}^{n+1}\widetilde{w}_i=1\) and, due to the independence of \(\left(X_1,Y_1\right),\ldots,\left(X_{n+1},Y_{n+1}\right)\) and \(\kappa\), it follows that
    \[\prob\Big(G^{\hat{\mathbb{P}}_{n+1}}_{X_{\kappa}}(Y_{\kappa})<u\Big| \hat{\mathbb{P}}_{n+1} \Big)
    =
    \sum_{j=1}^{n+1}\widetilde{w}_j\cdot \prob\Big(G^{\hat{\mathbb{P}}_{n+1}}_{X_{j}}(Y_{j})<u\Big|\hat{\mathbb{P}}_{n+1} \Big).\]
    Hence, for all $u \in (0,1)$, we obtain
    \begin{multline}
        \label{eq:secondlineA}
        \prob\Big(G^{\hat{\mathbb{P}}_{n+1}}_{X_{n+1}}(Y_{n+1}) < u \Big| \hat{\mathbb{P}}_{n+1} \Big) = \prob\Big(G^{\hat{\mathbb{P}}_{n+1}}_{X_{\kappa}}(Y_{\kappa}) < u \Big| \hat{\mathbb{P}}_{n+1} \Big)\\
        + \sum_{j=1}^{n+1}\widetilde{w}_j \Big[\prob\big(G^{\hat{\mathbb{P}}_{n+1}}_{X_{n+1}}(Y_{n+1}) < u \big| \hat{\mathbb{P}}_{n+1} \big) - \prob\big(G^{\hat{\mathbb{P}}_{n+1}}_{X_{j}}(Y_{j}) < u \big| \hat{\mathbb{P}}_{n+1} \big)\Big].
    \end{multline}
    All summands in the second line of \eqref{eq:secondlineA} are zero due to exchangeability. Conditional on $\hat{\mathbb{P}}_{n+1}$, the random variable $(X_\kappa,Y_\kappa)$ has distribution $\hat{\mathbb{P}}_{n+1} $, hence probabilistic calibration of $G$ yields that the above expression is less than or equal to $u$. The second inequality in part (i) is shown analogously.

For the proof of part (ii), we would like to show that, for each $y \in \mathbb{R}$,
    \[\E\Big[\one{\left\{Y_{n+1} >  y\right\}} \ \Big| \ \A\left(G^{\hat{\mathbb{P}}_{n+1}}_{X_{n+1}}\right)\Big]=1 - G^{\hat{\mathbb{P}}_{n+1}}_{X_{n+1}}(y).\]
    According to the definition of conditional expectations with respect to $\sigma$-lattices, see for example \citet[Definition 2.1]{ArnoldZiegel2023}, it suffices to show that, for any \(A\in \A(G^{\hat{\mathbb{P}}_{n+1}}_{X_{n+1}})\) and \(B\in \sigma(G^{\hat{\mathbb{P}}_{n+1}}_{X_{n+1}}(y))\), it holds that
    \begin{align*}
            \E\Big[\one{\left\{Y_{n+1}> y\right\}} \one_{A}\Big]&\leq \E\Big[(1-G^{\hat{\mathbb{P}}_{n+1}}_{X_{n+1}}(y))\one_{A}\Big],\\
            \E\Big[\one{\left\{Y_{n+1}> y\right\}} \one_{B}\Big]&= \E\Big[(1-G^{\hat{\mathbb{P}}_{n+1}}_{X_{n+1}}(y)) \one_{B}\Big].
    \end{align*}
    By \citet[Definition~3.1]{ArnoldEtAl2024}, it follows that \(A=\{G^{\hat{\mathbb{P}}_{n+1}}_{X_{n+1}}\in U\}\) for some Borel set $U$ (in the space of distributions on $\mathbb{R}$ equipped with the topology of weak convergence) that is an upper set with respect to the usual stochastic order.
    Thus,
    \begin{align}
        \label{eq:intermediate_1}
        \E\left(\one{\left\{Y_{\kappa}> y\right\}} \one{\left\{G^{\hat{\mathbb{P}}_{n+1}}_{X_{\kappa}}\in U\right\}}\right)
        &=\sum_{i=1}^{n+1} \widetilde{w}_i \E\Big(\one{\left\{Y_{i}> y\right\}} \one{\left\{G^{\hat{\mathbb{P}}_{n+1}}_{X_{i}}\in U\right\}}\Big)\nonumber \\
        &= \E\Big(\one{\left\{Y_{n+1}> y\right\}}\one{\left\{G^{\hat{\mathbb{P}}_{n+1}}_{X_{n+1}}\in U\right\}}\Big),
    \end{align}
    where the last equation holds because of exchangeability.
    By the definition of conditional expectations with respect to $\sigma$-lattices, we obtain
    \begin{align}
        \label{eq:intermediate_2}
        \E_{\hat{\mathbb{P}}_{n+1}}\Big[\one{\left\{Y_{\kappa}> y\right\}}\one{\left\{G^{\hat{\mathbb{P}}_{n+1}}_{X_{\kappa}}\in U\right\}}\Big]
        &\leq \E_{\hat{\mathbb{P}}_{n+1}}\Big[\E_{\hat{\mathbb{P}}_{n+1}}\Big(\one{\left\{Y_{\kappa}> y\right\}}\ \Big | \ \mathcal{A}\left(G^{\hat{\mathbb{P}}_{n+1}}_{X_{\kappa}}\right)\Big)\one{\left\{G^{\hat{\mathbb{P}}_{n+1}}_{X_{\kappa}}\in U\right\}}\Big]
    \end{align}
    Due to exchangeability, it holds that \(\left(\left(X_{\kappa},Y_{\kappa}\right)\mid \hat{\mathbb{P}}_{n+1}\right)\sim \hat{\mathbb{P}}_{n+1}\), where \(\kappa\) is the discrete random variable defined at the beginning of the proof.
    Since \(G\) is isotonically calibrated, it follows that
    \begin{equation}\label{eq:GisoHcal}
        \E_{\hat{\mathbb{P}}_{n+1}}\Big(\one{\left\{Y_{\kappa}> y\right\}}\ \Big | \  \A\left(G^{\hat{\mathbb{P}}_{n+1}}_{X_{\kappa}}\right)\Big)=1-G^{\hat{\mathbb{P}}_{n+1}}_{X_{\kappa}}(y),
    \end{equation}
    so right-hand side of \eqref{eq:intermediate_2} is equal to
    \begin{align*}
        \E_{\hat{\mathbb{P}}_{n+1}}\Big((1-G^{\hat{\mathbb{P}}_{n+1}}_{X_{\kappa}}(y))\one{\left\{G^{\hat{\mathbb{P}}_{n+1}}_{X_{\kappa}}\in U\right\}}\Big)
        &=\sum_{i=1}^{n+1} \Big(\widetilde{w}_j (1-G^{\hat{\mathbb{P}}_{n+1}}_{X_{j}}(y))\one{\left\{G^{\hat{\mathbb{P}}_{n+1}}_{X_{j}}\in U\right\}}\Big).
    \end{align*}
    Thus, from Equations \eqref{eq:intermediate_1}, \eqref{eq:intermediate_2}, it follows that
    \begin{align*}
        \E\Big(\one{\left\{Y_{n+1}> y\right\}} \one{\left\{G^{\hat{\mathbb{P}}_{n+1}}_{X_{n+1}}\in U\right\}}\Big)
        &=\E\left(\E_{\hat{\mathbb{P}}_{n+1}}\Big[\one{\left\{Y_{\kappa}>y\right\}}\one{\left\{G^{\hat{\mathbb{P}}_{n+1}}_{X_{\kappa}}\in U\right\}}\Big]\right)\nonumber \\
        &\leq \E\left[\sum_{i=1}^{n+1} \widetilde{w}_j  (1-G^{\hat{\mathbb{P}}_{n+1}}_{X_{j}}\pare{y})\one{\left\{G^{\hat{\mathbb{P}}_{n+1}}_{X_{j}}\in U\right\}}\right]\nonumber \\
        &=\E\Big[(1-G^{\hat{\mathbb{P}}_{n+1}}_{X_{n+1}}(y)) \one{\left\{G^{\hat{\mathbb{P}}_{n+1}}_{X_{n+1}}\in U\right\}}\Big],
    \end{align*}
    where the last step follows again from the exchangeability assumption.
    Now, we want to prove that
    \begin{equation*}
        \E\Big[\one{\left\{Y_{n+1}>y\right\}} \one_{B}\Big]= \E\Big[(1-G^{\hat{\mathbb{P}}_{n+1}}_{X_{n+1}}\left(y\right))\one_{B}\Big].
    \end{equation*}
    Let $V \in \mathcal{B}(\mathbb{R})$ such that $B = \{G^{\hat{\mathbb{P}}_{n+1}}_{X_{n+1}}(y) \in V\}$. Identity \eqref{eq:intermediate_1} continues to hold with $\{G^{\hat{\mathbb{P}}_{n+1}}_{X_{\bullet}} \in U\}$ replaced by $\{G^{\hat{\mathbb{P}}_{n+1}}_{X_{\bullet}}(y) \in V\}$ everywhere. With the same replacement, \eqref{eq:intermediate_2} now becomes an equality, where we need to use that \eqref{eq:GisoHcal} holds. Now, we can conclude as previously.

 The proof of part (iii) is similar to the proof of part (ii) but a bit more straightforward. 
    
    We would like to show that, for each $y \in \mathbb{R}$,
    \[\E\Big[\one{\left\{Y_{n+1}\leq y\right\}} \ \Big| \ \sigma\left(G^{\hat{\mathbb{P}}_{n+1}}_{X_{n+1}}\right)\Big]=G^{\hat{\mathbb{P}}_{n+1}}_{X_{n+1}}\left(y\right).\]
    According to the definition of conditional expectations, it suffices to show that, for any \(B\in \sigma\left(G^{\hat{\mathbb{P}}_{n+1}}_{X_{n+1}}\right)\), it holds that
    \begin{equation*}
            \E\Big[\one{\left\{Y_{n+1}\leq y\right\}} \one_{B}\Big]= \E\Big[G^{\hat{\mathbb{P}}_{n+1}}_{X_{n+1}}\left(y\right)\one_{B}\Big].
    \end{equation*}
    We have that \(B=\left\{G^{\hat{\mathbb{P}}_{n+1}}_{X_{n+1}}\in V\right\}\) for some Borel set $V$ in the space of distributions on $\mathbb{R}$ equipped with the topology of weak convergence.
    Due to exchangeability, it holds that \(\left(\left(X_{\kappa},Y_{\kappa}\right)\mid \hat{\mathbb{P}}_{n+1}\right)\sim \hat{\mathbb{P}}_{n+1}\), where \(\kappa\) is the discrete random variable defined at the beginning of the proof.
    Thus,
    \begin{align}
        \label{eq:intermediate_3}
        \E\left(\E \Big[\one{\left\{Y_{\kappa}\leq y\right\}}\one{\left\{G^{\hat{\mathbb{P}}_{n+1}}_{X_{\kappa}}\in V\right\}}\ \Big | \ \hat{\mathbb{P}}_{n+1}\Big]\right)
        &=\E\left(\E_{\hat{\mathbb{P}}_{n+1}}\Big[\one{\left\{Y_{\kappa}\leq y\right\}}\one{\left\{G^{\hat{\mathbb{P}}_{n+1}}_{X_{\kappa}}\in V\right\}}\Big]\right)\nonumber \\
        &=\sum_{i=1}^{n+1} \widetilde{w}_i  \E\Big(\one{\left\{Y_{i}\leq y\right\}}\one{\left\{G^{\hat{\mathbb{P}}_{n+1}}_{X_{i}}\in V\right\}}\Big)\nonumber \\
        &= \E\Big(\one{\left\{Y_{n+1}\leq y\right\}} \one{\left\{G^{\hat{\mathbb{P}}_{n+1}}_{X_{n+1}}\in V\right\}}\Big),
    \end{align}
    where the last equation holds because of exchangeability.
    From the definition of conditional expectations, we obtain
    \begin{align}
        \label{eq:intermediate_4}
        \E_{\hat{\mathbb{P}}_{n+1}}\Big[\one{\left\{Y_{\kappa}\leq y\right\}}\one{\left\{G^{\hat{\mathbb{P}}_{n+1}}_{X_{\kappa}}\in V\right\}}\Big]
        &= \E_{\hat{\mathbb{P}}_{n+1}}\Big[\E_{\hat{\mathbb{P}}_{n+1}}\Big(\one{\left\{Y_{\kappa}\leq y\right\}}\ \Big | \ \sigma\left(G^{\hat{\mathbb{P}}_{n+1}}_{X_{\kappa}}\right)\Big) \one{\left\{G^{\hat{\mathbb{P}}_{n+1}}_{X_{\kappa}}\in V\right\}}\Big]
    \end{align}
    Since \(G\) is auto-calibrated, it follows that
    \begin{equation*}
        \E_{\hat{\mathbb{P}}_{n+1}}\Big(\one{\left\{Y_{\kappa}\leq y\right\}}\ \Big | \  \sigma\left(G^{\hat{\mathbb{P}}_{n+1}}_{X_{\kappa}}\right)\Big)=G^{\hat{\mathbb{P}}_{n+1}}_{X_{\kappa}}(y),
    \end{equation*}
    so the quantity in \eqref{eq:intermediate_4} is equal to
    \begin{align*}
        \E_{\hat{\mathbb{P}}_{n+1}}\Big(G^{\hat{\mathbb{P}}_{n+1}}_{X_{\kappa}}(y)\one{\left\{G^{\hat{\mathbb{P}}_{n+1}}_{X_{\kappa}}\in V\right\}}\Big)
        &=\sum_{i=1}^{n+1} \Big(\widetilde{w}_j  G^{\hat{\mathbb{P}}_{n+1}}_{X_{j}}\pare{y}\one{\left\{G^{\hat{\mathbb{P}}_{n+1}}_{X_{j}}\in V\right\}}\Big).
    \end{align*}
    Thus, from Equations \eqref{eq:intermediate_3}, \eqref{eq:intermediate_4}, it follows that
    \begin{align*}
        \E\Big[\one{\left\{Y_{n+1}\leq y\right\}} \one{\left\{G^{\hat{\mathbb{P}}_{n+1}}_{X_{n+1}}\in V\right\}}\Big]
        &=\E\left(\E_{\hat{\mathbb{P}}_{n+1}}\Big[\one{\left\{Y_{\kappa}\leq y\right\}} \one{\left\{ G^{\hat{\mathbb{P}}_{n+1}}_{X_{\kappa}}\in V\right\}}\Big]\right)\nonumber \\
        &= \E\left[\sum_{i=1}^{n+1} \widetilde{w}_j  G^{\hat{\mathbb{P}}_{n+1}}_{X_{j}}\pare{y}\one{\left\{G^{\hat{\mathbb{P}}_{n+1}}_{X_{j}}\in V\right\}}\right]\nonumber \\
        &=\E\Big[G^{\hat{\mathbb{P}}_{n+1}}_{X_{n+1}}(y)\one{\left\{G^{\hat{\mathbb{P}}_{n+1}}_{X_{n+1}}\in V\right\}}\Big],
    \end{align*}
    where the last step follows again from the exchangeability assumption.
\end{proof}

\begin{lemma}\label{lem:monoPS}
    Let $z_1, \dots, z_m \in \mathcal{Z}$, with empirical distribution $\hat{\mathbb{P}}$, and let $x \in \mathcal{X}$. Suppose that
    \begin{equation}\label{eq:incr_cond}
    y \mapsto  A(\hat{\mathbb{P}} + (x, y),(x,y)) - A(\hat{\mathbb{P}} + (x, y), z)
    \end{equation}
    is increasing for any $z \in \{z_1, \dots, z_m\}$. Then, $y \mapsto G_{x}^{\hat{\mathbb{P}} + (x, y)}(y)$, and $y \mapsto G_{x}^{\hat{\mathbb{P}} + (x, y)}(y-)$ are increasing.
\end{lemma}

\begin{proof}
Writing out $G$ explicitly yields
\[
G_{x}^{\hat{\mathbb{P}} + (x, y)}(y) = \frac{1}{m+1} \left(\sum_{i=1}^m\one\{A(\hat{\mathbb{P}} + (x, y),z_i) \le A(\hat{\mathbb{P}} + (x, y),(x,y))\} + 1\right)
\]
For a given $y \in \mathbb{R}$, we define
    \begin{align*}
    I(y) &= \{i  \mid 0 \le  A(\hat{\mathbb{P}} + (x, y),(x,y)) - A(\hat{\mathbb{P}} + (x, y),z_i)\}.
    \end{align*}
    Then, $G_{x}^{\hat{\mathbb{P}} + (x, y)}(y) = (|I(y)|+1)/(m+1)$.
    For $y \le y'$, \eqref{eq:incr_cond} yields that $I(y) \subseteq I(y')$, which yields the claim. The argument for $G_{x}^{\hat{\mathbb{P}} + (x, y)}(\cdot -)$ works analogously.
\end{proof}

\begin{proof}[Proof of Proposition \ref{prop:bin_ac}]
Let $(X,Y) \sim \hat{\mathbb{P}}$. Since the support of $\hat{\mathbb{P}}$, $\supp(\hat{\mathbb{P}}) = \{z_1,\dots,z_m\}$ is finite, there are only $m' \le m$ different CDFs $G_{x_i}^{\hat{\mathbb{P}}}$, $i=1,\dots,m$, that $G_{X}^{\hat{\mathbb{P}}}$ can take as values. Denote these by $G_1,\dots,G_{m'}$. Define $B_i = \{j \in 1,\dots,m : G_{x_j}^{\hat{\mathbb{P}}} = G_i\}$, $i = 1,\dots,m'$. Then,
\[
\hat{\mathbb{P}}(Y \le y \mid G_{X}^{\hat{\mathbb{P}}}=G_i) = \frac{1}{|B_i|}\sum_{j\in B_i} \one\{y_j \le y\}, \quad y \in \mathbb{R}.
\]
Auto-calibration of $G$ yields the claim.
\end{proof}

\begin{proof}[Proof of Theorem \ref{thm:pred_int}]
    For any predictive system $\Pi$ constructed using the bounds at \eqref{eq:PB1} and \eqref{eq:PB2}, we have that
    \[
    \Pi_\ell[\hat{\mathbb{P}}_n, X_{n+1}](y) \le G_{X_{n+1}}^{\hat{\mathbb{P}}_{n+1}}(y) \le \Pi_u[\hat{\mathbb{P}}_n, X_{n+1}](y)
    \]
    for all $y \in \R$, and hence
    \[
    L_\alpha^G(X_{n+1}) \ge L_\alpha^\Pi (X_{n+1}), \quad  U_\alpha^G(X_{n+1}) \le U_\alpha^\Pi (X_{n+1})
    \]
    That is, the upper $\alpha/2$- and lower $1-\alpha/2$-quantiles of $G_{X_{n+1}}^{\hat{\mathbb{P}}_{n+1}}$ fall within $[L_\alpha^\Pi(X_{n+1}), U_\alpha^\Pi(X_{n+1})]$. Put differently, defining $C_{\alpha}^G(X_{n+1}) := [L_\alpha^G(X_{n+1}), U_\alpha^G(X_{n+1})]$, we have that $C_{\alpha}^G(X_{n+1}) \subseteq C_{\alpha}^\Pi(X_{n+1})$.

    For part (i), we have that
    \begin{align*}
         \mathbb{P}(Y_{n+1} \in C_\alpha^\Pi(X_{n+1}) ) & \ge  \mathbb{P}(Y_{n+1} \in C_\alpha^G(X_{n+1}) )\\
         &=  \mathbb{P}(Y_{n+1} \le U_\alpha^G(X_{n+1}) ) -  \mathbb{P}(Y_{n+1} <  L_\alpha^G(X_{n+1}) )\\
         & = \mathbb{P}\left(G_{X_{n+1}}^{\hat{\mathbb{P}}_{n+1}}(Y_{n+1}-)< 1-\frac{\alpha}{2}\right) - \mathbb{P}\left(G_{X_{n+1}}^{\hat{\mathbb{P}}_{n+1}}(Y_{n+1}) \le \frac{\alpha}{2}\right)\\
         \ge 1-\frac{\alpha}{2} - \frac{\alpha}{2} = 1-\alpha,
    \end{align*}
    where we used probabilistic calibration of $G_{X_{n+1}}^{\hat{\mathbb{P}}_{n+1}}$ for $Y_{n+1}$ in the last step, which holds according to Theorem \ref{thm:master}.

    For part (ii), we only show the first inequality. The second one is analogous. It is sufficient to show that for any Borel set $U \subset \mathbb{R}$, it holds that
    \begin{multline*}
    \mathbb{E}\left[\left(\one\{Y_{n+1} < L_{\alpha}^{\Pi}(X_{n+1})\} - \frac{\alpha}{2}\right) \one\{L_\alpha^G(X_{n+1}) \in U\}\right] \\\le \mathbb{E}\left[\left(\one\{Y_{n+1} < L_{\alpha}^{G}\} - \frac{\alpha}{2}\right) \one\{L_\alpha^G(X_{n+1}) \in U\}\right] \le 0.
    \end{multline*}
The first inequality is clear from the considerations above. For the second, we have
\begin{multline*}
    \mathbb{E}\left[\left(\one\{Y_{n+1} < L_{\alpha}^{G}(X_{n+1})\} - \frac{\alpha}{2}\right) \one\{L_\alpha^G(X_{n+1}) \in U\}\right] \\= \mathbb{E}\left[\mathbb{E}\left[\left(\one\{Y_{n+1} < L_{\alpha}^{G}\} - \frac{\alpha}{2}\right) \one\{L_\alpha^G(X_{n+1}) \in U\}\mid \hat{\mathbb{P}}_{n+1}\right]\right] \le 0,
\end{multline*}
    since $(X_{n+1},Y_{n+1})$ has distribution $\hat{\mathbb{P}}_{n+1}$ conditional on $\hat{\mathbb{P}}_{n+1}$ and IDR is in-sample quantile calibrated \citep[Proposition 5.2]{ArnoldZiegel2023}.

    For part (iii), by Theorem \ref{thm:master}, we know that $G_{X_{n+1}}^{\hat{\mathbb{P}}_{n+1}}$ is auto-calibrated for $Y_{n+1}$. Therefore,
    \begin{align*}
    \mathbb{P} & \left( Y_{n+1} \in C_\alpha^G(X_{n+1}) \mid G_{X_{n+1}}^{\hat{\mathbb{P}}_{n+1}} \right) \\
    &= \mathbb{P} \left( Y_{n+1} \le U_\alpha^G(X_{n+1}) \mid G_{X_{n+1}}^{\hat{\mathbb{P}}_{n+1}} \right) - \mathbb{P} \left( Y_{n+1} < L_\alpha^G(X_{n+1}) \mid G_{X_{n+1}}^{\hat{\mathbb{P}}_{n+1}} \right) \\
    &\ge \left( 1 - \frac{\alpha}{2} \right) - \frac{\alpha}{2} = 1 - \alpha.
    \end{align*}
    Since $C_{\alpha}^G(X_{n+1}) \subseteq C_{\alpha}^\Pi(X_{n+1})$, it follows that
    \[
    \mathbb{P} \left( Y_{n+1} \in C_\alpha^\Pi(X_{n+1}) \mid G_{X_{n+1}}^{\hat{\mathbb{P}}_{n+1}} \right) \ge \mathbb{P} \left( Y_{n+1} \in C_\alpha^G(X_{n+1}) \mid G_{X_{n+1}}^{\hat{\mathbb{P}}_{n+1}} \right) \ge 1 - \alpha. \qedhere
    \]
\end{proof}

\begin{proof}[Proof of Theorem \ref{thm:ConfIDR_thickness}]
    \citet{YangBarber2019} have introduced the sliding window norm for a random vector $x \in \R^{n+1}$, which depends on a function $\psi$ on the positive half-line that is increasing and such that $x \mapsto x/\psi(x)$ is concave. We use the choice $\psi(x)=x$. Then the sliding window norm is 
    \[
    \|x\|^{\mathrm{SW}} = \max_{1 \le i \le j \le n+1} \left|\sum_{\ell=i}^j x_\ell\right|, \quad x \in \R^{n+1}.
    \]
    For $y \in \R^{n+1}$, let $\mathrm{iso}(y)$ denote its isotonic projection, that is $\mathrm{iso}(y) = \argmin_{x \in \R^{n+1}}\{\|y-x\|^2 : x_1 \le \dots \le x_{n+1}\}$, where $\|\cdot\|$ is the Euclidean norm. \citet[Theorem 2]{YangBarber2019} shows that for any $x,y \in \R^{n+1}$ and $k=1,\dots,n+1$, we have
    \begin{align}
    \mathrm{iso}(x)_k - \mathrm{iso}(y)_k&\le \min_{1 \le m \le (n+1)-k+1} \left\{\overline{\mathrm{iso}(y)}_{k:(k+m-1)}- \mathrm{iso}(y)_k +\frac{\|x-y\|^{\mathrm{SW}}}{m}\right\},\label{eq:thm2}
    \end{align}
    Suppose now that $x_i = y_i$ for all $i \not= i_0$ and $x_{i_0} \ge y_{i_0}$. Then, $\|x-y\|^{\mathrm{SW}} = |x_{i_0} - y_{i_0}|$. There exists a partition of $\{1,\dots,n+1\}$ into intervals of indices $I_1^y,\dots,I_\ell^y$ such that $\mathrm{iso}(y)_k = \sum_{\ell \in I_j^y} y_\ell/|I_j^y|$ if $k \in I_j^y$, and analogously for $\mathrm{iso}(x)$. Suppose that $i_0 \in I_{j_0}^y \cap I_{\ell_0}^x$ . Then $\min I_{j_0}^y \le \min I_{\ell_0}^x$ and $\max I_{j_0}^y \le \max I_{\ell_0}^x$, which can be deduced from the min-max formula for isotonic regression. Taking $k = \min I_{j_0}^y$, we find from \eqref{eq:thm2} that
    \begin{equation}\label{eq:i_0bound}
    0 \le \mathrm{iso}(x)_{i_0} - \mathrm{iso}(y)_{i_0} \le \frac{|x_{i_0} - y_{i_0}|}{\max I_{j_0}^y - \min I_{\ell_0}^x + 1}.
    \end{equation}
   Define $\tilde{\mu} = \mu+\delta_{(X_{n+1},Y_{n+1})}$, $Z_i = H[\tilde{\mu},X_i]$, and let $\pi$ be the permutation of $1,\dots,n+1$ such that $Z_{(i)} = Z_{\pi(i)}$, $i=1,\dots,n+1$. Fix $y \in \R$. Define $y = (\one\{Y_{\pi(i)}> y\})_{i=1}^n$ and $x_i = x_i' = y_i$ for $\pi(i) \not= n+1$, $x_{i_0} = 1$, $x_{i_0}' = 0$ with $i_0 = \pi^{-1}(n+1)$. Now,
   \[
\mathbb{E}\left(\Pi_u[\hat{\mathbb{P}}_n,X_{n+1}](y)) - \Pi_{\ell}[\hat{\mathbb{P}}_n,X_{n+1}](y))\right) = \mathbb{E}(\mathrm{iso}(x)_{i_0} - \mathrm{iso}(x')_{i_0}).
\]
Conditional on $\hat{\mathbb{P}}_{n+1}$, the probability the $Z_{n+1}$ falls into any index interval $I \subset \{1,\dots,n+1\}$ is $|I|/(n+1)$. Split each index interval of the partition $I_1^y,\dots,I_\ell^y$ of $\{1,\dots,n+1\}$ into two consecutive index intervals $I_j'$ and $I_j''$ such that $I_j''$, which contains the larger indices, has $[|I_j^y|n^{-1/6}]$ elements. If $i_0$ falls in $I_{j_0}'$, then $I_{j_0}''\subset I_{j_0}^y \cap I_{\ell_0}^x = \{\min I_{\ell_0}^x, \dots, \max I_{j_0}^y\}$, hence $\max I_{j_0}^y - \min I_{\ell_0}^x + 1 \ge [|I_{j_0}^y|n^{-1/6}] \ge |I_{j_0}^y|n^{-1/6}$. Therefore,  
using \eqref{eq:i_0bound}, we obtain
\begin{align*}
0 \le \E(\mathrm{iso}(x)_{i_0} - \mathrm{iso}(y)_{i_0}|\tilde{\mu}) & \le \sum_{j=1}^\ell \E\left(\one\{i_0 \in I_j'\}\frac{n^{1/6}}{|I_j^y|}\;\Big|\;\hat{\mathbb{P}}_{n+1}\right)+ \E\left(\one\{i_0 \in I_j''\}\mid\hat{\mathbb{P}}_{n+1}\right)\\
&\le \sum_{j=1}^\ell \frac{|I_j^y|}{n}  \frac{n^{1/6}}{|I_j^y|} + \frac{|I_{j_0}^y|n^{-1/6} + 1}{n}\\
&= \ell n^{-1+1/6} + n^{-1/6} + \ell n^{-1} \le 3 n^{-1/6} + n^{-1/6} + 3 n^{-1/3} \le 7n^{-1/6},
\end{align*}
where the final inequality uses that $\ell \le 3n^{2/3}$ by \citet[Lemma 1]{DimitriadisDumbgenETAL2023}. An analogous argument can be applied to bound $0 \le \E(\mathrm{iso}(y)_{i_0} - \mathrm{iso}(x')_{i_0}|\hat{\mathbb{P}}_{n+1})$, which yields the claim.

\end{proof}

\section{Isotonic calibration}\label{app:isotoniccal}


A probabilistic forecast $F$ is auto-calibrated if $\L( Y \mid F) = F$ almost surely. That is, given that $F$ is issued as the forecast, the observations will arise according to this predicted distribution, meaning the forecast can be taken at face value. Auto-calibration is therefore an intuitive property that we would like our forecasts to satisfy. However, auto-calibration is generally prohibitively difficult to assess in practice: We require an estimate of the conditional distribution $\L( Y \mid F)$. Instead of conditioning on the $\sigma$-algebra generated by $F$, \cite{ArnoldZiegel2023} suggest conditioning on the $\sigma$-lattice generated by $F$, $\A(F) \subset \F$. They call a forecast isotonically calibrated if $F = \L(Y \mid \A(F))$.

A $\sigma$-lattice $\C \subset \F$ is a collection of subsets that contains $\varnothing$ and $\Omega$, and is closed under countable unions and intersections (but not necessarily under complements). Any $\sigma$-algebra is therefore also a $\sigma$-lattice, and the $\sigma$-lattice generated by a random variable is a subset of the $\sigma$-algebra it generates. In particular, while $\L(Y \mid F)$ represents the conditional distribution of $Y$ given all information about $F$, $\L(Y \mid \A(F))$ represents the conditional distribution of $Y$ given all information about $F$ under the additional assumption of isotonicity between $Y$ and $F$ \citep{ArnoldZiegel2023}. Isotonic calibration is therefore a strictly weaker requirement than auto-calibration. 

However, in contrast to auto-calibration, it is straightforward to assess isotonic calibration in practice. Isotonic distributional regression (IDR) \citep{HenziZiegelETAL2021} is uniformly consistent for the isotonic conditional law. Hence, given a set of forecast distributions $F_{1}, \dots, F_{n}$ and corresponding observations $y_{1}, \dots, y_{n}$, isotonic calibration can be assessed by applying IDR with the set of forecast distributions as covariates, and the observations as target variable. This is discussed in detail by \cite{ArnoldEtAl2024}.

\section{Probabilistic calibration}\label{app:PIT}

We show the following results that make our claims in Section \ref{sec:calibration} precise.

\begin{prop}\label{prop:PIT1}
Let $F$ be a deterministic CDF, and $Y$ a random variable. Suppose that
\[
\prob(F(Y) \le \alpha) \le \alpha \le \mathbb{P}(F(Y-) < \alpha), \quad \text{for all $\alpha \in (0,1)$.}
\]
Then, $Y$ has distribution $F$. If $Y$ has distribution $F$, then the above relation holds.
\end{prop}
\begin{proof}
Let $G$ be the CDF of $Y$. The proof that the inequalities imply $F=G$ is by contradiction.
\textit{Part 1:} Assume that $F(z) < G(z)$ for some $z$. Then, for $\alpha \in (F(z),G(z)) \subset (0,1)$, 
\[
\prob(F(Y) \le \alpha) \ge \prob(F(Y) \le F(z)) \ge \prob (Y \le z) = G(z) > \alpha,
\]
which is a contradiction to the first inequality above.

\noindent \textit{Part 2:} Assume that $F(z) > G(z)$ for some $z$. Let $\alpha \in (G(z),F(z)) \subset (0,1)$. Then,
\[
\prob(F(Y-) < \alpha) \le \prob(F(Y-) < F(z)) \le \prob(Y \le z) = G(z) < \alpha,
\]
which contradicts the second inequality. 

For the reverse statement, recall that, for any $u \in (0,1)$, it holds that $F(F^{-1}(u)) \ge u$ and $F(F^{-1}(u)-) \le u$. Let $U$ be a standard uniform random variable. Then, $F^{-1}(U) \sim F$. Therefore,
\[
\prob(F(Y) \le \alpha) = \prob(F(F^{-1}(U))\le \alpha) \le \prob(U \le \alpha) = \alpha,
\]
and
\[
\prob(F(Y-) < \alpha) = \prob(F(F^{-1}(U)-) < \alpha) \ge \prob(U < \alpha) = \alpha.
\]
\end{proof}

\begin{prop}\label{prop:PIT2}
Let the random CDF $F$ and the random variable $Y$ be defined on the same probability space, and suppose that
\[
\prob(F(Y) < \alpha) \le \alpha \le \mathbb{P}(F(Y-) \le \alpha), \quad \text{for all $\alpha \in (0,1)$,}
\]
holds. Then, (after possibly extending the probability space), there exists a random variable $\lambda$ on $[0,1]$ that may be dependent on $(F,Y)$ such that
\[
F(Y-) + \lambda (F(Y)-F(Y-))
\]
has a standard uniform distribution. 
\end{prop}

\begin{proof}
The random variable $F(Y)$ is stochastically larger than a standard uniform random variable and the random variable $F(Y-)$ is stochastically smaller than a standard uniform random variable. Therefore, by Strassen's theorem, we can find a probability space and random variables $Z_1, Z_2, U$ such that
\[
Z_1 \le U \le Z_2 \; \text{almost surely}, \quad \text{and} \quad Z_1 \overset{d}{=} F(Y-), U \sim \mathrm{UNIF}(0,1), Z_2 \overset{d}{=} F(Y).
\]
For each $\omega \in \Omega$ such that $Z_1(\omega) \le U(\omega) \le Z_2(\omega)$, define $\lambda'(\omega)$ such that $U(\omega) = Z_1(\omega) + \lambda'(\omega)(Z_2(\omega)-Z_1(\omega))$, that is, if $Z_2(\omega) - Z_1(\omega) > 0$, 
\[
\lambda'(\omega) = \frac{U(\omega)-Z_1(\omega)}{Z_2(\omega) - Z_1(\omega)} \in [0,1],
\]
and $\lambda'(\omega) = U(\omega)$ otherwise. Now, (after possibly extending the probability space), we can construct a random variable $\lambda$ on the original probability space such that $(F(Y-),F(Y),\lambda)$ has the same distribution as $(Z_1,Z_2,\lambda')$. 
\end{proof}

Finally, we show that the distributional regression procedure introduced in Example \ref{ex:procedure1} is probabilistically calibrated.

\begin{prop}
    \label{prop:proof_example_2.1}
   Let  \(z_1=\left(x_1,y_1\right),\ldots,z_m=\left(x_m,y_m\right) \in \mathbb{R}^p \times \mathbb{R}\) and $w_1,\dots,w_m \ge 0$ with $\sum_{i=1}^m w_i = 1$. Define \(\hat{\mathbb{P}}=\sum_{i=1}^m w_i\delta_{z_i}\).
    Let \(\widehat{y}_1,\ldots,\widehat{y}_m\) be any regression fit of \(\mathbf{y}=\left(y_1,\ldots,y_m\right)\) on \(\mathbf{x}=\left(x_1,\dots,x_m\right)\) and let \(\epsilon_i=y_i-\widehat{y}_i\) be the residuals.
    Define \(G_{x_k}^{\hat{\mathbb{P}}}(y)\) to be the weighted empirical distribution of the residuals centered at the regression fit, that is, for \(k=1,\ldots,m\),
    \begin{equation*}
        G_{x_k}^{\hat{\mathbb{P}}}(y)(y)=\sum_{j=1}^m w_j\one\left\{\widehat{y}_k+\widehat{\varepsilon}_j\leq y\right\},\ y\in \R.
    \end{equation*}
    Then, the procedure \(G\) is probabilistically calibrated.
\end{prop}

\begin{proof}
We first focus on the case where all the weights are equal.
Fix \(a\in (0,1)\) and suppose that \(k/m \le a <  (k+1)/m\) for some integer \(k\in \{0,\ldots,m-1\}\).
We first prove the left inequality, namely
\begin{equation*}
    \hat{\mathbb{P}}\big(G_{X}^{\hat{\mathbb{P}}}(Y)\le a\big)\leq a.
\end{equation*}
Since \(G_{X}^{\hat{\mathbb{P}}}(Y)\) only takes values in the set \(\{i/m\mid 0\leq i\leq m\}\), the event \(\{G_{X}^{\hat{\mathbb{P}}}(Y)\le a\}\) is equal to the event \(\{G_{X}^{\hat{\mathbb{P}}}(Y)\leq k/m\}\).
Similarly, since \(\hat{\mathbb{P}}\) takes values in the same set, the inequality we want to show is equivalent to
\begin{equation*}
    \hat{\mathbb{P}}\left(G_{X}^{\hat{\mathbb{P}}}(Y)\leq \frac{k}{m} \right)\leq \frac{k}{m}.
\end{equation*}
Suppose that this inequality is not true.
Then, it holds that \(\hat{\mathbb{P}}(G_{X}^{\hat{\mathbb{P}}}(Y)\leq k/m)\geq (k+1)/m\), which is equivalent to
\begin{equation*}
    \sum_{j=1}^m \one\{G_{x_j}^{\hat{\mathbb{P}}}(y_j)\leq k/m\}\geq k+1.
\end{equation*}
This means that there exist at least \(k+1\) indices \(i_1,\ldots,i_{k+1}\in \left\{1,\ldots,m\right\}\) such that
\begin{equation*}
    G^{\hat{\mathbb{P}}}_{x_{i_\ell}}(y_{i_\ell})\leq k/m,\text{ for all } \ell\in \{1,\ldots,k+1\}.
\end{equation*}
The condition \(G^{\hat{\mathbb{P}}}_{x_{i_\ell}}(y_{i_\ell})\leq k/m\) can be rewritten as
\begin{align*}
    \sum_{j=1}^m \one\{\widehat{y}_{i_\ell}+\widehat{\varepsilon}_j\leq y_{i_\ell}\}\leq k
    &\Leftrightarrow \sum_{j=1}^m \one\{\widehat{\varepsilon}_j\leq \widehat{\varepsilon}_{i_\ell}\}\leq k\\
    &\Leftrightarrow \sum_{j=1}^m \one\{\widehat{\varepsilon}_j>\widehat{\varepsilon}_{i_\ell}\}\geq m-k.
\end{align*}
This means that there exist at least \(m-k\) residuals that are strictly larger than \(\varepsilon_{i_\ell}\), and this is true for all \(\ell\in \{1,\ldots,k+1\}\).
Consider the set of indices
\begin{equation*}
    \Big\{1\leq j\leq m\mid \widehat{\varepsilon}_j>\max\{\widehat{\varepsilon}_{i_1},\ldots,\widehat{\varepsilon}_{i_{k+1}}\}\Big\}.
\end{equation*}
By definition, this set is a subset of \(\{1,\ldots,m\}\backslash \{i_1,\ldots,i_{k+1}\}\), so it has at most \(m-k-1\) elements.
However, due to the above observation, this set must have at least \(m-k\) elements.
This leads to a contradiction.

We now need to show the right inequality, namely \(a \leq \hat{\mathbb{P}}(G^{\hat{\mathbb{P}}}_{X}(Y-) < a)\).
Suppose that \(a\neq k/m\).
Since \(G\) and \(\hat{\mathbb{P}}\) take values in the set \(\{i/m\mid 0\leq i\leq m\}\), this inequality is equivalent to
\begin{equation*}
    \hat{\mathbb{P}}\left(G^{\hat{\mathbb{P}}}_{X}\left(Y-\right)\leq \frac{k}{m}\right)\geq \frac{k+1}{m}.
\end{equation*}
If \(a=k/m\), then the inequality is equivalent to
\begin{equation*}
     \hat{\mathbb{P}}\left(G^{\hat{\mathbb{P}}}_{X}\left(Y-\right)\leq \frac{k-1}{m}\right)\geq \frac{k}{m}.
\end{equation*}
The proof is exactly the same for both inequalities, so we only show that first one.
Suppose that it does not hold.
Then,
\begin{equation*}
    \sum_{j=1}^m \one\{G^{\hat{\mathbb{P}}}_{x_j}\left(y_j-\right)\leq k/m\}\leq k\Rightarrow \sum_{j=1}^m \one\{G^{\hat{\mathbb{P}}}_{x_j}\left(y_j-\right)> k/m\}\geq m-k,
\end{equation*}
so there exists (at least) \(m-k\) indices \(i_1,\ldots,i_{m-k}\) such that
\begin{equation*}
    G^{\hat{\mathbb{P}}}_{x_{i_\ell}}(y_{i_\ell}-)\geq \frac{k+1}{m},\text{ for all } \ell\in \left\{1,\ldots,m-k\right\}.
\end{equation*}
It holds that
\begin{equation*}
    G^{\hat{\mathbb{P}}}_{x_k}(y-)=\frac{1}{m}\sum_{j=1}^m \one\{\widehat{y}_k+\widehat{\varepsilon}_j< y\},
\end{equation*}
so the above inequality can be written as
\begin{align*}
    \sum_{j=1}^m \one\{\widehat{y}_{i_\ell}+\widehat{\varepsilon}_j<y_{i_\ell}\}\geq k+1
    &\Leftrightarrow \sum_{j=1}^m \one\{\widehat{\varepsilon}_j<\widehat{\varepsilon}_{i_\ell}\}\geq k+1.
\end{align*}
This means that there are at least \(k+1\) residuals that are strictly smaller than \(\widehat{\varepsilon}_{i_\ell}\), and this holds for all \(\ell\in \{1,\ldots,m-k\}\).
Consider the set of indices
\begin{equation*}
    \Big\{1\leq j\leq m\mid \widehat{\varepsilon}_j<\min\{\widehat{\varepsilon}_{i_1},\ldots,\widehat{\varepsilon}_{i_{m-k}}\}\Big\}.
\end{equation*}
By definition, this set is a subset of \(\{1,\ldots,m\}\backslash \{i_1,\ldots,i_{m-k}\}\), so it has at most \(k\) elements.
However, due to the above observation, this set must have at least \(k+1\) elements.
This leads to a contradiction, and it follows that $G$ is probabilistically calibrated.


We can now use the above argument to prove probabilistic calibration for any set of rational weights.
Suppose that for each \(i\in\left\{0,\ldots,m\right\}\) it holds that \(w_i=s_i/t_i\), where \(s_i,t_i\) are positive integers with \(s_i\leq t_i\).
If \(T\) is the least common multiple of \(t_1,\ldots,t_m\), then, for any \(k\in \left\{1,\ldots,m\right\}\),
\begin{equation*}
    G^{\hat{\mathbb{P}}}_{x_k}(y)=\frac{1}{T}\sum_{j=1}^m \widetilde{s}_j \one\{\widehat{y}_k+\widehat{\varepsilon}_j\leq y\},
\end{equation*}
where \(\widetilde{s}_1,\ldots,\widetilde{s}_m\) are positive integers.
By decomposing \(\widetilde{s}_j \one\{\widehat{y}_k+\widehat{\varepsilon}_j\leq y\}\) into \(\widetilde{s}_j\) equal terms, we can write \(G\) as a sum of terms with equal weights.
This shows that \(G\) is probabilistically calibrated.
The original argument for equal weights does not assume that \(\left(x_1,y_1\right),\ldots,\left(x_m,y_m\right)\) are distinct, so the fact that some terms of the sum are identical is not a problem.

Using a continuity argument, we can extend this to any set of non-negative weights.
Indeed, given a set of weights \(\{w_1,\ldots,w_m\}\) and an arbitrary \(\varepsilon>0\), we can define a measure \(\widetilde{\mathbb{P}}=\sum_{i=1}^m \widetilde{w}_i\delta_{z_i}\)
with rational weights, such that
\begin{equation*}
    \Big|G^{\hat{\mathbb{P}}}_{x_k}(y)-G^{\widetilde{\mathbb{P}}}_{x_k}(y)\Big|<\varepsilon,\ \Big|G^{\hat{\mathbb{P}}}_{x_k}(y-)-G^{\widetilde{\mathbb{P}}}_{x_k}(y-)\Big|<\varepsilon
\end{equation*}
and \(\left|\hat{\mathbb{P}}(A)-\widetilde{\mathbb{P}}(A)\right|< \varepsilon\) for all \(k\in \{1,\ldots,m\}\), \(y\in \R\) and all \(A\in \supp(\hat{\mathbb{P}})=\supp(\widetilde{\mathbb{P}})\).
For this measure it holds that
\begin{equation*}
    \widetilde{\mathbb{P}}\left(G^{\widetilde{\mathbb{P}}}_{X}(Y\right)<a)\leq a\leq \widetilde{\mathbb{P}}\left(G^{\widetilde{\mathbb{P}}}_{X}(Y-)\leq a\right)
\end{equation*}for all \(a\in (0,1)\),
so \(\hat{\mathbb{P}}(G^{\hat{\mathbb{P}}}_X(Y)<a-\varepsilon)\leq a+\varepsilon\)
and \(a-\varepsilon \leq \hat{\mathbb{P}}(G^{\hat{\mathbb{P}}}_X(Y-)\leq a+\varepsilon).\)
Measure continuity yields that
\begin{equation*}\hat{\mathbb{P}}\left(G^{\hat{\mathbb{P}}}_X(Y)<a\right)\leq a\leq \hat{\mathbb{P}}\left(G^{\hat{\mathbb{P}}}_X(Y-)\leq a\right),
\end{equation*}
which finishes the proof.

\end{proof}

\section{Crisp Conformal IDR predictive distribution}\label{app:crisp}

In the following, we discuss the choice of crisp CDF from conformal IDR predictive systems. The following proposition shows that the standard IDR fit is guaranteed to be within the conformal IDR bounds.

\begin{prop}\label{prop:cidr_crisp}
Assume that $\le$ is a partial order on $\mathcal{X}$. Then, for any $(X_1, Y_1), \dots, (X_n, Y_n), (X_{n+1}, y) \in \mathcal{X}\times \mathbb{R}$, the conformal IDR satisfies
\begin{align*}
 \Pi_{\ell}[\hat{\mathbb{P}}_n, X_{n+1}](y) & \leq \max\{G_{X_i}^{\hat{\mathbb{P}}_n}(y)\colon i = 1, \dots, n, \,  X_i \geq X_{n+1}\} \\
 & \leq \min\{G_{X_i}^{\hat{\mathbb{P}}_n}(y)\colon i=1,\dots,n, \, X_i \leq X_{n+1}\} \leq \Pi_{u}[\hat{\mathbb{P}}_n, X_{n+1}](y),
\end{align*}
where $\max \emptyset := 0$ and $\min \emptyset := 1$.
\end{prop}

\begin{proof}
The result follows from Lemma 2.1 (i) of \citet{HenziMoeschingDuembgen2022}.
More precisely, let
\[
  v \in [\max\{G_{X_i}^{\hat{\mathbb{P}}_n}(y)\colon X_i \geq X_{n+1}\},\, \min\{G_{X_i}^{\hat{\mathbb{P}}_n}(y)\colon X_i \leq X_{n+1}\}].
\]
The vector $\mathbf{z}$ can be defined as $\mathbf{z} = (s_i)_{i=1}^{n+1}$ 
with $s_i = \one\{Y_i \leq y\}$ for $i = 1, \dots, n$ and $s_{n+1} = v$, and $\mathbf{\tilde{z}}$ equal to $\mathbf{z}$ in its first $n$ components and with last component equal to $1$. Then $(G_{X_1}^{\hat{\mathbb{P}}_n}(y), \dots, G_{X_n}^{\hat{\mathbb{P}}_n}(y), v)$ is the antitonic regression of $\mathbf{z}$. That is, it minimizes
\[
  \sum_{i=1}^{n+1}(\theta_i - s_i)^2 = \sum_{i=1}^n (\theta_i - \one\{Y_i \leq y\})^2 + (\theta_{n+1} - v)^2
\]
over $(\theta_1, \dots, \theta_{n+1})$ such that $\theta_i \geq \theta_j$ if $X_i \leq X_j$, because it is defined as the IDR in its first $n$ components, and the last component does not violate the order constraints and has zero contribution to the error. Analogously, the antitonic regression of $\mathbf{\tilde{z}}$ yields a vector with $ \Pi_{u}[\hat{\mathbb{P}}_n, X_{n+1}](y)$ in its last component, and so $v \leq \Pi_{u}[\hat{\mathbb{P}}_n, X_{n+1}](y)$ by Lemma 2.1 (i) of \citet{HenziMoeschingDuembgen2022}.
The inequality for $\Pi_{\ell}[\hat{\mathbb{P}}_n, X_{n+1}]$ 
follows by setting the last component of $\mathbf{\tilde{z}}$ to $0$ and 
interchanging the roles of $\mathbf{z}$ and $\mathbf{\tilde{z}}$.
\end{proof}

Alternatively, the crisp CDF could be selected to optimize a scoring rule, inspired by the work of \citet{VovkPetejETAL2015}, who considered conformal IDR for binary outcomes under the name of Venn-Abers predictors. In this case, conformal IDR consists of two predictive probabilities $f_0(X)$ and $f_1(X)$ (corresponding to the prediction obtained from IDR when the $n+1$-th observation in the training data is assumed to be zero and one respectively). Depending on how large $n$ is, and how central $X_{n+1}$ is within $X_1,\dots,X_n$, the two probabilities may or may not differ a lot. This is analogous to conformal IDR for real-valued outcomes; see Figure \ref{fig:2}. 

If one wants to combine the two probability predictions into one, \citet{VovkPetejETAL2015} suggest the following strategy. Take a proper scoring rule $S$ \citep{GneitingRaftery2007}, and use $p$ to denote the combined probability prediction. If the future outcome is $Y=\ell\in \{0,1\}$ and we predict $p$ instead of $f_\ell(X)$, we lose
$S(p,\ell) - S(f_\ell(X),\ell)$. Therefore, choose $p$ to minimize the maximal possible loss
\[
\max_{\ell \in \{0,1\}}S(p,\ell) - S(f_\ell(X),\ell).
\]
If $S$ is the logarithmic score, $S(p, \ell) = -\log |p + \ell - 1|$, we obtain $p = f_1(X)/(1 + f_1(X) - f_0(X))$, and $f_0(X) \le p \le f_1(X)$.
With respect to the Brier score, $S(p, \ell) = (p - \ell)^2$, one obtains
\begin{equation}\label{eq:Brier}
p = f_1(X) - \frac{1}{2}f_1(X)^2 + \frac{1}{2}f_0(X)^2.
\end{equation}

For choosing a crisp CDF within conformal IDR, we would like to apply the same strategy using the continuous ranked probability score (CRPS) as the proper scoring rule, defined as
\begin{equation}\label{def:crps}
\operatorname{CRPS}(F,y) = \int\left(\one\{y \le z\} - F(z)\right)^2\diff z,
\end{equation}
for a CDF $F$ and $y \in \R$.
However, we cannot apply the same strategy as before with the CRPS. To see this, let $G$ be the IDR procedure and $\hat{\mathbb{P}}_n=\sum_{j=1}^n \delta_{(X_j,Y_j)}$. Then, for any possible future outcome $Y=y \in \R$, we have
\begin{align}
\operatorname{CRPS}(F,y) &- \operatorname{CRPS}(G_{X_{n+1}}^{\hat{\mathbb{P}}_n + (X_{n+1}, y)}(,y)\nonumber\\ &= \int \left[\left(\one\{y \le z\}-F(z)\right)^2 - \left(\one\{y \le z\} -G_{X_{n+1}}^{\hat{\mathbb{P}}_n + (X_{n+1}, y)}(z)\right)^2  \right]\diff z.\label{eq:CRPSdiff}
\end{align}
We would like to minimize the supremum of this expression over any CDF $F$ with $\Pi_\ell[\hat{\mathbb{P}}_n,X_{n+1}](z) \le F(z) \le \Pi_u[\hat{\mathbb{P}}_n,X_{n+1}](z)$, for all $z \in \R$, where $\Pi$ is the conformal IDR predictive system. However, the integral at \eqref{eq:CRPSdiff} diverges to $\infty$ as $y \to \infty$, since then $G_{X_{n+1}}^{\hat{\mathbb{P}}_n + (X_{n+1}, y)}(z) \to \Pi_\ell[\hat{\mathbb{P}}_n,X_{n+1}](z)$ for every $z \in \R$, $F(z) \geq \Pi_{\ell}\left[\hat{\mathbb{P}}_n, X_{n+1}\right](z)$, and $\lim_{z \rightarrow \infty}\Pi_{\ell}\left[\hat{\mathbb{P}}_n, X_{n+1}\right](z) < 1$. 

Instead, we follow a pointwise approach and minimize the maximal possible loss of the integrand for each fixed $z$. We have from \eqref{eq:Brier} that
\begin{equation*}
F(z) = \Pi_u[\hat{\mathbb{P}}_n, X_{n+1}](z) - \frac{1}{2} \Pi_u[\hat{\mathbb{P}}_n, X_{n+1}](z)^2 + \frac{1}{2} \Pi_{\ell}[\hat{\mathbb{P}}_n, X_{n+1}](z)^2.
\end{equation*}
Since the function $x - (1/2)x^2$ is increasing on $[0,1]$, $F(z)$ is increasing in $z$. In order to obtain a CDF, we set $F$ equal to zero and one beyond some large interval $[D_1,D_2]$, respectively. This provides an alternative approach with which to construct the crisp conformal IDR CDF. 

\section{Additional simulation study results}\label{app:sim}

\begin{figure}
    \centering
    \includegraphics[width=0.9\linewidth]
    {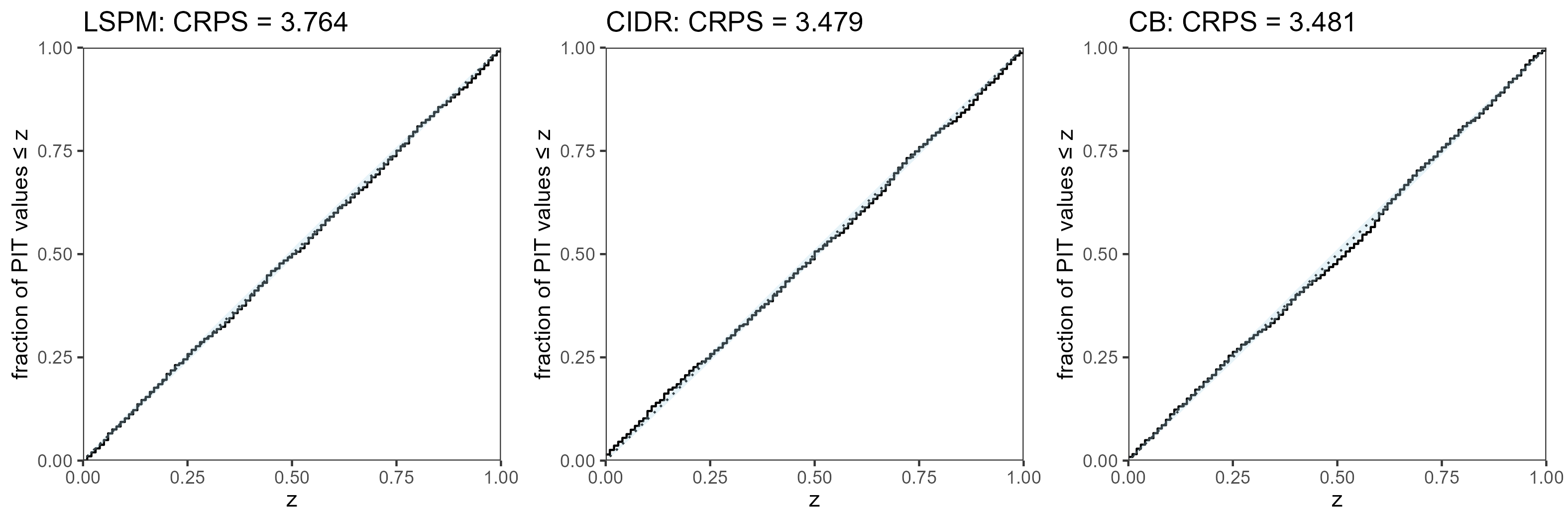}
    \includegraphics[width=0.9\linewidth]{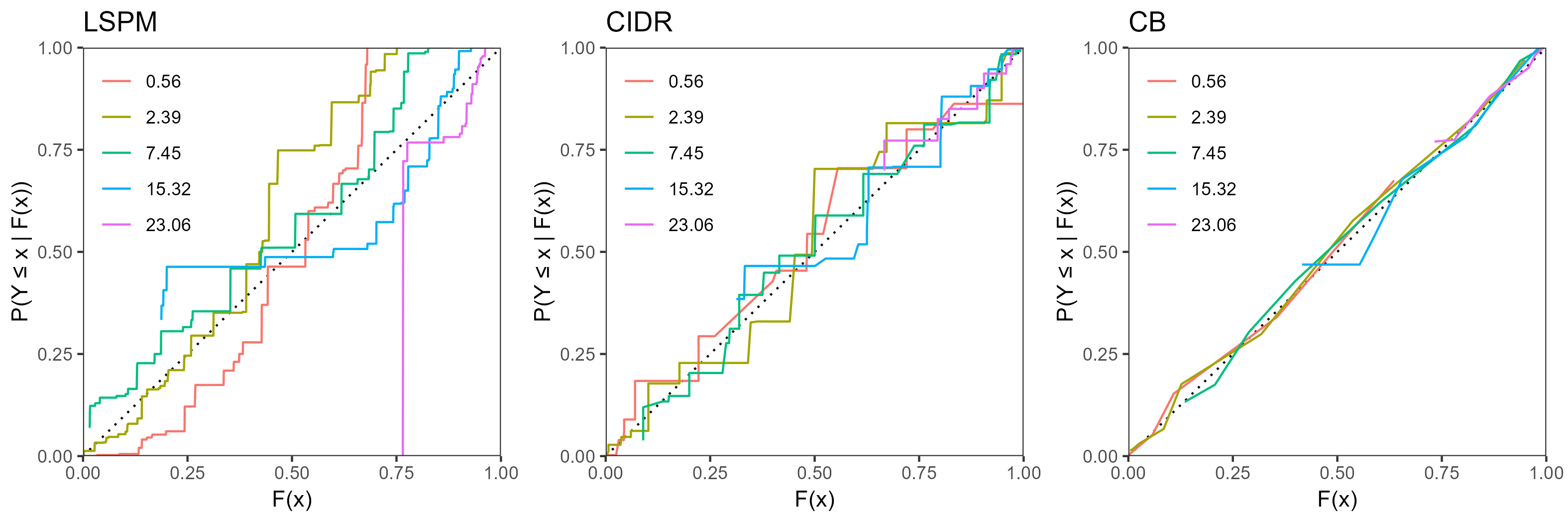}
    \caption{p-p plots of PIT values (top row) and threshold calibration reliability diagrams (bottom) for the least squares prediction machine (LSPM), conformal IDR (CIDR), and conformal binning (CB) when predicting the data generated from \eqref{eq:simstudy_data}. The p-p plots contain a 90\% consistency interval around the diagonal, as well as average CRPS values for each method. Threshold calibration is displayed at five thresholds $x$, corresponding to the 10$^{th}$, 25$^{th}$, 50$^{th}$, 75$^{th}$, and 90$^{th}$ percentiles of the observations.}
    \label{fig:simstudy}
\end{figure}

\begin{figure}
    \centering
    \begin{subfigure}{0.24\textwidth}
        \includegraphics[width=\linewidth]{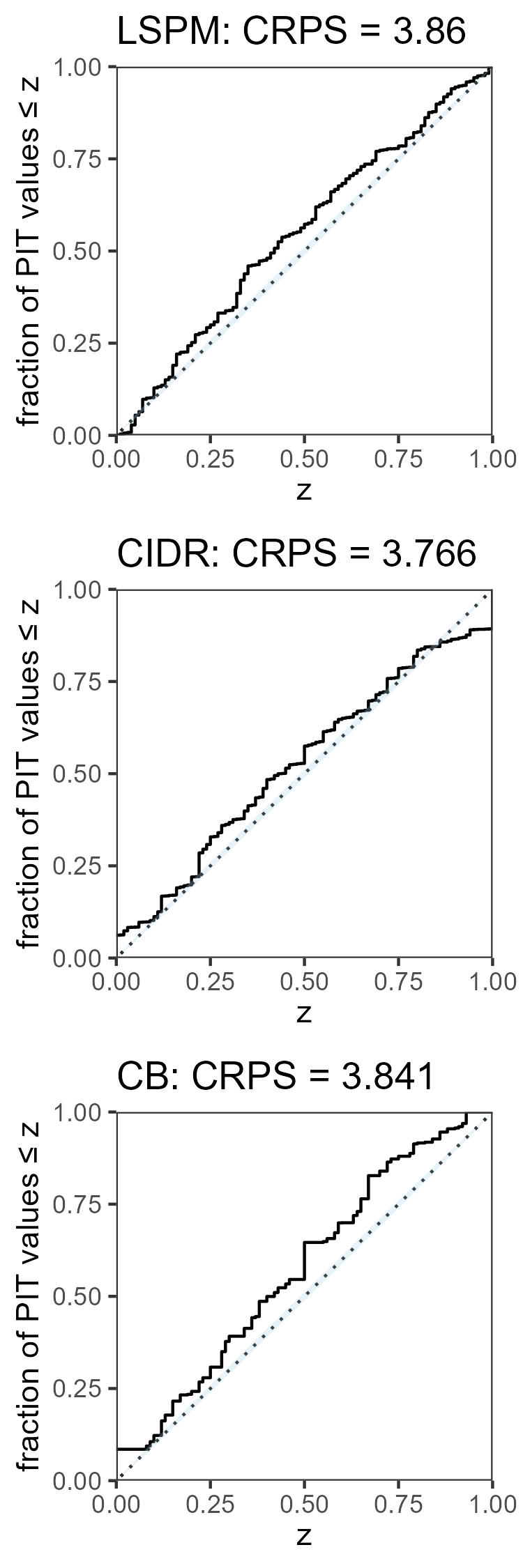}
    \caption{$n = 100$}
    \end{subfigure}
    \begin{subfigure}{0.24\textwidth}
        \includegraphics[width=\linewidth]{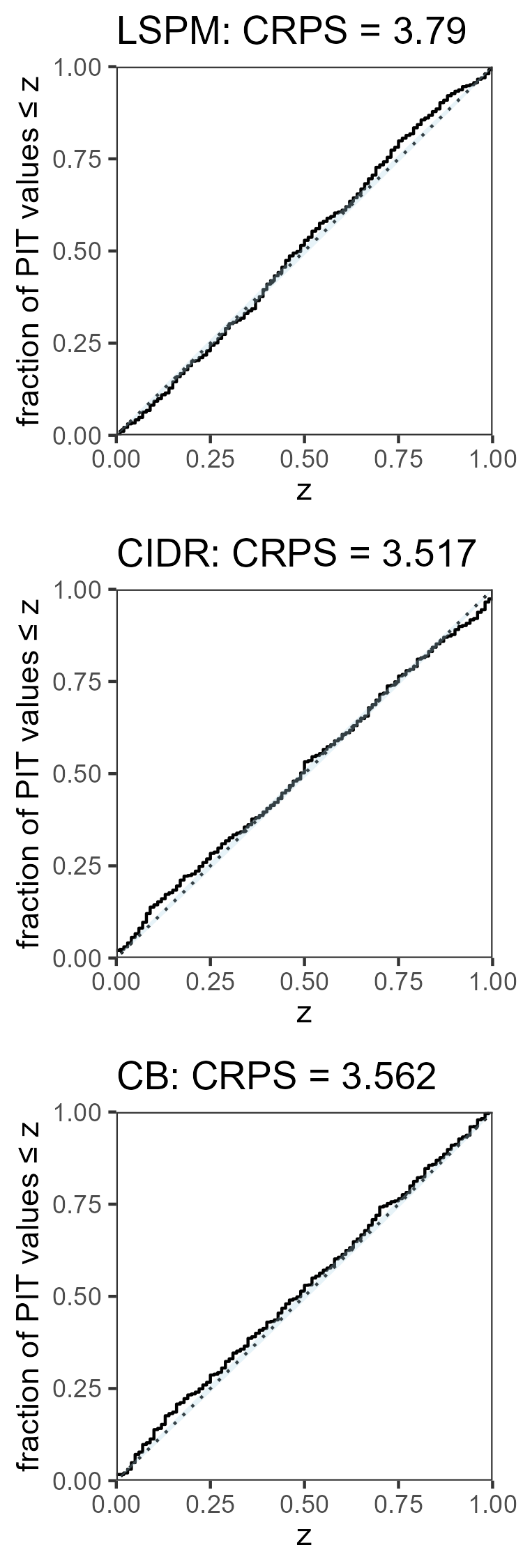}
    \caption{$n = 500$}
    \end{subfigure}
    \begin{subfigure}{0.24\textwidth}
        \includegraphics[width=\linewidth]{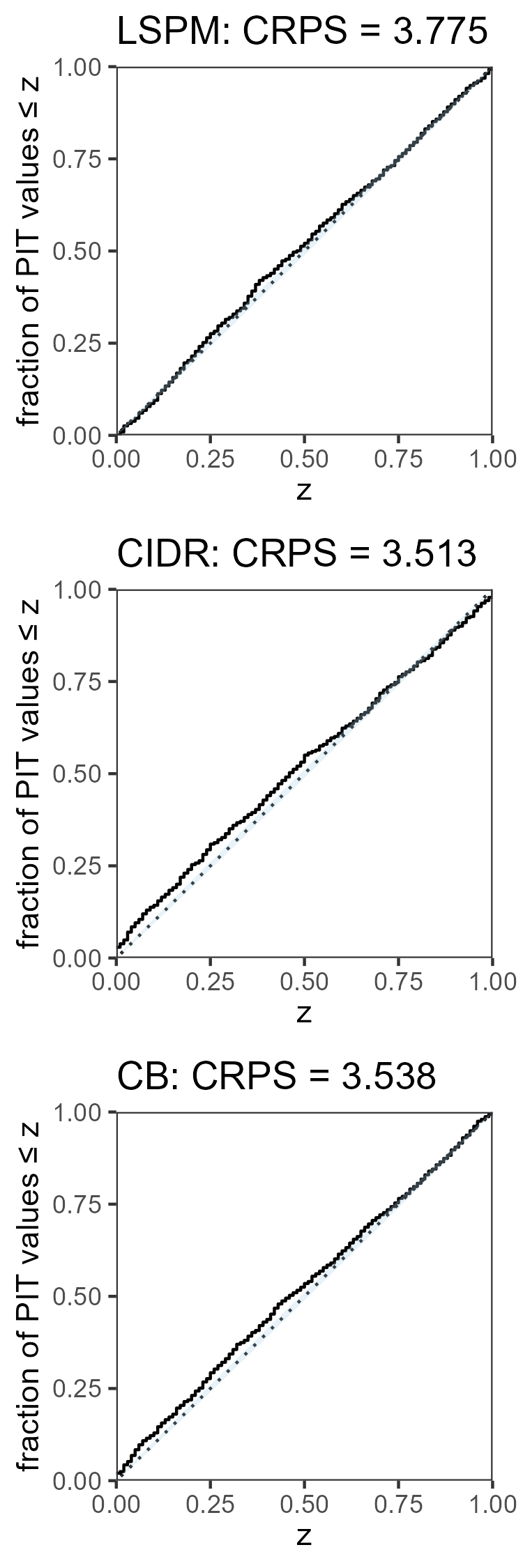}
    \caption{$n = 1000$}
    \end{subfigure}
    \begin{subfigure}{0.24\textwidth}
        \includegraphics[width=\linewidth]{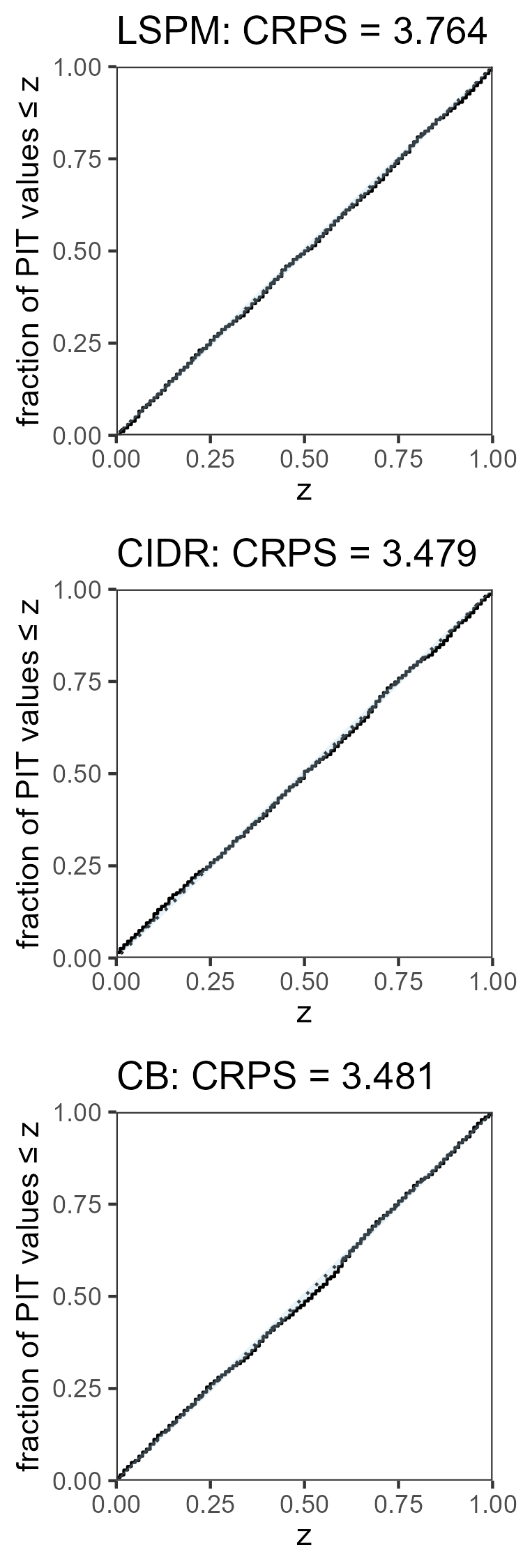}
    \caption{$n = 2000$}
    \end{subfigure}
    \caption{p-p plots of PIT values for the least squares prediction machine (LSPM, first row), conformal IDR (CIDR, second row), and the conformal binning (CB, thrid row) approach when trained using $n = 100, 500, 1000$ and 2000 independent draws from (\ref{eq:simstudy_data}). A 90\% consistency interval is shown around the diagonal in light blue. Average CRPS values for each method are also displayed.}
    \label{fig:simstudy_app}
\end{figure}

Figure \ref{fig:simstudy} presents PIT p-p plots and average CRPS values for the crisp CDF of the LSPM, conformal IDR, and the conformal binning approach when applied to the simulated data generating process in Section \ref{sec:sims}. PIT p-p plots display the empirical CDF of the PIT values; since these should be uniformly distributed, deviations from the diagonal are indicative of miscalibration. All predictive systems appear probabilistically calibrated. Conformal IDR and conformal binning outperform the LSPM, resulting in CRPS values around 8\% lower on average. The conformal binning approach is more sensitive to small sample sizes (Figure \ref{fig:simstudy_app}), though this could be circumvented by varying the number of bins depending on the amount of training data. 

Figure \ref{fig:simstudy} also displays threshold calibration reliability diagrams for the three predictive systems; these are based on the CORP approach proposed by \cite{DimitriadisGneitingETAL2021} and \cite{GneitingResin2023}. For conformal IDR and conformal binning, the curves lie roughly on the diagonal for all thresholds considered, confirming that these methods are threshold calibrated. In contrast, the curves deviate from the diagonal for the LSPM: For small thresholds, where there is relatively low uncertainty in the data (Figure \ref{fig:simstudy_data}), the predictions are under-confident and overestimate the uncertainty, while the opposite is true for larger thresholds; these errors average out over all thresholds. 


\begin{figure}
    \centering
    \begin{subfigure}{0.3\textwidth}
        \includegraphics[width=\linewidth]{Figures/simstudy_data.png}
    \caption{Isotonic}
    \end{subfigure}
    \begin{subfigure}{0.3\textwidth}
        \includegraphics[width=\linewidth]{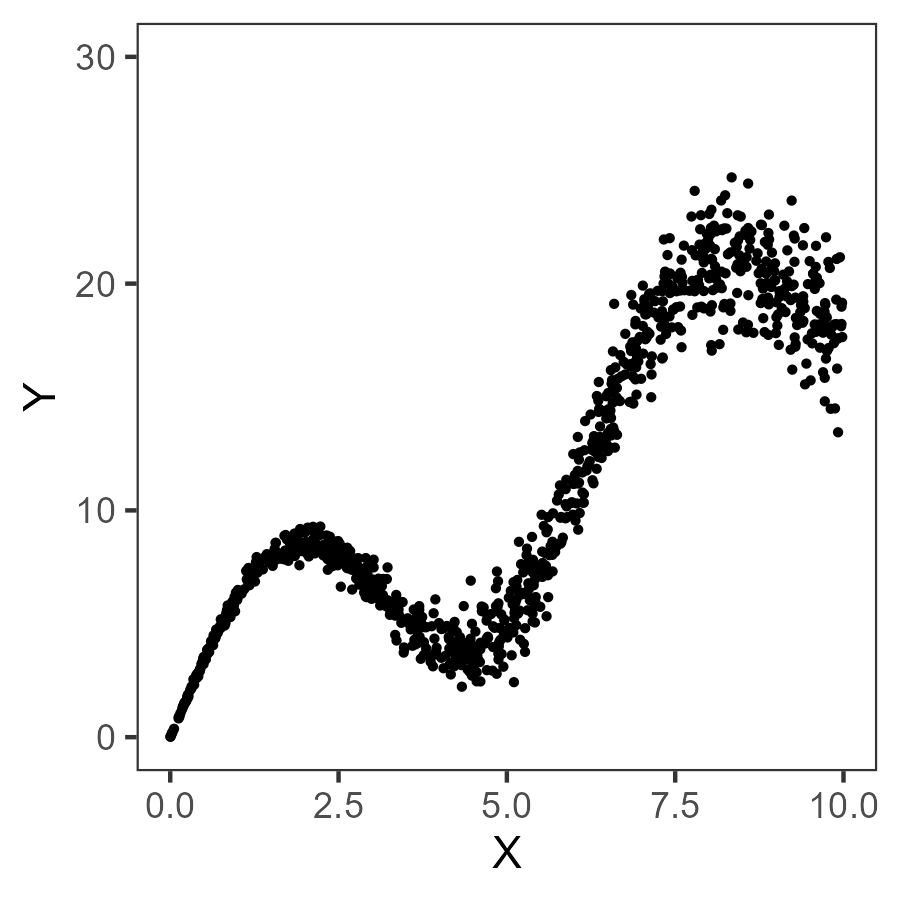}
    \caption{Less isotonic}
    \end{subfigure}
    \caption{1000 points generated using (a) the model at \eqref{eq:simstudy_data}, and (b) at \eqref{eq:simstudy_data_less}.}
    \label{fig:simstudy_data2}
\end{figure}

In \eqref{eq:simstudy_data}, there is a (weak) isotonic relationship between the covariate and the outcome, thereby satisfying the key assumption underlying IDR. Instead, consider the data generating process
\begin{equation}\label{eq:simstudy_data_less}
    Y \mid X \sim \mathcal{N}(2X + 5\sin(X), (X/5)^{2}),
\end{equation}
for which the assumption of isotonicity between $X$ and $Y$ is questionable. Data drawn from this model are displayed in Figure~\ref{fig:simstudy_data2}. 

\begin{figure}[t]
    \centering
    \includegraphics[width=0.49\linewidth]{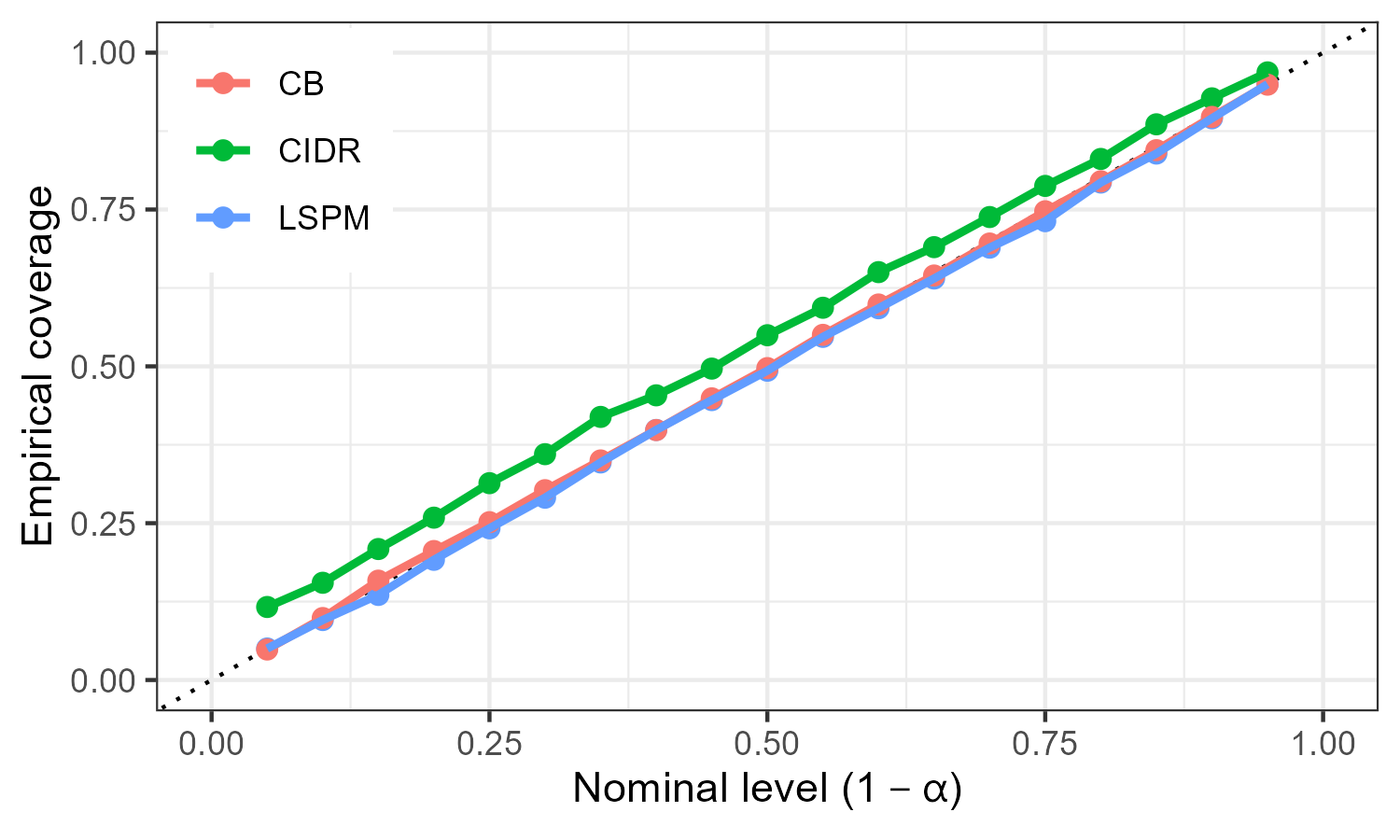}
    \includegraphics[width=0.49\linewidth]{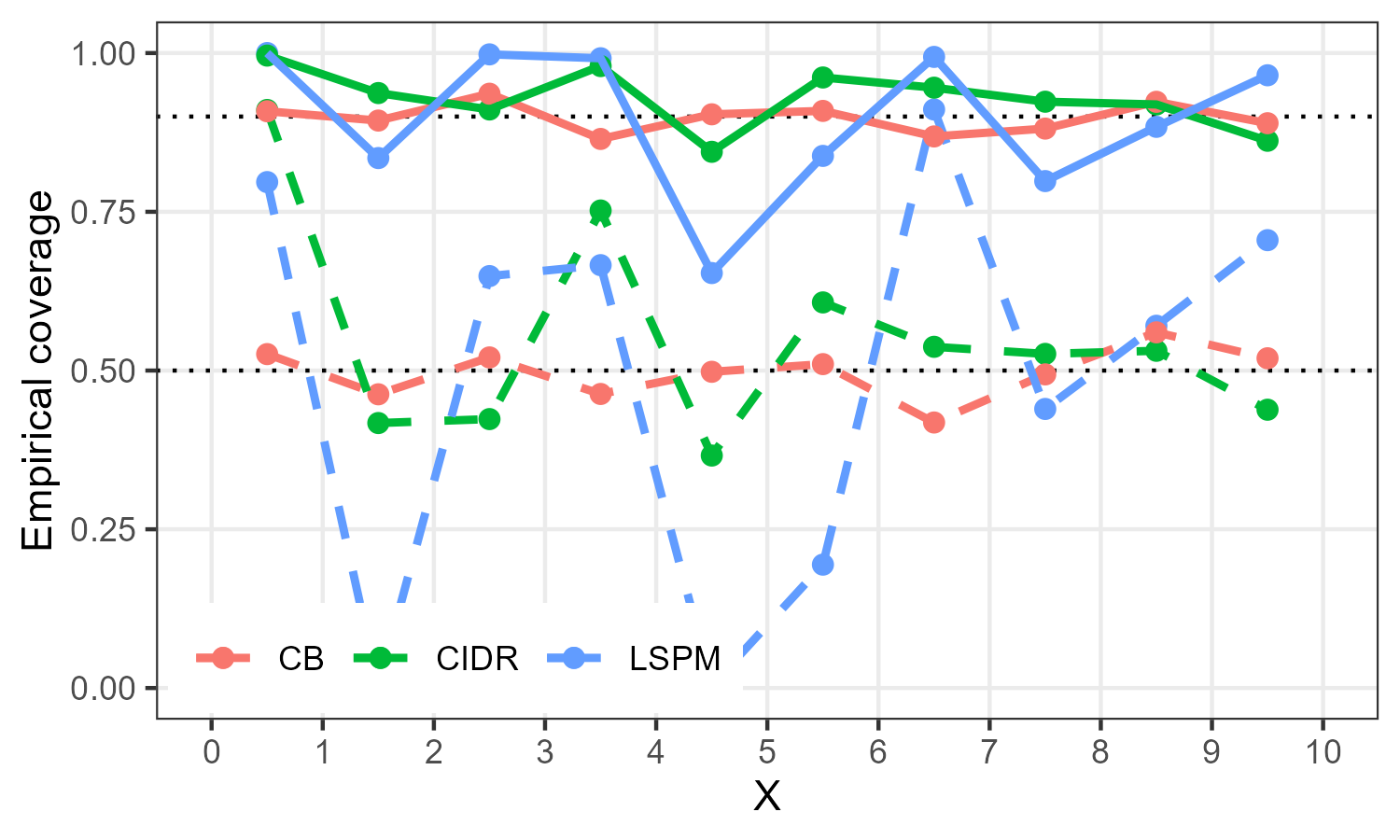}
    \includegraphics[width=0.49\linewidth]{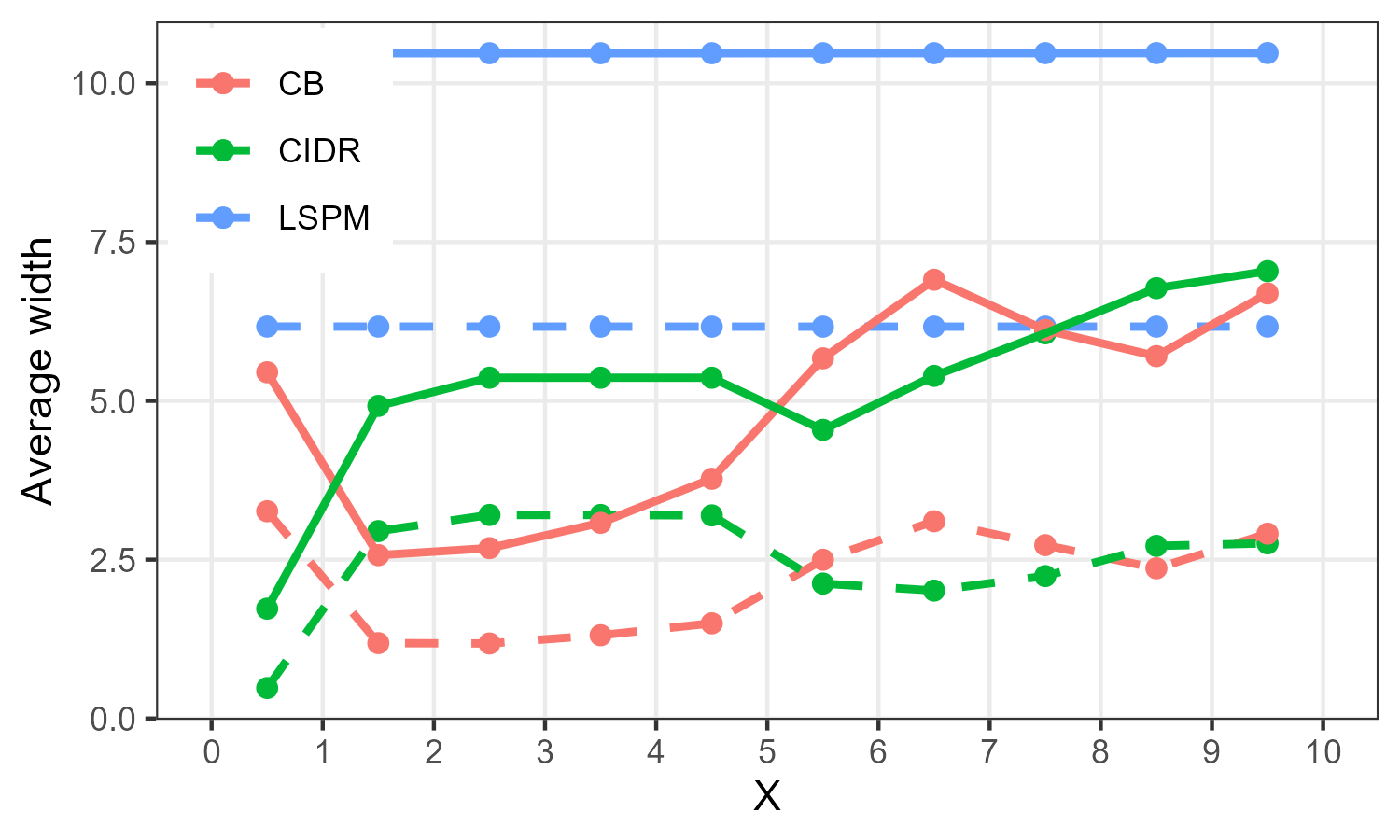}
    \includegraphics[width=0.49\linewidth]{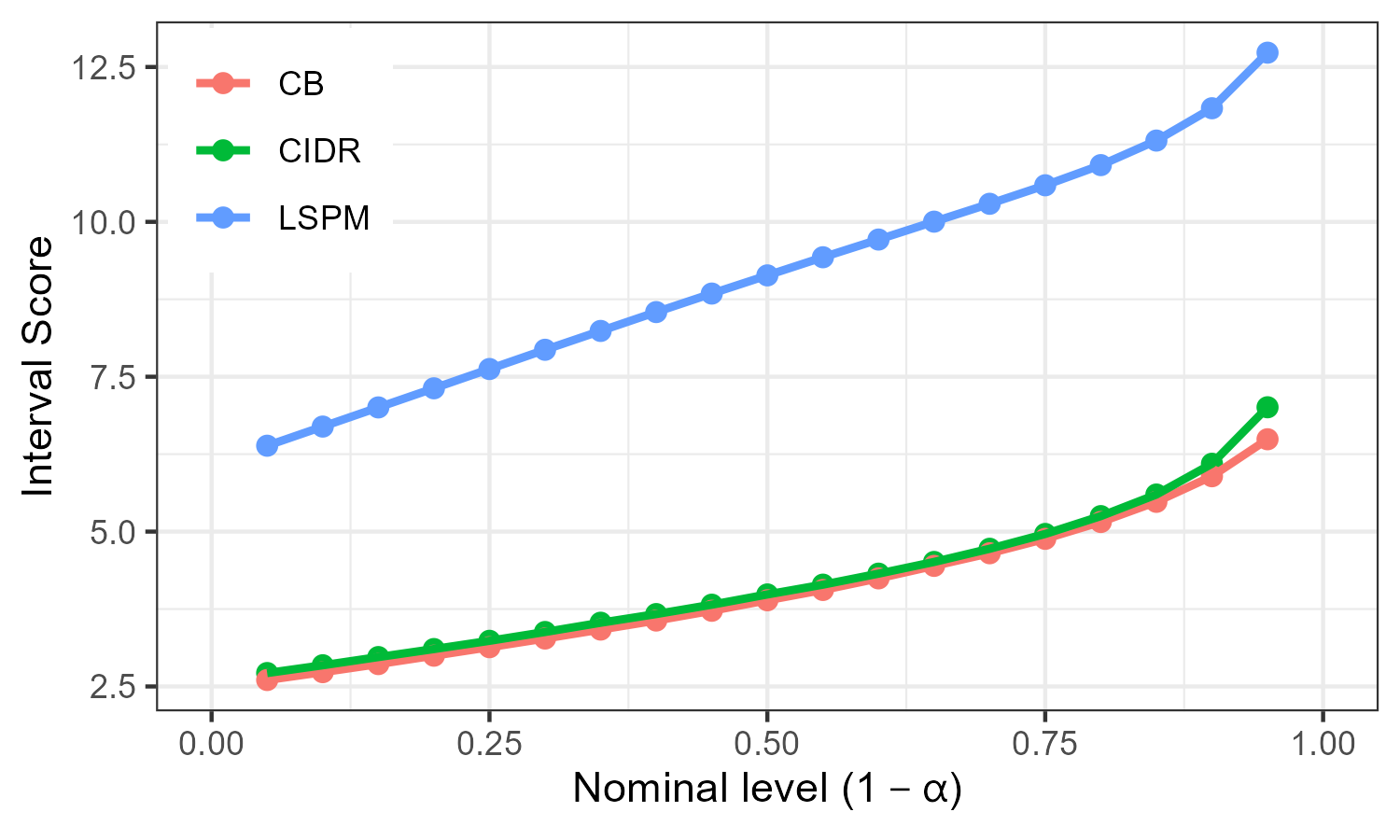}
    \caption{As in Figure \ref{fig:simstudy_cov} but with data generated as at \eqref{eq:simstudy_data_less}.}
    \label{fig:simstudy_le}
\end{figure}

Plots of empirical coverage, average width, and interval score for the prediction intervals generated by conformal binning, conformal IDR, and the LSPM are shown in Figure \ref{fig:simstudy_le}. While the three predictive systems all issue prediction intervals with the correct unconditional coverage, the LSPM prediction intervals are clearly do not satisfy the stronger notion of conditional calibration given the covariate information. Conformal IDR and conformal binning therefore outperform the LSPM when evaluated using the interval score. Since the degree of isotonicity is weaker, conformal IDR also performs marginally worse than conformal binning. Interestingly, the performance of the LSPM also deteriorates when the isotonicity is weakened.

\end{document}